%% file: main.tex
\documentclass[letterpaper, 12pt]{report}

\AtBeginDocument{\addtocontents{toc}{\protect\thispagestyle{headings}}} 
\AtBeginDocument{\addtocontents{lof}{\protect\thispagestyle{headings}}} 
\AtBeginDocument{\addtocontents{lot}{\protect\thispagestyle{headings}}}

\usepackage[english]{babel}

\usepackage{amsmath}

\usepackage{times}

\usepackage{graphicx}

\usepackage{subfig}

\usepackage[table]{xcolor}

\usepackage{pstricks}

\usepackage{setspace}

\usepackage{booktabs}

\usepackage{multirow}

\usepackage{lineno}

\usepackage{array}
\makeatletter
\g@addto@macro{\endtabular}{\rowfont{}}
\makeatother
\newcommand{\rowfonttype}{}
\newcommand{\rowfont}[1]{
   \gdef\rowfonttype{#1}#1%
}
\newcolumntype{L}{>{\rowfonttype}l}
\newcolumntype{C}{>{\rowfonttype}c}

\newcommand{\oic}{(Original In Color).}

\graphicspath{{truck/image/geant_study/}{truck/image/}{truck/image/shms_blueprints/}{truck/image/introduction/}{truck/image/reflectivity/}{truck/image/mirror_selection/}{truck/image/mirror_selection/setup/}{truck/image/mirror_selection/result/}{truck/image/mirror_selection/sample_fit_measurement_plot/}{truck/image/detector_image/}}

\newcommand{\inchsign}{^{\prime\prime}}

\makeatletter
\let\insertdate\@date
\makeatother

\makeatletter
\title{Heavy Gas \v{C}erenkov Construction for Hall C at Thomas Jefferson National Accelerator Facility}

\author{Wenliang Li \\ \\
Supervisor: Dr. Garth Huber}

\begin{document}

\pagestyle{headings}


\pagenumbering{roman}




\input{truck/titlepage.tex}

\doublespace

\chapter*{Abstract}
\thispagestyle{headings}
\input{truck/abstract.tex}

\chapter*{Acknowledgements}
\thispagestyle{headings}
\input{truck/acknowledgements.tex}

\tableofcontents
\listoffigures
\listoftables 


\chapter{Introduction}

\pagenumbering{arabic}
\thispagestyle{headings}



\input{truck/introduction.tex}

\chapter{\v{C}erenkov Radiation}
\thispagestyle{headings}
\input{truck/cerenkov_radiation.tex}

\chapter{HGC Specification and Design}
\thispagestyle{headings}
\input{truck/shms_hgc.tex}


\chapter{Mirror Selection}
\thispagestyle{headings}
\input{truck/mirror_selection.tex}

\chapter{Mirror Reflectivity Measurement}
\thispagestyle{headings}
\input{truck/reflectivity.tex}

\chapter{Geant Simulation}
\thispagestyle{headings}
\input{truck/g4_simulation.tex}

\chapter{Summary and Future Remarks}
\thispagestyle{headings}
\input{truck/summary.tex}

\input{truck/bibliography.tex}

\appendix
\thispagestyle{headings}
\input{truck/appendix.tex}

\end{document}

%% file: truck/titlepage.tex
{

\begin{titlepage}
\begin{center}
\thispagestyle{empty}
\renewcommand{\baselinestretch}{0}%
\textrm{\LARGE Heavy Gas Cherenkov Detector Construction\\[5mm]for Hall C at \\[5mm] 
Thomas Jefferson National Accelerator Facility\\ } 
\vspace{1.5cm}
\renewcommand{\baselinestretch}{0}%
\textrm{\Large A Thesis \\[5mm]
Submitted to the Faculty of Graduate Studies and Research \\[5mm]
In Partial Fulfilment of the Requirements \\[5mm]
for the Degree of \\[5mm]
Master of Science \\[5mm]
in Physics \\[5mm]
University of Regina}\\
\vspace{4cm}
\textrm{\large By \\[5mm]
Wenliang Li \\[5mm]
Regina, Saskatchewan\\[5mm]
\insertdate\\[1cm]
\copyright 2012: Wenliang Li
}
\end{center}
\end{titlepage}

}

%% file: truck/abstract.tex
{

The Thomas Jefferson National Accelerator Facility (JLab) has undertaken the 12 GeV Upgrade to double the accelerating energy of its electron beam. This attracts many interesting proposals to probe the quark-gluon nature of nuclear matter at higher energy, therefore a new set of equipment is required. Experimental Hall C of JLab has planned to construct a new Super High Momentum Spectrometer (SHMS) to replace the existing Short Orbit Spectrometer (SOS). The University of Regina is assigned to construct the Heavy Gas \v{C}erenkov (HGC) Detector as part of the SHMS focal plane detectors. This detector will be used as critical component to provide reliable $\pi/K$ separation between 3-11 GeV/c central momenta in the SHMS experimental program. In this thesis, we will report the design, quality control studies and simulated expected performances of the HGC detector.

}

%% file: truck/acknowledgements.tex
{

I would like to thank my supervisor Garth Huber for everything. I could not wish for a better supervisor and am extremely excited about the opportunity to continue my study under his meticulous and patient guidance.  

It was such a pleasure to work with Keith Wolbaum, Derek Gervais, Lee Sichello, Paul Selles and Alex Fischer as a member of the HGC detector construction team. Collaborators such as Donal Day and Tanja Horn gave much generous assistance regarding the detector construction and Geant4 simulation. I would like to acknowledge Chris Ng at the University of Alberta for his brilliant effort in completing the detector construction drawing. I am very grateful for the detector construction fund and research assistantships provided by NSERC of Canada, teaching assistantships provided by the Department of Physics, and conference travel support provided in part by the Faculty of Graduate Studies and Research.

My appreciation to the Hall C scientists: Howard Fenker, Dave Gaskell, Mark Jones, Dave Mack, Steve Wood, and engineers: Michael Fowler, Steve Furches, Steve Lassiter, Bert Metzger, Eric Sun, for their fantastic support. The permanent reflectivity setup at Jefferson Lab was a successful collaborative work between the University of Regina and Jefferson Lab. My gratitude goes to Brad Sawatzky, Michelle Shinn, Carl Zorn, and other Detector group and FEL staffs.

Special thanks to all the faculty members and graduate students of the Department of Physics at the University of Regina. Their suggestions and criticisms have helped tremendously to achieve the final goal. Many thanks to the hospitality of the Morrison family and their relatives. 

I would like to thank my family and my friends for the support they provided, in particular Miss Fei Yang, for her understanding and encouragement.

}

%% file: truck/introduction.tex
{
\label{introduction}

The purpose of this thesis is to describe the design, mirror quality control and projected performance of the Heavy Gas \v{C}erenkov (HGC) detector constructed at the University of Regina. The HGC detector is an important part of the 12~GeV upgrade installation, used for particle identification in experimental Hall C at the Thomas Jefferson National Accelerator Facility (Jefferson Lab).

In this chapter, a brief description of Jefferson Lab and the 12~GeV upgrade project is given. The Hall C experimental apparatus and its upgrade plan is then described in more detail. Chapter~2 contains a discussion of \v{C}erenkov radiation in terms of classical electrodynamics. Chapter~3 describes the specification and design of the HGC detector. The mirror quality control studies are presented in Chapters~4 and 5. Chapter~4 gives the detailed description on mirror selection methodology and Chapter~5 presents the aluminized mirror reflectivity results. Chapter~6 discusses the computer simulation of the detector performance, and Chapter~7 concludes with a summary and outlook.


\section{CEBAF and the 12~GeV Upgrade}

The Thomas Jefferson National Accelerator Facility (Jefferson Lab) houses the world's largest superconducting radio frequency (RF) linear accelerator named CEBA (Continuous Electron Beam Accelerator). Its reliable continuous electron beam has become the most effective method to probe the quark-gluon nature of nuclear matter at high energy.

The CEBA consists of an electron injector, a pair of superconducting linear accelerators and several bending arcs. The electrons are generated in the injector, then are grouped into micro-bunches and released into the north Linac at an energy of 45~MeV with 0.667~ns separation, which corresponds an RF repetition rate of 1497~MHz. The north and south Linacs have identical design, each hosting 20 cryomodules with an accelerating gradient of 5~MeV/m. The Linacs are connected by arcs at both ends so that the entire setup looks like a racetrack ring (see Fig. \ref{pic_cebaf}). Each electron may complete as many as five circles in the racetrack ring before reaching the experimental area; the maximum energy gain is 1.2~GeV per pass, thus the maximum energy gain for an electron is 6~GeV. The electron bunches are in turn delivered to each of the three experimental halls every 2~ns. The current setting of CEBA produces 6~GeV maximum energy at with beam current of 200~mA.

Jefferson Lab currently has three experimental halls. Each of the experimental halls have different scientific objectives and therefore are equipped with different experimental apparatus. A short description of each is given below:
\begin{description}
\item[HALL A:] A high resolution hall, equipped with a pair of High Resolution Spectrometers (HRS) which are optimized to study nuclear structure with high precision.
\item[HALL B:] A large solid angle hall, designed to provide full 4$\pi$ solid angle coverage and broad momentum range for capturing produced charged and neutral particles produced in nuclear interactions. 
\item[HALL C:] A multi-purpose high luminosity hall, equipped with a High Momentum Spectrometer (HMS) and a Short Orbit Spectrometer (SOS). The specialty of Hall C is its capability to study rare interactions at high event rate. The probability of the event detection is directly proportional to the maximum beam current, where in Hall C it can reach as high as 180 mA which is 90\% of the total beam current. 
\end{description}

Jefferson Lab is currently undertaking a 12~GeV upgrade project to double its accelerating energy and to improve the instrumentation in each of the experimental halls. The upgrade project will replace three of the existing cryomodules and install five new cryomodules in each of the accelerators and improve the energy gain from 1.2~GeV to 2.2~GeV per pass, thus delivering up to 11~GeV beam to the existing halls. As a result, the maximum electron beam current after the 12~GeV upgrade will be halved to 100~mA to conserve the total power output of the accelerator. A new experimental hall, namely Hall D, will be built to extend the physics program at Jefferson Lab to search principally for exotic hybrid mesons. A new bending arc is installed to guide the electron beam to Hall D. Thus, this hall will make use of 5.5 pass electron beam, with a maximum beam energy of 12~GeV. 

\begin{figure}
  \centering
  \includegraphics[width=0.75\textwidth]{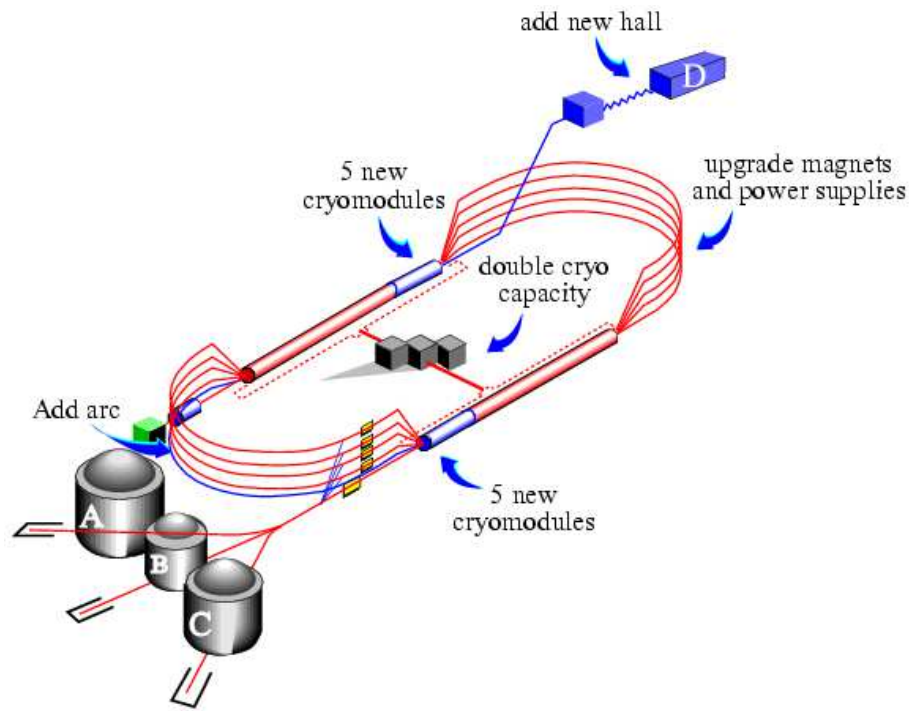}
  \caption[Schematic Diagram of the Jefferson Lab]{Schematic diagram of the Jefferson Lab and the 12~GeV upgrade \cite{jlabpic}. \oic}
  \label{pic_cebaf}
\end{figure}

\section{Physics Program at Hall C After 12 GeV Upgrade}

Hall C will play a vital role in the overall Jefferson Lab physics program after 12 GeV upgrade. The planned physics experiments in Hall C require spectrometers with acceptance for very forward-going particles with momenta approaching that of the incoming beam (11 GeV/c). Their focal plane detectors must provide excellent particle identification even at these high energies. They must be capable of rapid, accurate changes to the kinematic settings with well understood acceptances allowing experiments to efficiently cover broad regions of phase space, enabling, for example, precise L/T separations. And they must possess efficient, highly time-resolved trigger systems and target and data-acquisition systems suitable for running at high luminosity.


These features are essential for studies such as the pion form factor experiment~\cite{fpi}, which requires precise L/T separations. The long-term interest in measuring the charged pion form factor ($F_{\pi}$) is due to the rigorous and unique prediction made by the theory of Quantum Chromodynamics at asymptotic values of Q$^2$, which indicates the energy limitation where valence quarks behave as free particles. The pion is ideal choice for this study because the smaller number of valence quarks in the pion means that the asymptotic regime will be reached at lower values for $F_{\pi}$ than for the nucleon form factors. The high quality, continuous electron beam of Jefferson Lab makes it the only place to seriously pursue these measurements. Other important examples of the Hall C physics program are color transparency~\cite{transparency}, duality~\cite{duality} and nucleon form factor measurements~\cite{fnucleon,fproton}. These experiments will contribute to improve our current understanding of the nucleon structure.

\section{Experimental Hall C}

The current permanent setup of Hall C consists of a High Momentum Spectrometer (HMS) and a Short Orbit Spectrometer (SOS). A Super High Momentum Spectrometer (SHMS) currently under construction will replace the SOS, becoming the standard experimental setup coupled with the HMS after the 12~GeV Upgrade. 

The Hall C target station is located at the fixed central bearing of the two spectrometers. The target ensemble consists of a three-loop cryogenic target stack together with an optics target assembly. The latter is designed for the calibration of the optics of the spectrometers. The target ensemble is mounted inside a vacuum scattering chamber in such a way that the stack of cryogenic cells and optics target can be moved up and down as a whole. The cylindrical scattering chamber has an inner radius of 61.6~cm and a height of 150~cm. The spectrometers are not vacuum-coupled to the scattering chamber.


\begin{figure}[t]
  \centering
  \includegraphics[width=0.75\textwidth]{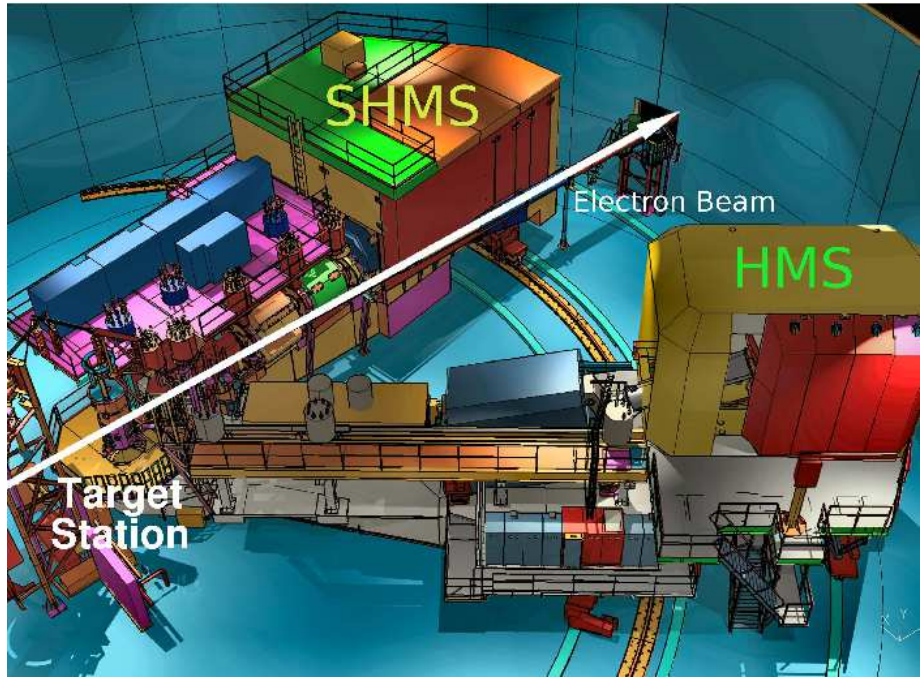}
  \caption[Hall C 12~GeV Overhead View]{Overhead view of standard experiment configuration for Hall C after 12~GeV upgrade. Key components at labeled on the figure \cite{fowler}. \oic}
  \label{pic_12gev_hallc}
\end{figure}

\subsection{Spectrometer Coordination Definition}
\label{coordination}
The coordinate system used throughout the thesis is based on the axes convention for charged particle transport in dispersive magnetic systems. The central ray particles in the SHMS will be bent by 18.4$^{\circ}$ (Table~\ref{tab_hms_shms_specification}) vertically by the superconducting dipole before reaching the detector stack, where the $z$ axis direction will remain parallel to the central ray after the bending, so that the $z$ direction is changed by the dipole bending angle. With a constant dipole magnetic field, the bend angle is smaller for particles with higher momenta assuming the same particle masses. As a result, when particles exit the dipole, they will form an envelope with lower momenta particles on the top quadrants and higher momenta particles on the bottom quadrants. The direction of the $x$ axis follows the increase in particle momenta (low to high) according to the transport convention. Thus, $+x$ points down with respect to the HGC detector frame, which is 18.4$^\circ$ from the vertical axis of the global spectrometer frame. From the direction of the $x$ and $z$ axes. It can be deduced that the $y$ axis ($\vec{y}=\vec{x}\times\vec{y}$) points away from the electron beam line (right to left with respect to the beam envelope after bending).

\subsection{Super High Momentum Spectrometer (SHMS)}

The SHMS is an 18.4$^{\circ}$ vertical bend spectrometer and has a maximum central momentum of 11~GeV/c. The optical design of the SHMS is similar to the HMS and consists of three super-conducting magnetic quadrupoles ($Q1, Q2, Q3$) and one vertically bending dipole ($D$). The purpose of the dipole $D$ is to select the incoming particles depending on their electrical charge and momentum. The purpose of the quadrupoles is to focus the particles into the dipole, improving the angular acceptance (solid angle) of the spectrometer. The combined action of the quadrupoles and dipole is to disperse the charged particles according to momenta at the focal plane, which is located 18.1 m from the target station.

The SHMS horizontal scattering angle is as small as 5.5$^\circ$ with an acceptance of $\pm$1.3$^\circ$, thus the smallest detectable angle is 4.3$^\circ$ from the electron beam path. In order to achieve such a small scattering angle, the SHMS is in addition equipped with a horizontal bending super-conducting magnet ($d$) in front of the first quadrupole $Q1$ and behind of the target station, which can bend most produced changed particles by 3$^\circ$. Thus, the main spectrometer assembly is centered about 8.5$^\circ$. To summarize, the optical configuration of the SHMS is $dQQQD$. The vertical bending angle is 18.4$^\circ$ of the $D$ dipole. The important performance specifications are listed in Table~\ref{tab_hms_shms_specification}. A side view of the SHMS is shown in Fig. \ref{pic_shms}.

Due to the spatial constraints, the vertical bending dipole $D$ extends into the shield house of SHMS. The shield house has an asymmetrical trapezoid shape and is sub-divided into a detector hut and an electronic hut. The detector hut hosts the focal plane detector stack and the electronic hut protects the critical electronic modules used for the data acquisition. The two huts are separated by a 50~cm concrete wall and 2.3 mm of boron/lead; the back and right sides of the shield house are protected by 64~cm concrete and 2.3 mm boron/lead; the top and bottom are protected by 50~cm and 70~cm concrete, respectively, and 2.3mm lead/boron; the left side (close to beam line) has 90~cm concrete and the front side has 100~cm concrete wall. In addition, the left and front sides have 5~cm lead and boron shield. The top shield bricks are removable to allow overhead crane access to the detector and electronic huts. 

\begin{figure}[t]
  \centering
  \includegraphics[width=0.75\textwidth]{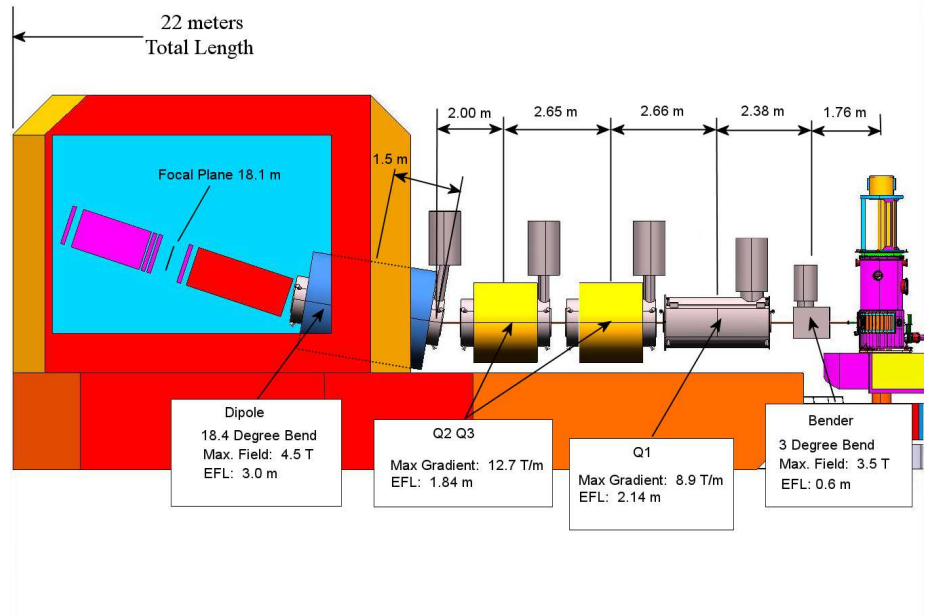}
  \caption[SHMS Side View]{Side view of SHMS: $dQQQD$ superconducting magnets configuration and detector hut geometry \cite{fenker}. The horizontal bending bending dipole $d$ is labeled as `bender' in the schematic. \oic }
  \label{pic_shms}
\end{figure}

\subsection{High Momentum Spectrometer (HMS)}
The HMS has a $QQQD$ superconducting magnet configuration and has been used for experiments in Hall C since 1994. The spectrometer has a 25$^\circ$ bend in vertical direction for the central ray particles. The maximum central momentum is 7.3~GeV/c. The horizontal scattering angle range is between 10.5$^\circ$ and 85$^\circ$ with $\pm$1.576$^\circ$ horizontal acceptance. This corresponds to a smallest measurable angle of 8.92$^\circ$.

All the magnets are supported by a single carriage, which can be moved on rails around a fixed central bearing near the target station. The quadrupoles can be moved as a group along the optical axis ($z$ axis). The shield house is on a separate carriage than the magnets. However, the two carriages are coupled together. Compared to the SHMS, the HMS shield has a rectangular shape with much more space for the detectors. The shield roof is not removable. The left side (away from the beam line) of the shield house is made of concrete strips, which can be removed to allow access to the detector stack. The important performance specifications are listed in Table~\ref{tab_hms_shms_specification}.

%

\begin{table}[t]
\centering
\caption[HMS and SHMS Specification]{HMS and SHMS specification \cite{hallc}.}
\renewcommand{\arraystretch}{1.4}
\begin{tabular}{lcc}
\toprule%
Quantity                       & \multicolumn{2}{c}{Specification}                         \\
		                       & HMS                          & SHMS                        \\
\toprule                
Dipole Bend Angle              & 25$^\circ$                   & 18.4$^\circ$                \\     
Maximum Central Momentum       & 7.4~GeV/c                    & 11~GeV/c                     \\
Focal Length                   & 26.0 m                       & 18.1 m                       \\
Scattering Angular Range       & 10.5$^\circ$ to 85$^\circ$   & 5.5$^\circ$ to 40$^\circ$   \\
Momentum Acceptance            & $\pm$10\%                    & -10\%$<\delta<$+22\%        \\
Momentum Resolution            & $<$0.1\%                     & 0.03\%-0.08\%               \\
Solid Angle Acceptance         & 6.7 msr                      & 4.0 msr                      \\
Horizontal Acceptance          & $\pm$27.5 mrad               & $\pm$ 24 mrad                \\
Vertical Acceptance            & $\pm$70 mrad                 & $\pm$ 40 mrad                \\
Horizontal Resolution          & 0.8 mrad                     & 0.5-1.2 mrad                 \\
Vertical Resolution            & 0.9 mrad                     & 0.3-1.1 mrad                 \\
Target Vertex Length           & $\pm$7 cm                    & $\pm$15 cm                   \\
Target Vertex Reconstruction Accuracy & 1 mm                         & 0.1-0.3 mm                   \\
\bottomrule
\end{tabular}
\label{tab_hms_shms_specification}
\end{table}

\begin{figure}[h]
  \centering
  \includegraphics[width=0.75\textwidth]{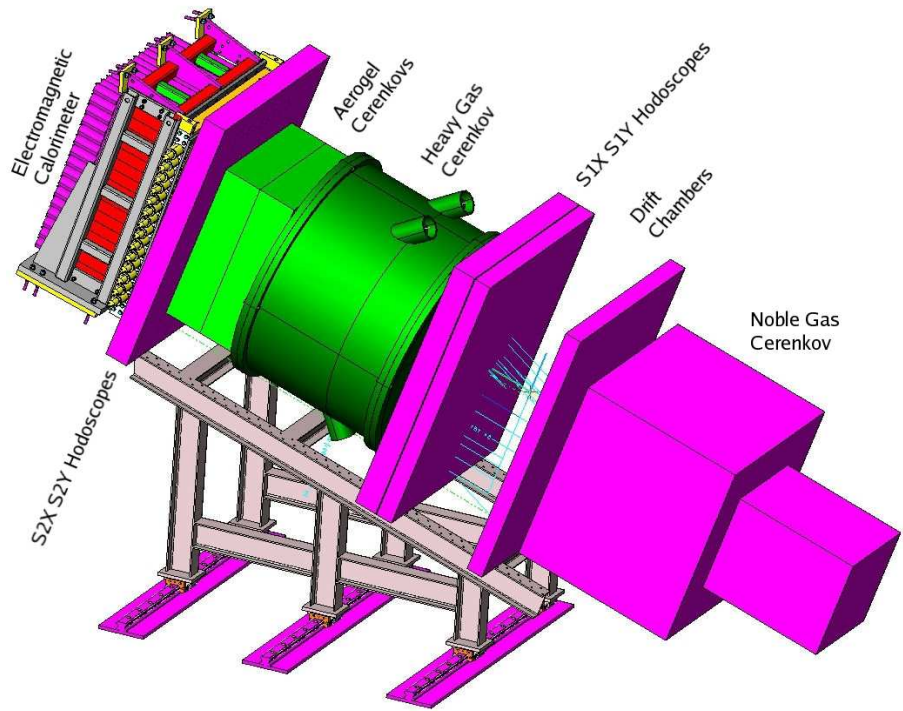}
  \caption[SHMS Detector Stack]{SHMS detector stack layout inside of the detector hut after the superconducting dipole. Charged particles focused by the SHMS magnetic elements traverse the stack from right to left in this figure \cite{metzger}. \oic}
  \label{pic_SHMS_stack}
\end{figure}

\subsection{SHMS and HMS Detector Packages}

The focal plane instrumentation of the SHMS and HMS consists of a series of detectors, which are designed to work together to reliably reconstruct the particle identity and trajectory information of all tracks intercepting them over a broad momentum range. These individual detectors will be described in the order that they are traversed by an incident particle.

\subsubsection{Noble Gas \v{C}erenkov Detector (NGC)}
The noble gas \v{C}erenkov (NGC) detector is only used in the SHMS to separate electrons from heavier charged particles at high central momentum where $p>$ 6~GeV/c. Due to the spatial constraints, it is located in front of the focal plane, whereas ideally it should be placed behind, so as to not adversely affect the reconstructed momentum resolution. For experiments with lower central momenta, the NGC is replaced with a vacuum tank of the same length to eliminate a source of multiple scattering. The detector is filled with Ar gas at momentum $p<$ 5.5~GeV/c and He gas at $p>$ 5.5~GeV/c, with the operating pressure being 1~atm over the full momentum range. The detector housing is unique due to its rectangular body shape and continuous gas circulation is required to eliminate O$_2$ contamination. The \v{C}erenkov radiation is reflected by four curved mirrors and focused onto four PMTs near the top and bottom of the detector. Due to the good UV transmission characteristics of noble gasses, the \v{C}erenkov emission band can reach as low as 140~nm wavelength. Such deep UV light is extremely difficult to detect since the PMT quantum efficiency falls dramatically below 200~nm wavelength. One possibility being explored is to use a wavelength shifting technique to increase the \v{C}erenkov radiation wavelength from the deep UV to a PMT detectable region. The NGC detector is currently being constructed at the University of Virginia.

\subsubsection{Drift Chambers}
The drift chambers are used to measure the horizontal and vertical angles and positions of the charged particles before and after the focal plane, in order to determine their momentum and trajectory. The basic operation principle is as follows: charged particles induce ionization of the gas atoms inside the chamber and the free electrons produced due to the ionization process are captured by sense wires. Good spatial resolution is achieved by measuring the electron drift time. The electric field inside of the chambers needs to be of a very specific configuration, which is achieved by surrounding the sense wires with non-sensed wires at high voltage. The trajectory information of the two chambers is combined to determine the track of the charged particles through the focal plane. 

Both the SHMS and HMS are equipped with a pair of drift chambers. The focal plane is sandwiched in between the chambers. Each drift chamber contains six planes of sense wires. In the SHMS, the wire planes are ordered $u$, $u^\prime$, $x$, $x^\prime$, $v$, $v^\prime$. There are no $y$ planes, and $u$ and $v$ plane wires are at $\pm$30$^\circ$ with respect to $x$ plane. The cell spacing is 1~cm and the position resolution is approximately 200 $\mu$m per plane. As a result, the $y$ resolution of SHMS detector is better than in the HMS. The two chambers are placed at a distance of 40~cm before and after the focal plane, respectively. The SHMS drift chambers are currently under construction at Hampton University.

In the HMS, the wire planes are ordered $x$, $y$, $u$, $v$, $y^\prime$, $x^\prime$. The $x$ and $y$ planes measure the vertical and horizontal track position. $u$ and $v$ plane wires are at $\pm$15$^\circ$ with respect to the $x$ plane. The cell spacing is 1~cm and the position resolution is approximately 150 $\mu$m per plane. The HMS has better resolution in the $x$ direction. The two chambers are placed at a distance of 40~cm before and after the focal plane.

\subsubsection{Hodoscopes}
In the SHMS, the hodoscopes are only used to generate the trigger for the data acquisition system. The S1 hodoscopes (see Fig. \ref{pic_SHMS_stack}) are located before the heavy gas \v{C}erenkov detector and the S2 are positioned after the Aerogel \v{C}erenkov detector. Each hodoscope consists of two planes, each plane is read out by PMTs at both ends. The first plane of each hodoscope is in the horizontal (S1X, S2X), and second plane is in the vertical direction (S1Y, S2Y). S1X, S1Y and S2X are made of plastic scintillator and S2Y is made of quartz. This configuration is used to provide additional confirmation on the particle identification and reduce the background signal. High momentum neutron background from the beam dump or beam pipe has high probability to scatter a proton free, then this secondary proton will generate a strong signal inside the scintillator detectors through ionization process. However, in the quartz hodoscope, the secondary proton will have low probability to generate a \v{C}erenkov signal which exceeds the threshold. The quartz hodoscope consists of 21 quartz bars, each measuring 125~cm $\times$ 5.5~cm $\times $2.5~cm. There is 0.5~cm overlap between each quartz bar and the total active area is 205.5~cm$\times$ 115~cm. Each S1 scintillator hodoscope panel consists of 13 paddles, each measuring 100~cm $\times$ 9.8~cm $\times$ 0.5~cm. The S2 scintillator hodoscope panel consists of 14 paddles of the same dimension. The overlap of the scintillator paddles are 0.5~cm for the S1 and S2 hodoscopes. Scintillation hodoscopes are being constructed at James Madison University and the quartz hodoscope is being constructed at the North Carolina A\&T University.

In the HMS, the order of the planes is reversed. Each of the hodoscope panels in HMS is 1.0~cm thick and 8~cm wide, with 0.5~cm overlap. The HMS hodoscopes serve two purposes: generating the trigger for the data acquisition system and measuring the time-of-flight (TOF) to determine the particle velocity. At low central momenta, the particles' velocity deviates significantly depending on the mass. However, at higher momenta the deviation is much smaller as $v$ approaches the light speed, therefore the difference in the TOF would be less than the hodoscope timing resolution for most particles. For this reason, the TOF is not expected to be very useful in the SHMS. The data acquisition systems in the SHMS and HMS are triggered by the coincidence signal from at least three out of four hodoscope planes. This is to allow the hodoscope inefficiency to be measured.

\subsubsection{Heavy Gas \v{C}erenkov Detector (HGC)}

%

The SHMS HGC detector is used to distinguish charged pions from heavier charged particles. It is filled with C$_4$F$_8$O gas having a refractive index of 1.0014 at a pressure of 1 atm. The detector pressure will vary depending on the central momentum ($p$) setting of the SHMS for the incoming particles: for 3$<p<$7~GeV/c, detector pressure is set to 0.95 atm; for $p>7$~GeV/c the detector pressure is reduced. This will be discussed in more detail in Chapter \ref{sec_hgc}. The \v{C}erenkov radiation is reflected by four curved mirrors and focused onto four PMTs placed on top and bottom of the detector. This detector is being constructed at the University of Regina.

The role of the HMS HGC detector is different from that in the SHMS, where it is used to distinguish electrons from other charged particles and is filled with C$_4$F$_{10}$ gas at a pressure of 0.78 atm. The resulting refractive index of 1.0011 corresponds to a pion threshold of 3~GeV/c. The \v{C}erenkov light is reflected by two parabolic mirrors and focused onto two PMTs on the top and bottom of the detector.

\subsubsection{Aerogel \v{C}erenkov Detector (ACD)}
In the SHMS, the Aerogel \v{C}erenkov Detector (ACD) can be used for pion/kaon ($\pi/K$) separation at low momenta or kaon/proton ($K/p$) separation at high momenta. The aerogel panel dimension is 90~cm $ \times$ 60~cm with a thickness of 5-10~cm and is constructed of small aerogel bricks. Two sets of aerogel panels with different refractive indices ($n$=1.030 and $n$=1.015) will be used for the particle identification at different momenta. The signal readouts are on both sides. The overall detector dimension is 110~cm $\times$ 100~cm $\times$ 30~cm. The detector is being constructed jointly by the Catholic University of America, the University of South Carolina, Florida International University and the Yerevan Physics Institute.

In the HMS, the aerogel is used for pion-proton separation. The aerogel panel dimension is 117~cm $\times$ 67~cm with a thickness of 9.5~cm and is being constructed of 650 small aerogel bricks. The overall detector dimension is 120~cm $\times$ 70~cm $\times$ 34.5~cm.

\subsubsection{Electromagnetic Calorimeter (ECAL)}

The electromagnetic calorimeters are comprised of lead-glass and are used to provide additional confirmation on the electron-hadron identification. The SHMS calorimeter consists of two parts: pre-shower and shower. The pre-shower is constructed using 28 lead glass blocks formerly used in the SOS, each measuring 10~cm$ \times $10~cm$ \times $70~cm. The PMT signal readouts are on both sides. The shower is constructed with 224 lead glass blocks formerly used in the HERMES experiment at DESY, each measuring 9~cm$ \times$ 9~cm$ \times$ 50~cm. The shower blocks are stacked perpendicularly to the pre-shower blocks and the signal readouts are on the backside. The whole SHMS calorimeter is 140~cm wide, 144~cm tall and 60~cm thick; total number of readout channels are 252. The calorimeter will be constructed jointly by the Yerevan Physics Institute and Jefferson Lab.

The HMS calorimeter is built out of blocks of lead glass, which have dimensions 10~cm$ \times$ 10~cm$ \times$ 70~cm. The detection stack is four layers deep and 13 layers tall. The signal readout is only on one side.

}

%% file: truck/cerenkov_radiation.tex
{

\label{chap_cerenkov}

All equations in this chapter expressed are in Gaussian Units.

The interactions of charged particles with material take many forms. In this chapter the properties of \v{C}erenkov radiation are discussed in detail. Several of these properties are highly relevant for the design and projected performance of the HGC detector.

\v{C}erenkov radiation is the electromagnetic radiation emitted by charged particles when their velocity exceeds the light velocity ($c/n$) in a dielectric medium with refractive index $n$. The moving charged particles electrically polarize the molecules inside the medium, which then turn back rapidly to the ground state. The polarization-depolarization process generates an oscillating electric dipole of angular frequency $\omega$, which emits electromagnetic radiation as a result.

To determine the energy loss between the charged particle and the dielectric medium, the fields in the medium are calculated assuming that the mass and charge of the atoms are uniformly and continuously distributed with a macroscopic relative permittivity $\epsilon_r(\omega)$. $\epsilon_r(\omega)$ can be defined as 
$$\epsilon_r(\omega) = \frac{\epsilon(\omega)}{\epsilon_0}$$
where $\epsilon(\omega)$ is the absolute permittivity, the measure of the resistance that is encountered when an electric field forms in a medium, and $\epsilon_0$ is the vacuum permittivity.

The problem of finding the electric field in the medium due to the incident fast particle moving with the constant velocity can be solved by Fourier transforms. If the potential component $A_{\mu}(x)$ and the source density component $J_{\mu}(x)$ are transformed in space and time according to the general rule:
\begin{equation}
F(\vec{x}, t) = \frac{1}{(2\pi)^2} \int d^3k \int d \omega\, F(\vec{k}, \omega) \, e^{\,i \vec{k} \cdot \vec{x} - i \omega t} \,
\end{equation}
where $k$ is the wavenumber. The transformed wave equations become
\begin{equation}
\begin{array}{l}
\displaystyle \left[ k^2 - \frac{\omega^2}{c^2} \epsilon_r (\omega) \right] \Phi(\vec{k}, \omega) = \frac{4 \, \pi}{\epsilon_r (\omega)} \, \rho(\vec{k}, \omega) \\[5mm]
\displaystyle \left[ k^2 - \frac{\omega^2}{c^2} \epsilon_r (\omega) \right] \vec{A}(\vec{k}, \omega) = \frac{4 \, \pi}{c} \,  \vec{J}(k, \omega) \,.
\label{eqn_wave_equation}
\end{array}
\end{equation}
where the $\Phi$ is the scalar potential, $\rho$ is the change density, $\vec{A}$ is the vector potential and $\vec{J}$ is the current density. The Fourier transforms of 
\begin{equation} 
\begin{array}{l}
\displaystyle \rho (\vec{x}, t) = ze \, \delta(\vec{x} - \vec{v} t) \\ [5mm]
\displaystyle \vec{J}(\vec{x}, t) = \vec{v} \, \rho ( \vec{x}, t)\,.
\end{array} 
\end{equation}
are readily found to be 
\begin{equation} 
\begin{array}{l}
\displaystyle \rho (\vec{k}, \omega) = \frac{ze}{2 \pi} \, \delta(\omega - \vec{k} \cdot \vec{v}) \\ [5mm]
\displaystyle \vec{J}(\vec{k}, \omega) = \vec{v} \, \rho ( \vec{k}, \omega)\,
\end{array} 
\end{equation}
where $\vec{v}$ is the velocity of the charged particle.

From (\ref{eqn_wave_equation}), the Fourier transforms of the potentials are
\begin{equation}
\begin{array}{l}
\displaystyle \Phi(\vec{k}, \omega) = \frac{2ze}{\epsilon_r(\omega)} \, \frac{\delta \, (\omega - \vec{k} \cdot \vec{v})}{k^2 - \frac{\omega^2}{c^2} \, \epsilon_r(\omega)}\\ [5mm]
\displaystyle \vec{A}(\vec{v}, \omega) = \epsilon_r(\omega) \, \frac{\vec{v}}{c} \, \Phi (\vec{k}, \omega)
\end{array}
\label{eqn_phi_a}
\end{equation}
The electromagnetic fields in terms of the potentials are
\begin{equation}
\begin{array}{l}
\displaystyle \vec{E} = -\overrightarrow{\nabla} \Phi - \frac{\partial \vec{A}}{\partial t}  \\
\displaystyle \vec{B} =  \overrightarrow{\nabla} \times \vec{A}\,,
\end{array}
\end{equation}
substitute (\ref{eqn_phi_a}) and obtain:  
\begin{equation}
\begin{array}{l}
\displaystyle \vec{E}(\vec{k}, \omega) = i \left[ \frac{\omega\epsilon_r(\omega)}{c} \, \frac{\vec{v}}{c}  -\vec{k} \right] \Phi (\vec{k}, \omega) \\[5mm]
\displaystyle \vec{B}(\vec{k}, \omega) = i \, \epsilon_r(\omega) \vec{k} \times \frac{\vec{v}}{c} \, \Phi(\vec{k}, \omega) \,.
\end{array}
\label{eqn_Ekw}
\end{equation}

In calculating the energy loss, the Fourier transform in time of the electromagnetic fields is calculated at a perpendicular distance $b$ from the path of particle moving along $z$ axis. The required electric field is of the form:
\begin{equation}
\vec{E} = \frac{1}{(2\pi)^{3/2}} \int d^3k \, \vec{E} (\vec{k}, \omega) \, e^{ibk_2}\,,
\label{eqn_Ew}
\end{equation}
where the observation point has coordinates (0, $b$, 0). Substitute (\ref{eqn_phi_a}) and (\ref{eqn_Ekw}) into (\ref{eqn_Ew}), and integrate over first component $dk_1$ to obtain:

\begin{equation}
\begin{array}{lcl}
\displaystyle E_{1}(\omega) &=& \displaystyle  -\frac{ize\omega}{\sqrt{2 \pi} \, v^2} \left[ \frac{1}{\epsilon_r(\omega)} - \beta^2 \right] \, \int_{-\infty}^\infty \! \, \, \frac{e^{i b k_2}}{(\lambda^2 + k_2^2)^{1/2} } \, dk_2 \\ [5mm]
\displaystyle		        &=& \displaystyle   -\frac{izew}{v^2} \left(\frac{2}{\pi}\right)^{1/2} \left[ \frac{1}{\epsilon_r(\omega) - \beta^2}\right] K_0(\lambda b) \\ [5mm]
\displaystyle E_{2}(\omega) &=&  \displaystyle  \frac{ze}{v} \, \left( \frac{2}{\pi} \right) ^{1/2} \frac{\lambda}{\epsilon_r(\omega)} \, K_1(\lambda b)\\ [5mm]
\displaystyle B_{3}(\omega) &=&  \displaystyle  \epsilon_r(\omega) \, \beta \, E_{2} (\omega)
\end{array}
\label{eqn_real_field}
\end{equation}
where $\beta^2 = v^2/c^2$, $K_0(\lambda b)$ and $K_{1}(\lambda b)$ are modified Bessel functions. $\lambda^2$ can be written as
\begin{equation}
\lambda^2 = \frac{\omega^2}{v^2} - \frac{\omega^2}{c^2}\epsilon_r(\omega) = \frac{\omega^2}{v^2} [1-\beta \epsilon_r(\omega)] \,.
\label{eqn_lambda}
\end{equation}

If we apply the far field approximation, where $|\lambda b| \gg 1$, the Bessel functions can be estimated by their asymptotic forms, and the fields (\ref{eqn_real_field}) become
\begin{equation}
\begin{array}{l}
\displaystyle E_{1}(\omega ,b) \rightarrow i \, \frac{ze\omega}{c^{2}} \left[ 1-\frac{1}{\beta^2\,\epsilon_r(\omega) }\right] \, \frac{e^{-\lambda b}}{\sqrt{\lambda b}} \\ [5mm]
\displaystyle E_{2}(\omega, b) \rightarrow \frac{ze}{v\,\epsilon_r(\omega)} \, \sqrt{\frac{\lambda}{b}}\,e^{-\lambda b}\\ [5mm]
\displaystyle B_{3}(\omega, b) \rightarrow \rho \, \epsilon_r(\omega) \, E_{2} (\omega, b) \,.
\label{eqn_farfield}
\end{array}
\end{equation}

The energy deposited per unit length along of the charged particle path is
\begin{equation}
\left( \frac{d\rm{E}}{dx}\right)_{b>a} = -ca~\textrm{Re} \int_0^{\infty} B_3^{\ast}(\omega)E_1(\omega)\,d\omega
\end{equation}
where $a$ is the cylinder radius around the path of the incident particle. Take the RHS integral in the far field limit and obtain
\begin{equation}
\left( \frac{d\rm{E}}{dx}\right)_{b\rightarrow\infty} \rightarrow \frac{z^2 e^2}{c^2} \left( -i \sqrt{ \frac{\lambda ^ \ast}{\lambda} } \right) \omega \left[ 1-\frac{1}{\beta\epsilon_r(\omega)}\right] e^{-(\lambda+\lambda^{\ast})a}
\label{eqn_energy_loss}
\end{equation}
where $\lambda^{\ast}$ is the complex conjugate of $\lambda$. The real part of this expression gives the energy deposited far from the path of the particle. If $\lambda$ has a positive real part, as is generally true \cite{Jackson}, the exponential factor in (\ref{eqn_energy_loss}) will cause the expression to vanish rapidly at large distances, therefore all the energy is deposited near the path. This is not true when $\lambda$ is purely imaginary. In this case, the exponential term is 1 and the expression is independent of distance $a$; some energy escapes to infinity as electromagnetic radiation. From (\ref{eqn_lambda}) it can be seen that $\lambda$ can be purely imaginary if $\epsilon_r(\omega)$ is real (no absorption) and $\beta^2\epsilon_r(\omega) > 1$, which can be expressed in the more transparent form 
\begin{equation}
v > \frac{c}{\sqrt{\epsilon_r(\omega)}} = \frac{c}{n(\omega)} \,,
\label{eqn_velocity_condition}
\end{equation}
since the refractive index of a medium can be written as
\begin{equation}
n=\sqrt{\epsilon_r(\omega)\mu_r}
\label{eqn_refractive_index}
\end{equation}
where $\mu_r$ is the relative permeability of the material, and $\mu_r=1$ for most natural existing materials at optical frequencies \cite{Yaroslav}. Equation (\ref{eqn_velocity_condition}) shows that the velocity of the particle must be larger than the phase velocity of the electromagnetic fields at frequency $\omega$ in order to emit  \v{C}erenkov radiation at that frequency. 

To determine the frequency dependence of the \v{C}erenkov radiation, the relative permeability can be expressed as \cite{Jackson}
\begin{equation}
\epsilon_r(\omega) \simeq 1+ \frac{4\pi N e^2}{m} \sum_{j} \, \frac{f_j }{\omega_j^2 - w^2 - i\gamma_j \omega} \,,
\label{eqn_epsilon_r}
\end{equation}
where $N$ is the number of molecules per unit volume, $\omega_j$ is the natural resonant frequency of the molecule, $\gamma_j$ is the damping ratio of electromagnetic field in each molecule and $f_{j}$ is the number of electrons that have the resonant frequency $\omega_j$ and damping ratio $\gamma_j$. By substituting Equation (\ref{eqn_epsilon_r}) into Equation (\ref{eqn_refractive_index}) and applying the Taylor series expansion one obtains 
\begin{equation}
n = \sqrt{\epsilon_r(\omega)} \simeq 1+ \frac{2\pi N e^2}{m} \sum_{j} \, \frac{f_j (\omega_j^2 - \omega^2)}{(\omega_j^2 - \omega^2)^2 + \gamma_j^2 \omega^2} \,.
\label{eqn_index}
\end{equation}

Fig. \ref{pic_n_omega} shows a sketch of the real part of Equation (\ref{eqn_index}) versus $\omega$. The \v{C}erenkov radiation emission band is shaded in red, where $n > \beta^{-1}$. The absorption coefficient ($\alpha$) of the dielectric medium also depends on $\omega$ and can be written as \cite{griffiths}
\begin{equation}
\alpha \simeq \frac{4\pi N e^2\omega^2}{mc} \sum_{j} \, \frac{f_j \gamma_j}{(\omega_j^2 - \omega^2)^2 + \gamma_j ^2 \omega^2} \,.
\end{equation}
The absorption curve will become dominant at $\omega > \omega_1$ as shown in Fig .\ref{pic_n_omega}, where the emitted \v{C}erenkov radiation will be absorbed by the medium
\begin{figure}[t]
\centering
\includegraphics[width=0.95\textwidth]{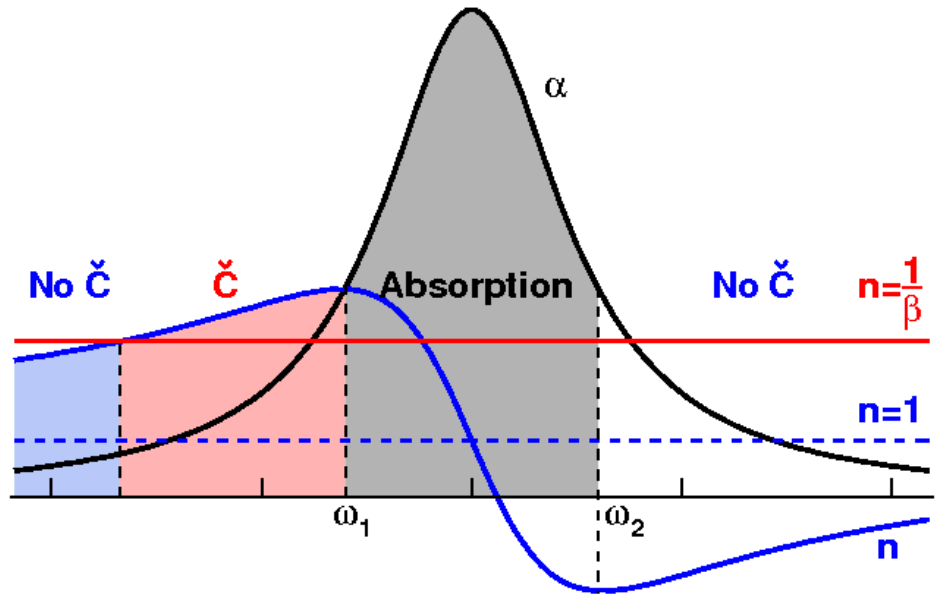}
\caption[Refractive Index and Aborption vs Frequency Sketch]{Sketch of the index of refraction ($n$) and absorption ($\alpha$) versus angular frequency $\omega$. The index of refraction curve is in blue; the absorption curve is in black; the \v{C}erenkov radiation band is shaded in red; the absorption band is shaded in grey; the red solid line indicates $n=\beta^{-1}$; the dashed blue line indicates $n=1$. $\omega_1$ and $\omega_2$ represent the lower and upper boundary for anomalous dispersion region, where the $n$ drops sharply. Note that the anomalous dispersion region coincides with the maximum absorption region. \oic}
\label{pic_n_omega}
\end{figure}

The direction of propagation of the \v{C}erenkov radiation is given by $\vec{E} \times \vec{B}$. The emission angle relative to the velocity of the charged particle is given by:
\begin{equation}
\tan \theta_C = - \frac{E_1}{E_2} \,.
\label{eqn_tan}
\end{equation}
Substitute far fields Equation (\ref{eqn_farfield}) into Equation (\ref{eqn_tan}) to obtain
\begin{equation}
\cos \theta_C = \frac{1}{\beta \sqrt{\epsilon_r(\omega)}} = \frac{1}{\beta n} \,.
\label{eqn_cerenkov_angle}
\end{equation}

Fig. \ref{pic_cerenkov_wavefront} is a sketch of one set of spherical wavelets radiated by a charged particle traveling faster than the speed of light in the medium. For $v > c / \sqrt{\epsilon_r}$, a synchronized electromagnetic ``shock" wavefront occurs, moving in the direction given by the \v{C}erenkov angle. The black circles indicate the expanding wavefronts of electromagnetic radiation emitted at previous times. When the particle travels faster than the speed of light in the medium, the shock fronts coherently add along the \v{C}erenkov cone with angle $\theta_C$.

The energy radiated by the \v{C}erenkov process per unit distance along the path of the charged particle is \cite{Jackson}: 
\begin{equation}
\left( \frac{d\rm{E}}{dx}\right)_{rad} = \frac{(ze)^2}{c^2} \int_{\epsilon_r(\omega)>(1/\beta^2)} \omega \left( 1 - \frac{1}{\beta^2\epsilon_r(\omega)} \right) d \omega \,.
\label{epsilon_r_omega}
\end{equation}

\begin{figure}[t]
\centering
\includegraphics[width=2.9in]{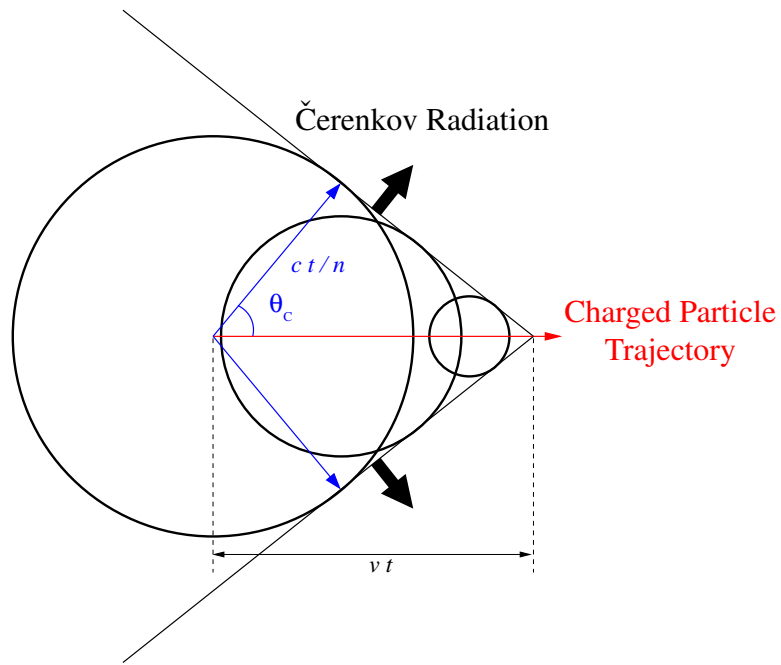}
\begin{pspicture}(0,0)(0,0)
\rput(-2, 7.3){\v{C}erenkov Radiation}
\rput(1.35,4.4){\red 
\begin{tabular}{c} 
Charged Particle \\Trajectory
\end{tabular}
}
\end{pspicture}
\caption[\v{C}erenkov Radiation]{\v{C}erenkov radiation. The black arrows indicate the direction of the emitted \v{C}erenkov radiation wave front; the red arrow indicates the direction of the traveling charged particle in the dielectric medium. The black circles indicate the expanding wavefronts of electromagnetic radiation emitted at previous times. When the particle travels faster than the speed of light in the medium, the shock fronts coherently add along the \v{C}erenkov cone with angle $\theta_C$. \oic}
\label{pic_cerenkov_wavefront}
\end{figure}

}

%% file: truck/shms_hgc.tex
\label{sec_hgc}

\section{Detector Design}

The interaction between the accelerated electrons and hydrogen nuclei inside of the target chamber produce a shower of particles such as: electrons (e), pions ($\pi$), kaons (K), and protons. A small fraction of the produced particles at specific angles and momenta travels through the apertures inside the SHMS quadrupoles and dipoles, and eventually reach the detector hut. 

Following the 12~ GeV upgrade, the charged particles inside of the spectrometers will have higher momenta and their velocities will approach the speed of light, so the TOF will be nearly within the resolving time for all charged particles. The SHMS is designed to detect particles at higher momenta than the HMS, therefore, more particle identification detectors with different refractive indices are required to have reliable particle identification over a wide momentum range. 

The purpose of the Heavy Gas \v{C}erenkov (HGC) detector is to provide good $\pi^\pm$ identification over a momentum range of 3-11~GeV/c. Recall from Chapter \ref{chap_cerenkov} the \v{C}erenkov radiation emission threshold condition
\begin{equation}
v > c/n \, ,
\end{equation}
which can be rewritten in terms of $\beta$ and $n$ as:
\begin{equation}
n>\frac{1}{\beta} \,.
\label{eqn_threshold}
\end{equation}

According to special relativity, the momentum ($p$) and energy ($E$) of relativistic particles are defined as
\begin{eqnarray}
\displaystyle             p&=&\gamma m v     \nonumber   \\ 
\displaystyle             &=&\gamma m \beta c     \nonumber   \\ 
\displaystyle            E&=&\gamma m c^2   \nonumber   \\ 
\displaystyle \frac{p}{E} &=& \frac{\gamma m \beta c}{\gamma m c^2} \nonumber \\ \nonumber \\
\displaystyle \Rightarrow \beta &=& \frac{pc}{E} \,,
\label{eqn_gamma_e_p}
\end{eqnarray}
where $\gamma$ is known as the Lorentz factor, which can be expressed as
$$\gamma = \frac{1}{\sqrt{1- v^2/c^2}} = \frac{1}{\sqrt{1-\beta^2}}\,.$$
Since the energy $E$ can also be written as
\begin{equation}
E= \sqrt{p^2c^2+m^2c^4}\,,
\label{eqn_beta}
\end{equation} 
where $m$ is the mass of the charged particle, expressions (\ref{eqn_threshold}) and (\ref{eqn_gamma_e_p}) become
\begin{eqnarray}
\beta &=& \frac{pc}{\sqrt{p^2c^2 + m^2c^4}} \\
\nonumber \\
n &>& \frac{\sqrt{p^2c^2 + m^2c^4}}{pc}\,.
\label{eqn_n_beta}
\end{eqnarray}

Table~\ref{tab_particle_refractive_index} shows the \v{C}erenkov threshold refractive indices for different particles over the SHMS momentum range (3-11~GeV/c). The $n$ values are directly calculated using Equation (\ref{eqn_n_beta}). For 3~GeV/c momentum $\pi^{\pm}$, the threshold $n$ is 1.00108 and for K is 1.01345. In order to separate them, the \v{C}erenkov material inside the $\pi$ identification detector has to have a refractive index value of 1.00108$<n_{detector}<$1.01345. Since the particle threshold $n$ decreases as the particle central momentum ($p$) increases, the $n_{detector}$ needs to be adjusted accordingly.

\begin{table}[t]
\centering
\caption[Threshold Refractive Indices Table]{Threshold refractive indices at different particles with momenta as listed. The particle masses are listed in the second row of the table header.}
\renewcommand{\arraystretch}{1.4}
\begin{tabular}{LCCCC}
\toprule%
Momentum   &    $n_e$         &  $n_\pi$ &   $n_K$        &  $n_{Proton}$  \\
\rowfont{\footnotesize}%
(GeV/c)    &  (0.5 MeV/c$^2$)  &  (139.57 MeV/$c^2$) &   (493.67 MeV/$c^2$) &  (938.27 MeV/c$^2$) \\
\toprule                                                   
\rowfont{\normalsize}%
  3        &  1.01435    &    1.00108    &    1.01345 &     1.04778 \\
  5        &  1.00519    &    1.00039    &    1.00486 &     1.01745 \\
  7        &  1.00265    &    1.00020    &    1.00248 &     1.00894 \\
  9        &  1.00160    &    1.00012    &    1.00150 &     1.00542 \\
  11       &  1.00107    &    1.00008    &    1.00101 &     1.00363 \\
\bottomrule
\end{tabular}
\label{tab_particle_refractive_index}
\end{table}

\begin{table}[t]
\centering
\caption[Material Refractive Indices Table]{List of the refractive indices for different materials \cite{Rindex, jochen, c4f10,swaniclci}. STP stands for Standard Temperature Pressure.}
\renewcommand{\arraystretch}{1.4}
\begin{tabular}{llcc}
\toprule%
                         &  Material              & $n$       &   $\rho$        \\
						 &                        &           &   (kg/m$^{-3}$) \\
\toprule                                                        
						 &  Vacuum                & 1.000000  &     0.000       \\
\midrule                                                                       
\multirow{3}{*}{Gases}   &  Air (STP)             & 1.000293  &     1.200       \\
                         &  Helium (STP)	      & 1.000036  &     0.179       \\
                         &  Hydrogen (STP)        & 1.000132  &     0.090       \\
                         &  C$_4$F$_{10}$ (STP)   & 1.001400  &     11.21       \\
                         &  C$_4$F$_8$O (STP)     & 1.001389  &     9.190       \\
\midrule                                                                       
\multirow{2}{*}{Liquids} &  Water (20$^\circ$)    & 1.333000  &     1,000       \\
                         &  Ethanol (20$^\circ$)  & 1.360000  &     789.0       \\
\midrule                                                                       
\multirow{4}{*}{Solids}  &  Ice 	              & 1.309000  &     916.7       \\
                         &  Fused Silica (Quartz) & 1.460000  &     2,203       \\
                         &  Crown Glass  	      & 1.520000  &     2,500       \\
                         &  Diamond 	          & 2.420000  &     3,500       \\
\bottomrule
\end{tabular}
\label{tab_refractive_index_tab}
\end{table}

From Table~\ref{tab_refractive_index_tab}, the desired $n_{detector}$ is similar to the refractive indices for gases. The C$_4$F$_8$O gas at 0$^\circ$C and 1 atmospheric (atm) pressure has $n$ value of 1.001389, this makes it a suitable \v{C}erenkov medium candidate for $\pi$-K separation. Note that C$_4$F$_{10}$ has $n$ value of 1.0014, it is the historically used gas (e.g. HMS \v{C}erernkov detector). Unfortunately, it has become too expensive, otherwise the HGC detector would use it again, as its $n$ is even better than C$_4$F$_8$O.

The C$_4$F$_8$O is known as octafluorotetra-hydrofuran, it is a form of freon and often referred to as heavy gas. The C$_4$F$_8$O gas density at STP is much higher than normal atmosphere (air), and it has one of the larger $n$ values for a gas material. It has been used widely as a \v{C}erenkov medium and its stability was extensively studied at FermiLab. The conclusion of the study states `no measurable amounts of reaction product were observed between the heavy gas and other materials in the period of 8 years' \cite{c4f8o}, which proves the C$_4$F$_8$O gas is stable.

The connection between $n$ and the gas pressure ($P$) can be written as 
\begin{equation}
P = \frac{(n-1)}{(n_{\,1\,\bf{atm}} -1)}
\label{eqn_n_pressure}
\end{equation}  
where $n_{\,1\,\bf{atm}}$ is the refractive index at 1~atm pressure, and $P$ is pressure measured in atm. By controlling the operating pressure ($P$), $n$ can be adjusted to any desired value up to the vapor pressure of the gas. Note that the vapor pressure of the heavy gas is around 2~atm, above this limit the gas will start to condense into liquid form.

\begin{figure}[t]
	\centering
	\includegraphics[scale=0.5, angle=90]{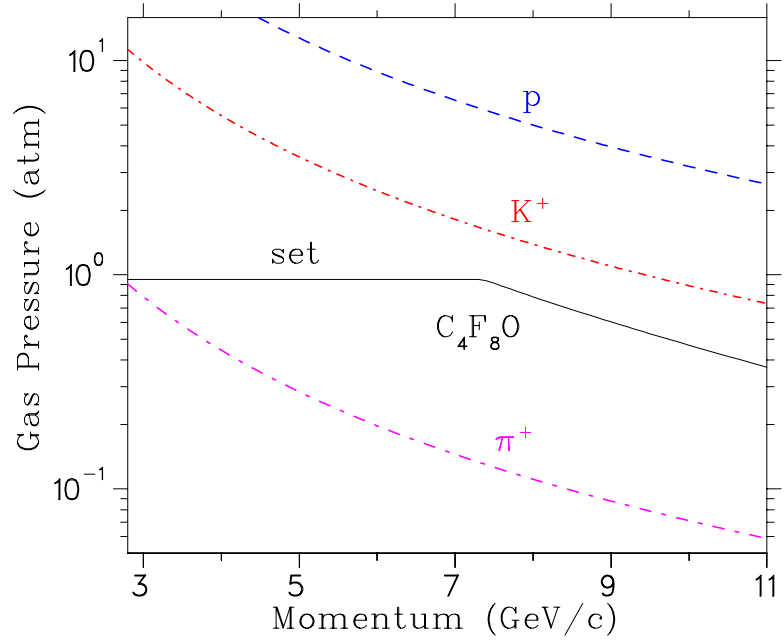}
	\caption[Momentum vs Pressure Plot]{Momentum vs pressure plot. The black curve is the C$_4$F$_8$O gas pressure of the HGC detector. The colored curves represent the \v{Cerenkov} threshold pressure for different particles \cite{garth}. \oic }
	\label{fig_pressure_momentum}
\end{figure}

Fig.~\ref{fig_pressure_momentum} shows the \v{C}erenkov threshold pressure for C$_4$F$_8$O ($P$) vs the particle momentum ($p$). From Equation (\ref{eqn_n_pressure}) $n$ and $P$ are directly proportional, therefore the shape of the $P$ vs $p$ should be same as the $n$ vs $p$ curve. The black curve is our recommended operating pressure of the HGC detector. Between 3 and 7~GeV/c central momentum, the detector is operated at 0.95~atm ($n\approx$1.0001389), which is slightly lower than the atmospheric pressure, to ensure the atmosphere is acting inwards against the vessel. For particles with $p>$ 7~GeV/c, the detector pressure has to be reduced to separate $\pi$ and K since $n$ is reduced.

From Table~\ref{tab_hms_shms_specification}, the SHMS has momentum acceptance of $-13\%<\delta<22\%$, which means that at a central momentum ($p_0$) of 8~GeV/c, the accepted particle momentum can range from 6.96 to 9.76~GeV/c. Therefore, it is possible for a 6.96~GeV/c pion and 9.76~GeV/c Kaon to both be within the SHMS momentum acceptance. The $n_{detector}$ at 8~GeV/c momentum must be set lower than the Kaon threshold $n$ at 9.76~GeV/c to keep this Kaon from generating \v{C}erenkov radiation. In addition, $(n_{detector}-1)$ is scaled down by an extra 10$\%$ as a safety factor (in case the detector pressure is mis-set). At 11~GeV/c momentum, the detector pressure is around 0.35~atm, which implies the HGC detector must be designed as a vacuum vessel to withstand the pressure difference.

%
%
%
%
%

\begin{figure}[t]
	\centering
	\includegraphics[scale=0.5]{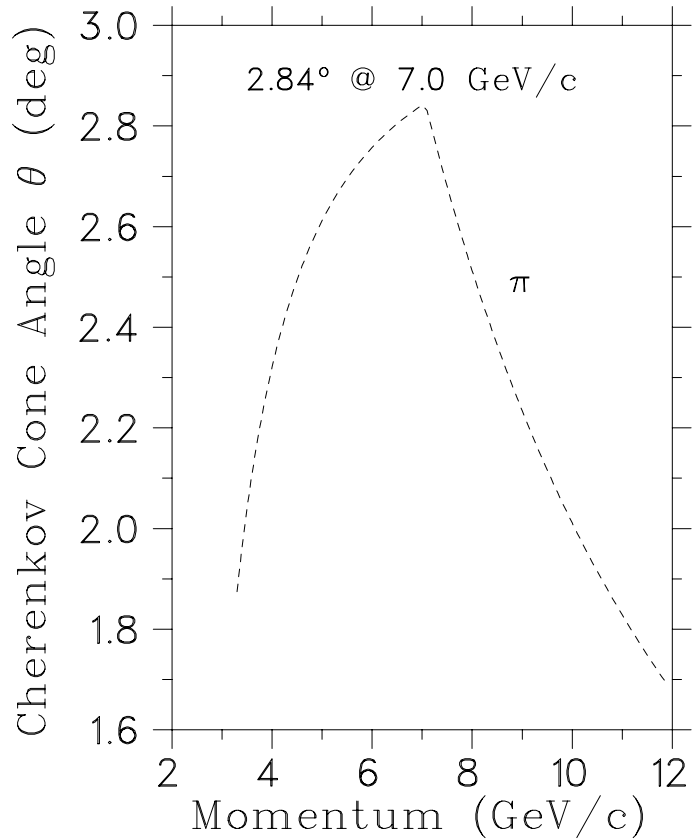}
	\caption[\v{C}erenkov Angle Plot]{\v{C}erenkov angle vs the momentum plot \cite{garth}.}
	\label{fig_angle_momentum}
\end{figure}

\section{\v{C}erenkov Radiation Angle}

The corresponding pion \v{C}erenkov radiation angle $\theta_C$ is calculated using Equation (\ref{eqn_cerenkov_angle}) with $n$ values corresponding to the black curve in Fig.~\ref{fig_pressure_momentum}. Fig.~\ref{fig_angle_momentum} shows the $\theta_C$ vs momentum plot. $n_{detector}$ is kept constant between 3 and 7~GeV/c, therefore $\theta_C$ gradually increases to a maximum of 2.84$^\circ$ at 7~GeV/c; when $p >$ 7~GeV/c, $n_{detector}$ starts to decrease therefore $\theta_C$ also decreases. Since the larger $\theta_C$ results in a larger \v{C}erenkov envelope and more divergent light rays, it is more difficult for the mirrors to focus all of the light onto the PMTs. Therefore, all optical alignment studies were carried out at 7~GeV/c momentum, where the $\theta_C$ is maximum.

\section{Detector Structure}

The HGC detector is the fourth component in the focal plane detector stack of the SHMS behind the Noble Gas \v{C}erenkov, the Drift Chambers and the Hodoscope S1. It is followed by the Areogel \v{C}erenkov, Hodoscope S2 and Lead-glass Calorimeters. The front of the HGC detector is at 18.8~m from the target chamber in optical ($z$) axis from the target chamber and thus 0.7~m downstream of the focal plane. Once all beam envelope clearance and vessel mechanical issues are taken into account, the detector diameter is 1.829 m. Its length is 1.3 m, to allow a sufficient length of \v{C}erenkov radiator gas, and room for the light collection optics to operate efficiently \cite{lee}. Due to the dipole configuration in front of the SHMS detector hut, the HGC detector has 5~cm offset in $+x$ direction, therefore the detector center in 3D is at (5~cm, 0, 19.476 m). The detector vacuum tight specification is 10$^{-9}$~atm and the operating temperature is 20-25$^\circ$C.

\begin{figure}[t]
	\centering
	\includegraphics[width=0.90\textwidth]{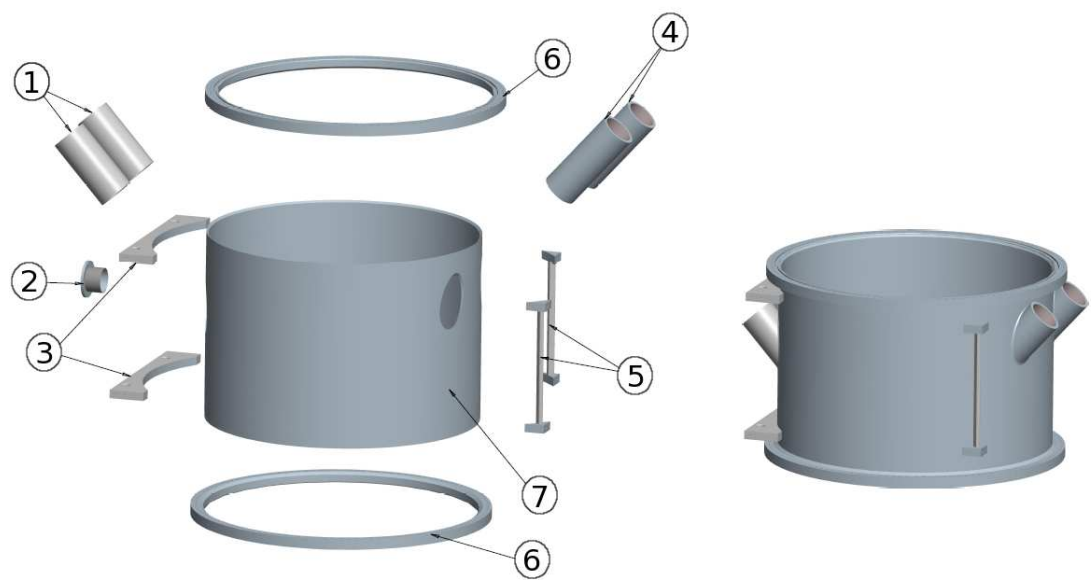}
	\caption[HGC Vessel Structure]{The HGC vessel structure. Components as labelled: 1 and 4. PMT sleeves; 2. Gas port; 3. Cradle brackets; 5. Lifting lugs; 6. Window flanges; 7 Vessel cylinder \cite{ng}. The particles go through the detector vessel from top to bottom in this picture. \oic}
    \label{exploded_tank_blue_prints}
\end{figure}

The HGC vessel structure is shown in Fig.~\ref{exploded_tank_blue_prints}. The components include: vessel cylinder, window flanges, lifting lugs, cradle brackets, gas port and PMT sleeves. The diameter of vessel cylinder is 1.675~m and the wall thickness is 12.7~mm. There are four PMT sleeves on the top and bottom of the vessel: the top two sleeves are tilted at 42$^\circ$ and bottom two are at $-42^{\circ}$ with respect to the vertical axis.

\section{Mirrors}

The dimension of the resultant \v{C}erenkov envelope at 7~GeV/c momentum is 90~cm $\times$ 80~cm in the $x$-$y$ plane at the mirror location, and a detailed investigation was carried out to optimize the mirror-PMT arrangement. In the end, a design of four concave reflecting mirrors with four 5" PMTs was chosen. Each mirror has dimension: 60~cm$\times$55~cm with radius of curvature of 110~cm and thickness of 3~mm \cite{lee}. In order to prevent any possible gaps at the joint location, the mirrors are interleaved in the order of mirror \#: 4, 3, 2, 1, where mirror \#4 is in $-x$, $-y$ quadrant, \#3 is in $-x$, $+y$ quadrant, \#2 is in $+x$, $-y$ quadrant and \#1 is in $+x$, $+y$ quadrant. The closest mirror to mirror approach is between 7-10~mm. There is a 5~cm overlap between mirror \#1 and \#2, the same for mirror \#3 and \#4 in the $y$ direction. The mirrors are manufactured by Sinclair Glass \cite{sinclair} using the slumping method. A detailed description on the mirror slumping and quality control method is given in Chapter \ref{mirror_selection}.

The generated \v{C}erenkov photon wavelength range is between 200 to 600 nm, therefore the mirrors need be coated with aluminum grain to reflect 70\% deep UV (200nm) photons. The mirror aluminization vendor is ECI \cite{eci}, and the reflectivity test results of the sample mirrors are presented in Chapter \ref{reflectivity}.

\begin{figure}[t]
	\centering
	\includegraphics[width=0.4\textwidth]{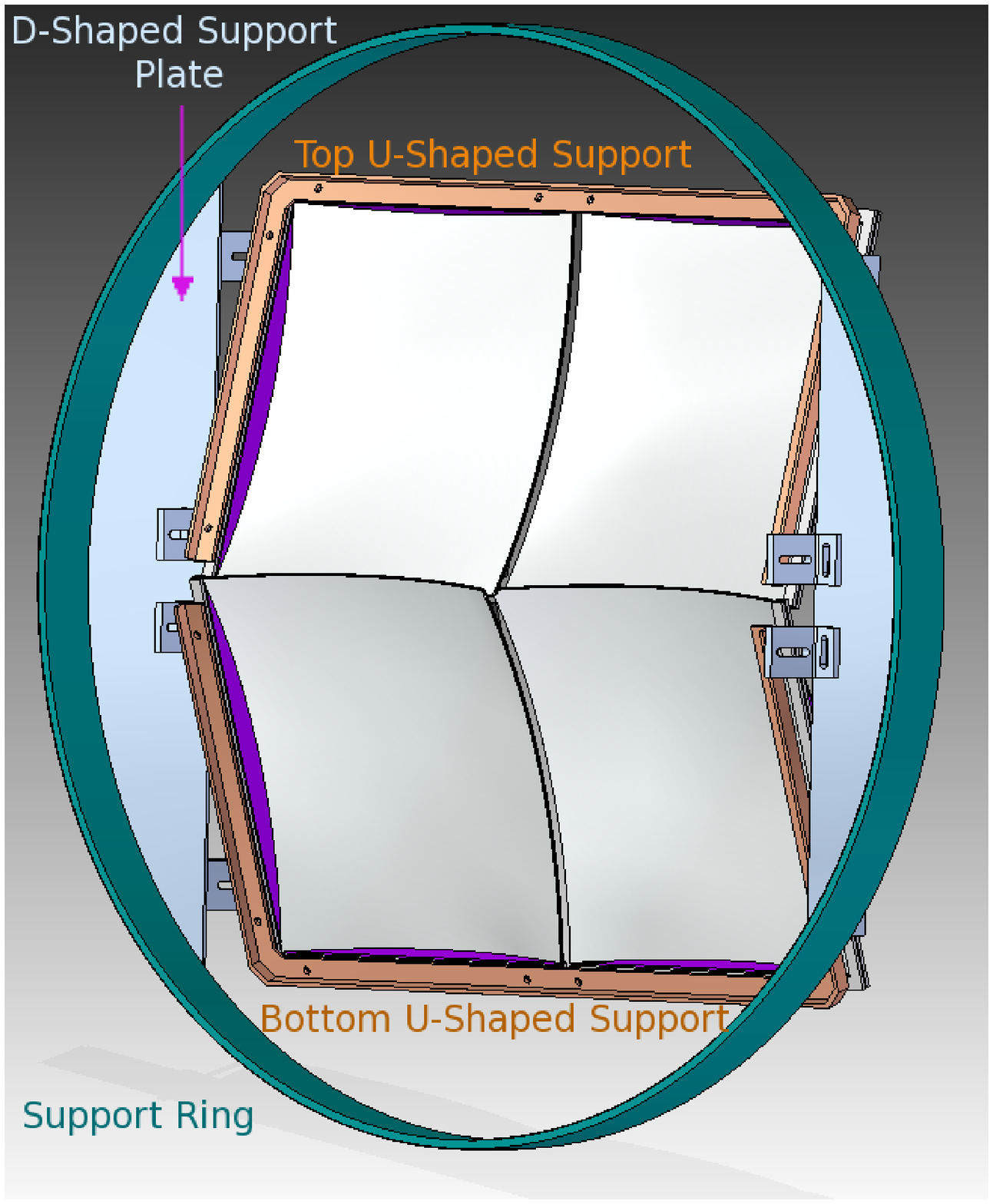}
	\caption[Mirror Support Structure]{Diagram of the mirror support structure \cite{leepic}. \oic}
	\label{cad2}
\end{figure}

\subsection{Window Mounting Scheme}


Each mirror is supported along its two outer edges, with one free corner near the center of the detector overlapping the others, in order to minimize the amount of material inside of the beam path. The supporting edges are sandwiched between flexible gaskets, and then locked into the metal clamps by setscrews. The metal clamps are bolted onto the U-shaped supports which are shown in Fig.~\ref{cad2}. The clamps will have long bolt holes to allow small mirror position and angle adjustments.

\begin{figure}[t]
	\centering
	\includegraphics[width=0.75\textwidth]{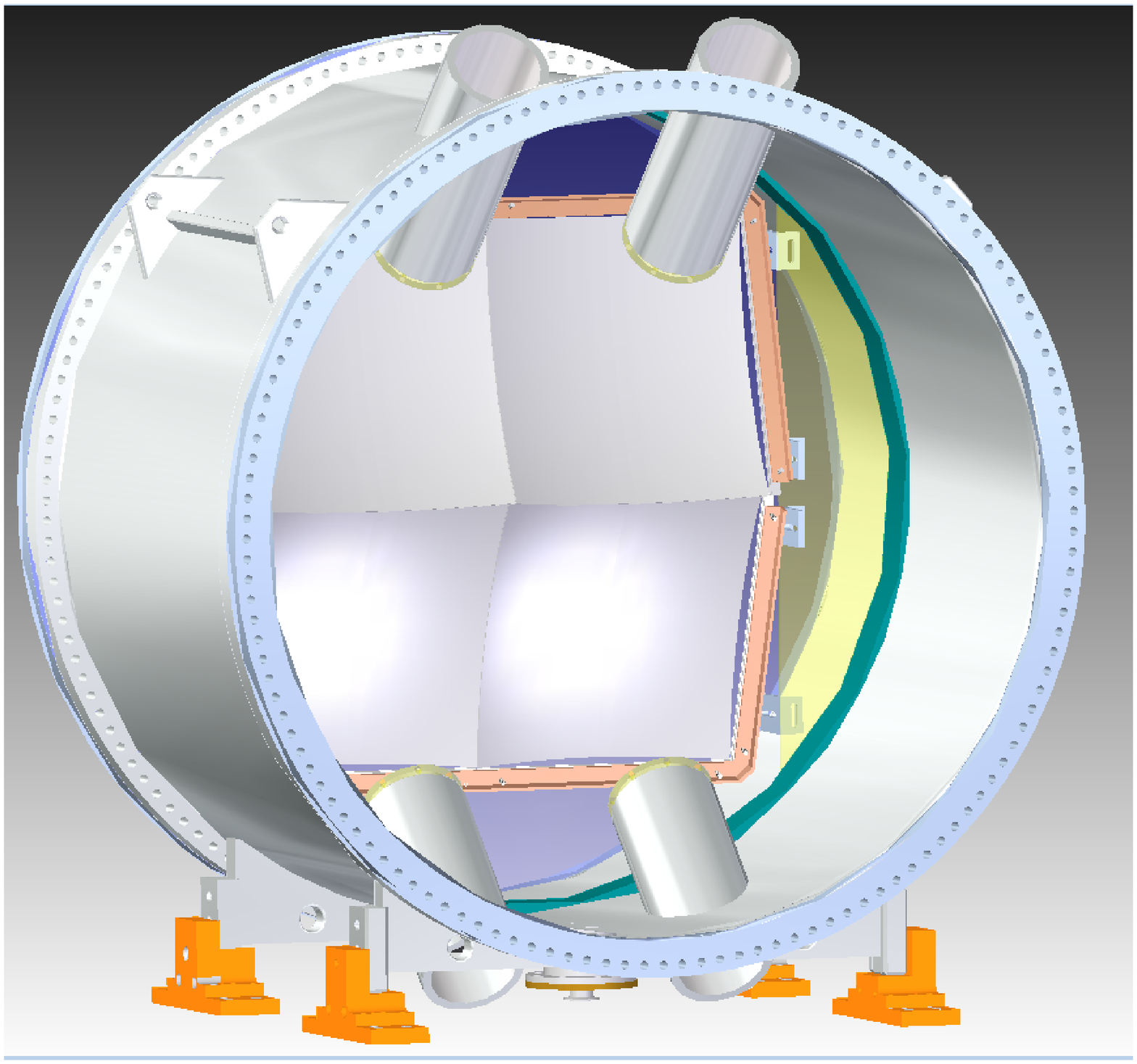}
	\caption[Front View of the HGC Detector]{Front view of the Heavy Gas \v{C}erenkov Detector \cite{leepic}. \oic}
    \label{pic_front_tank}
\end{figure}

A half-inch-thick support ring is introduced to hold the upper and lower U-shaped supports as shown in Fig.~\ref{cad2}; the ring can be bolted onto the inner surface of the vessel. On the edge of the support ring, two D-shaped half-inch-thick plates are attached. Each of the plates has four right angle brackets bolted onto it (two on each side), the U-shaped supports are then bolted to the brackets. The bolt holes on the brackets are milled to allow adjustment on the U-shaped support. A front view of the detector structure, including the vessel cylinder and mirror support ring assembly are shown in Fig. \ref{pic_front_tank}.

Much effort was spent investigating the possibility of using carbon fiber material to pre-form a backing and then glue it onto the back of the mirror. Such a scheme would allow to grip the edges of the carbon fiber backings instead of the mirrors to provide extra strength and stability, since the strength needed to grip the mirrors directly must be gentle and protective. However, this scheme raised the following complications:

\begin{description}
\item[Out-gassing:] During the heavy gas filling process, the HGC detector is pumped down to a pressure of 10$^{-6}$~atm. The out-gassing level of the carbon fiber epoxy needs to be carefully studied to estimate the possibility for contamination. When the carbon fiber backing is glued to the back of a mirror using the epoxy, some air will be trapped inside. It is not obvious how the trapped air bubbles would affect the mirror-backing bond as they expand under a high vacuum condition.
\item[Radiation hardness:] When material is exposed to high level of radiation, its properties often change. Studies are required under intensive radiation environment, to understand the changes in the epoxy properties and mirror-backing bonding strength.
\item[Beam contamination:] The mirror thickness used in HGC detector is 3~mm; the thickness of prototype mirror backing is also 3~mm. This indicates the overall mirror assemble thickness would be 6mm, which would increase beam contamination probability for the down stream detectors. 
\end{description}  
     
For these reasons, we have since decided to support the mirrors with clamps.

\section{PMT Hosting Assembly}
\label{sec_pmt_host_assem}

Each mirror focuses the \v{C}erenkov radiation onto the corresponding 5" PMT and there are four PMTs in total. The PMT front surface is made of UV transparent glass with a radius of curvature of 13~cm, and are manufactured by Hamamastu Photonics \cite{hamamastu}.

To avoid the variable gas pressure causing mechanical strain on the delicate PMTs, they are designed to be outside of the \v{C}erenkov medium, and are housed inside the aluminum sleeves wielded on the top and bottom of the vessel cylinder. Each PMT views the \v{C}erenkov gas enclosure through 1~cm thick quartz windows, and thus requires a quartz adaptor that has flat surface on one side to press against the quartz window and a concave surface on the other side to couple the curved PMT front surface. The center thickness of the adapter is 0.5~cm. In addition, there is a thin (1-2~mm) cookie made of room temperature vulcanizing (RTV) silicone compound to fill the air gaps in between the quartz adaptor and PMT surface. Silicon grease is used to couple the smooth surfaces. 

The quartz windows and adaptors are made of Corning \cite{corning} 7980 quartz, and machined by Hardin Optical Company \cite{hardin}. The RTV silicone compound is produced by Momentive Company \cite{momentive}.

%
%


%

\section{Window Design}

The pressure of the HGC detector is purposely set to be 0.95~atm pressure at low momentum, due to the consideration of the window design. The HGC gas filling scheme is to pump the vessel to 10$^{-6}$~atm pressure and then fill the heavy gas to 0.95~atm pressure. If the detector windows at both ends are flat, the force induced by the pressure difference will deform and destroy the windows, however, if the windows are pre-deformed to the desired curvature, their strength will be increased by an approximate factor of two and they will not deform further due to the pressure difference. The detector must have pressure lower than 1~atm during operation to protect the pre-deformed window curvature. The HGC windows will be made of 1~mm thick 2024-T4 alloy aluminum and have a diameter of 1.829~m. This soft aluminum alloy is chosen to ensure the thin windows will rip slowly, rather than shatter, if they are punctured. The windows will be pre-deformed to a radius of curvature of 4.78~m using the hydro-forming method at the University of Regina.

%% file: truck/mirror_selection.tex
{
\label{mirror_selection}

\section{Introduction}
15 HGC mirrors manufactured by Sinclair Glass \cite{sinclair} were received in June 2011. The specified mirror dimension is: 60~cm $\times$ 55~cm, thickness is 3~mm and radius of curvature of 110~cm.  

The method that Sinclair Glass used to make the mirrors is known as the slumping technique, and its procedure is shown in Fig.~\ref{fig_slumping}. A large steel mold of dimension 1 m $\times$ 1 m $\times$ $40$~cm, was machined to a concaved surface with radius of curvature of 110~cm. Then, the concave surface was coated with a layer of release agent to make sure the mirror would not bond to the mold. A flat soda-lime glass mirror was pre-cut to a rectangular shape but with slight curvature along the edge, so that the square corners and edges could form after slumping. After the flat mirror was gently placed onto the mold, it was moved into an oven where overhead heating was applied to soften the glass. Then the flat mirror would gradually slump towards the pre-formed mold curvature. 

Air holes were drilled at 10~cm separation on the pre-formed surface. All the air hole tunnels are inter-jointed into a main gas outlet, which is connected to a vacuum pump to evacuate the air between the glass and mold. This eliminates air bubbles trapped by the slumping glass, however, some sequential imperfections in the form of small dimples were formed because of the air holes. In order to avoid the imperfections on the mirror back surface, we specified that the HGC mirrors should not be slumped to completely touch the metal mold. The ideal curvature would be slightly parabolic with fewer imperfections.

The level of slumped curvature is dictated by many factors, such as the heating temperature and time, etc. Therefore, individual mirror qualities could differ dramatically. We had to develop systematic methods to understand the quality in term of the radius of curvature and shape for each mirror to select the final detector mirrors. The methodologies and results for coordinate measurement and optical test are given in this chapter. The overall conclusions on mirror quality are drawn using combined results of the two methods.

\begin{figure}[t]
	\centering
    \raisebox{-1.5mm}{\includegraphics[width=0.4\textwidth]{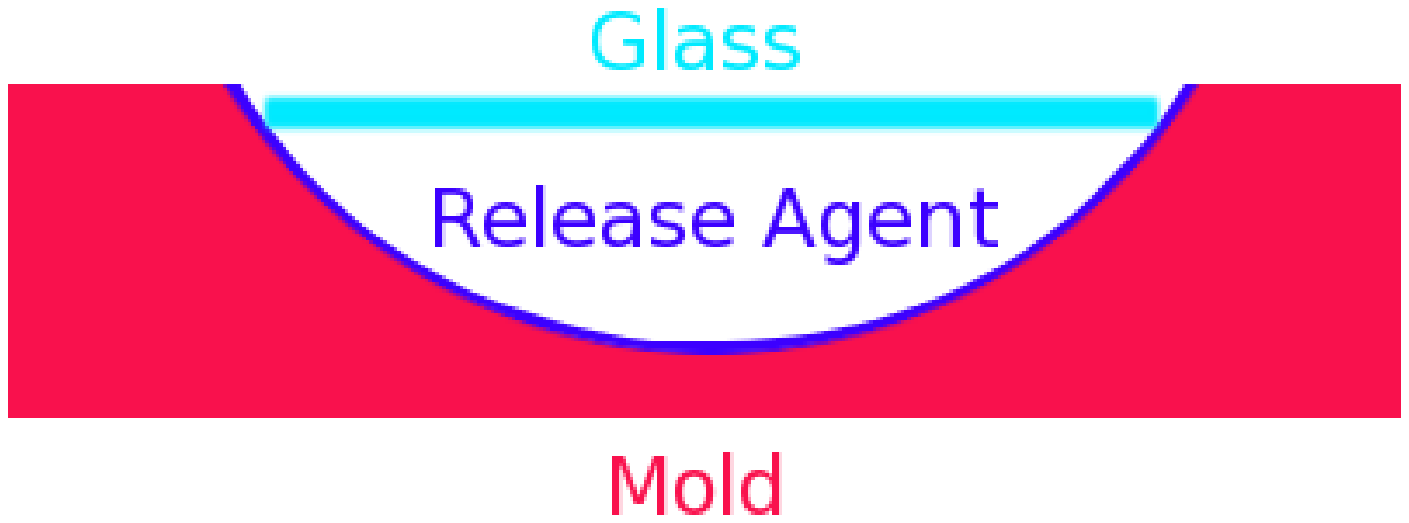}} \raisebox{0.5cm}{\bf \huge$\rightarrow$}
    \includegraphics[width=0.43\textwidth]{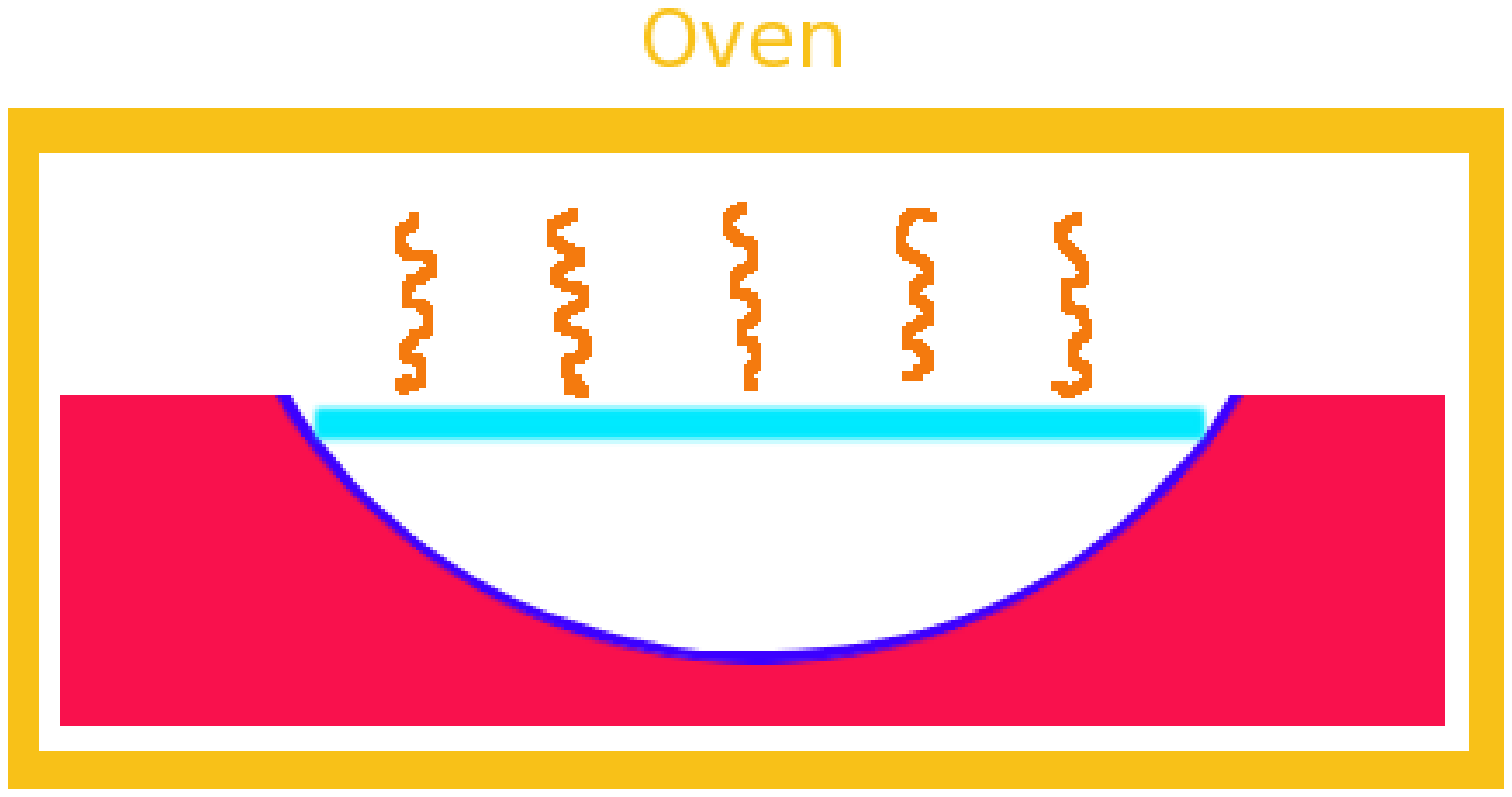}\\ {\quad\qquad\bf \huge$\swarrow$}\\ 
	\includegraphics[width=0.4\textwidth]{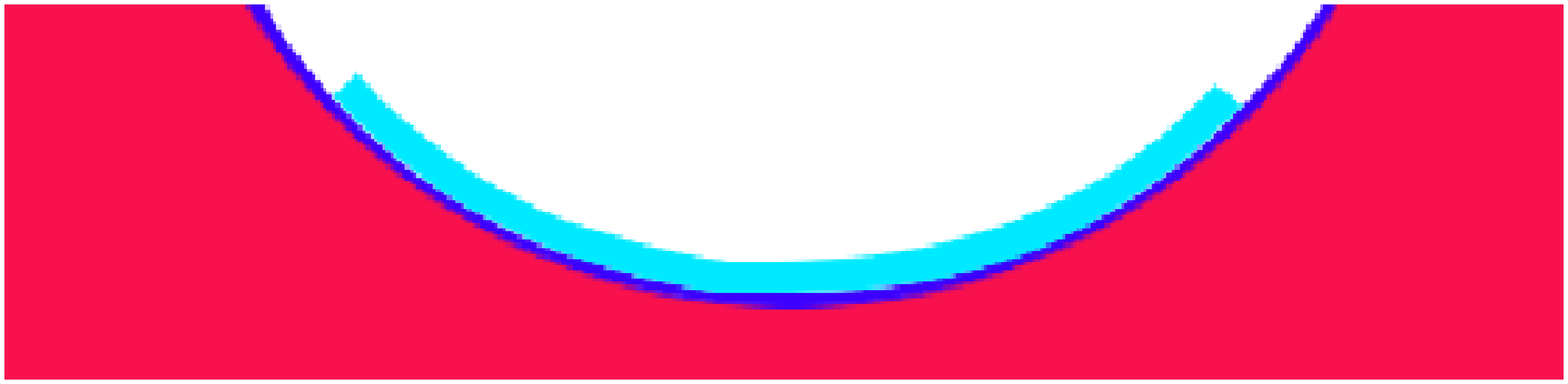}
	\caption[Slumping Method Demonstration]{Slumping method procedure demonstration. \oic} 
	\label{fig_slumping}
\end{figure}



\section{$X$, $Y$, $Z$ Coordinate Measurement}
\subsection{Methodology}
\label{mirror_selection_methodology}
A local company named Dumur \cite{dumur} was hired to map out the $x$, $y$, $z$ coordinates for all mirrors. Their setup is shown in Fig.~\ref{Faroarm} and Fig.~\ref{mirror_place}. The equipment in Fig.~\ref{Faroarm} is called a FaroArm~\cite{faroarm}; it is a high precision 3D coordinate measuring system. During the measurement, the mirror was placed horizontally on top of a grid map and FaroArm sensor was zeroed at a reference corner of the mirror.

The $x$, $y$ grid map contains 20 $x$ direction and 18 $y$ direction measuring points, for a total of 360 points with each measuring point being 3~cm apart in the $x$ and $y$ directions as shown in Fig.~\ref{mirror_place}. The measuring points did not extend to the corners and edges of the mirror and the maximum coverage area was 90\% of the central region. For each measurement, the sensor was manually placed above a measuring point and descended until it gently touched the mirror surface, then the FaroArm system was triggered manually to record the $x$, $y$, $z$ coordinates. Because the sensor was placed manually the grid point spacing was slightly irregular.

\begin{figure}[t]
	\centering
	\includegraphics[width=0.5\textwidth]{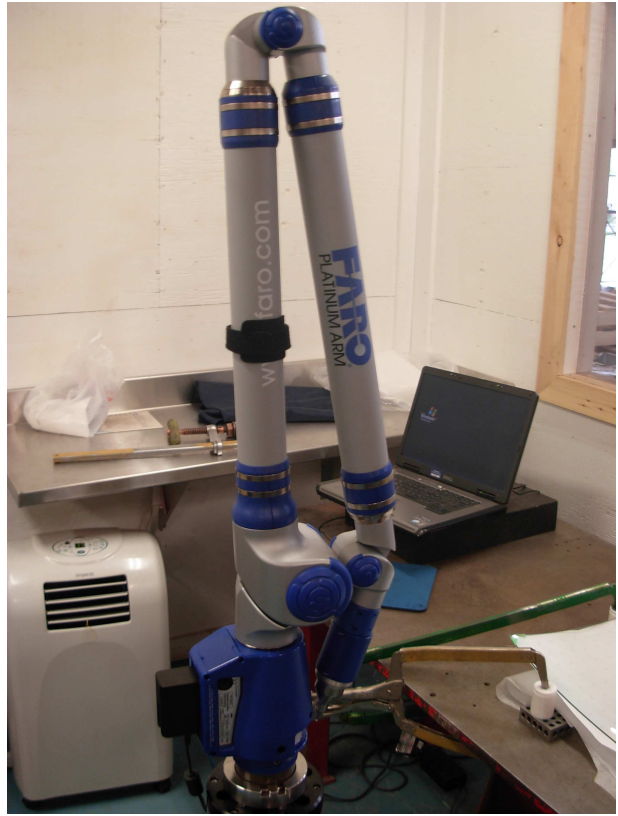}
	\caption[FaroArm Instrument]{Sensor equipment used by Dumur to measure a grid of ($x$, $y$, $z$) coordinates for each mirror. \oic}
	\label{Faroarm}
\end{figure}

\begin{figure}[t]
	\centering
	\includegraphics[width=0.8\textwidth]{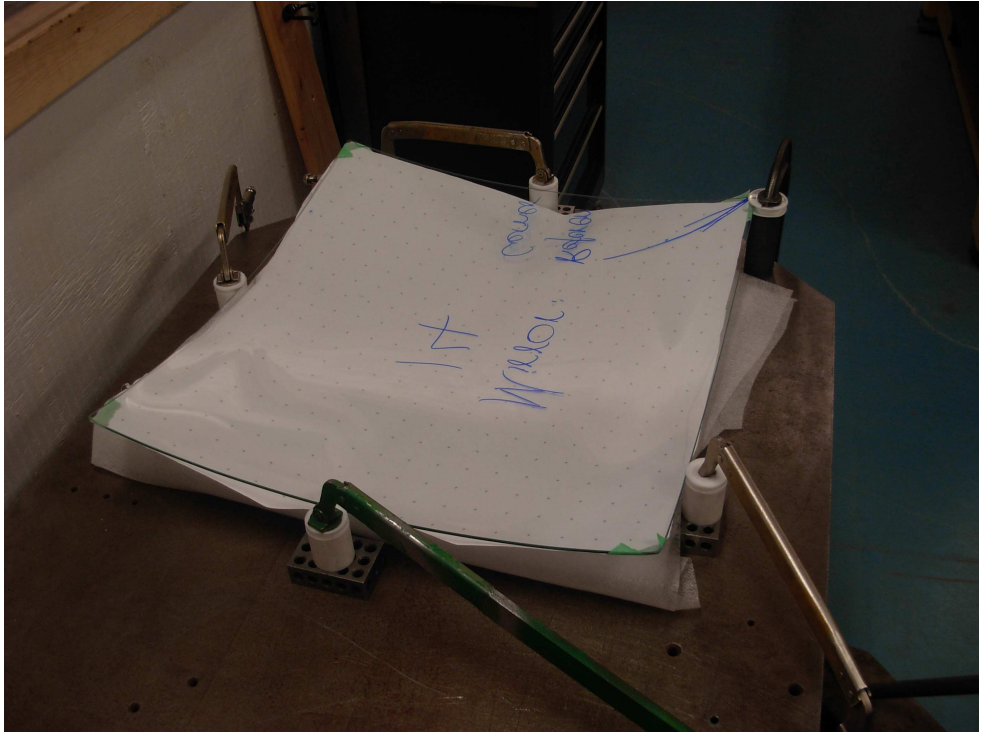}
	\caption[Mirror Placement During Coordinate Measurement]{Mirror placement during the $x$, $y$, $z$ coordinate measurement. \oic}
	\label{mirror_place}
\end{figure}

\begin{figure}[h]
	\centering
	\includegraphics[width=0.9\textwidth]{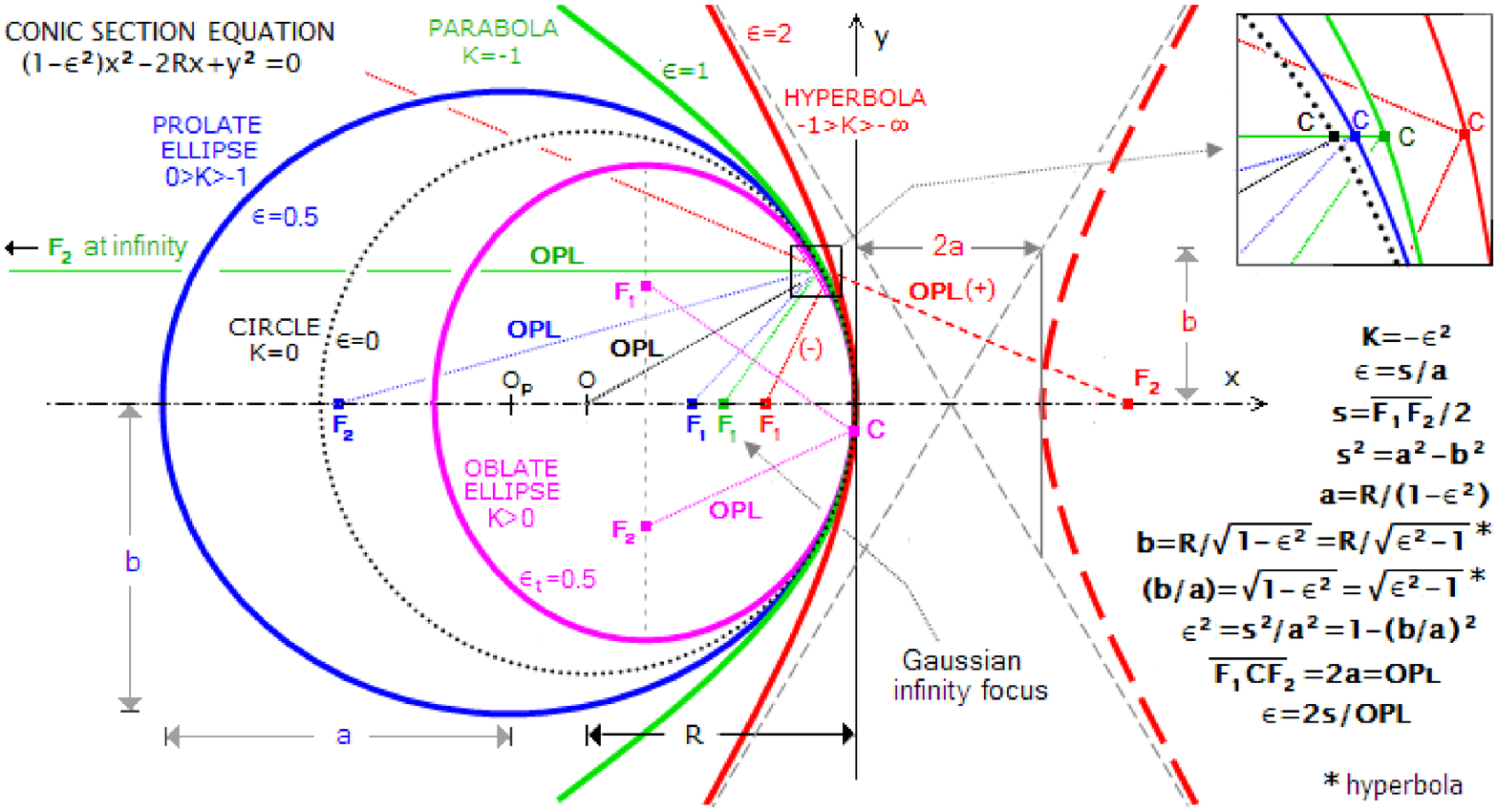}
	\caption[Curvatures At Different Conic Constants]{Different types of curvature for various values for conic constant \cite{conic}. \oic} 
	\label{conic_plot}
\end{figure}

Once a coordinate map is produced, the conic constant formula is used to fit the data, where the conic constant formula is given as \cite{conic}:

\begin{equation}
	z= \frac{(x-x_{off})^2+(y-y_{off})^2}{R+ \sqrt{R^2-(1-\kappa)[(x-x_{off})^2+(y-y_{off})^2]}} +z_{off}
	\label{conic_equ}
\end{equation}
where $x$ $y$ $z$ represent the measured values, $x_{off}$ $y_{off}$ $z_{off}$ are the offset parameters, $\kappa$ and $R$ are conic constant and radius. The function will fit $x_{off}$, $y_{off}$, $z_{off}$, $R$ and $\kappa$. $R$ and $\kappa$ will directly indicate the mirror quality.

The conic constant $\kappa$ characterizes the curvature shape, as indicated in Fig.~\ref{conic_plot}:

\begin{description}
\item[$ 0<\kappa<1: $] 	 The curvature is an oblate ellipsoid.
\item[$ \kappa = 1:  $] The curvature is a sphere.
\item[$ -1<\kappa< 0: $] The curvature is a prolate ellipsoid.
\item[$ \kappa = -1: $] The curvature is a paraboloid.
\item[$ \kappa < -1: $] The curvature is a hyperboloid.
\end{description}

In order to optimize the focusing and avoid imperfections, we requested that the detector mirrors be made in a paraboloid shape if possible. Thus, the expected fitting results are $\kappa\approx-1$ and $R$=110~cm.

For the quality categorization, five sets of fits were applied to the coordinate data, therefore five sets $R$ and $\kappa$ values are obtained for each mirror:

\begin{description}
\item[90\%:] Fits all 360 measuring points, which is close to 90\% of the central area (1$<x<$57~cm and 1$<y<$53~cm).
\item[75\% Inner Fit:] Fits 75\% of the central area (4$<x<$ 56~cm and 4$<y<$51~cm).
\item[75\% Outer Fit:] Fits 25\% of the area round the edge ($x<$4~cm or $x>$ 56~cm or $y<$4~cm or $y>$51~cm).
\item[50\% Inner Fit:] Fits 50\% of the central area (7.5$<x <$52.5~cm and 8.5$<y<$ 46.5~cm).
\item[50\% Outer Fit:] Fits 50\% of the area round the edge ($x<$7.5~cm or $x>$52.5~cm and $y<$8.5~cm or $y>$46.5~cm).
\end{description}

\subsection{Fitting Results}

\begin{table}[t]%
\small
\centering

\caption[Coordinate Measurement Fitting Results]{Radius ($R$) and Conic Constant ($\kappa$) fitting results for 90\%, 75\% inner, 75\% outer, 50\% inner and 50\%  outer fit, for 15 mirrors. Note: gray highlighted mirrors have the worst fitting results, magenta highlighted mirrors have the best fitting results. \oic}

\begin{tabular}{c|cr|cr|cr|cr|cr}
\hline%
\hline%
\multicolumn{1}{c|}{Mirror}         &
\multicolumn{2}{c|}{90\% Fit }      &
\multicolumn{2}{c|}{75\% Inner Fit} &
\multicolumn{2}{c|}{75\% Outer Fit} &
\multicolumn{2}{c|}{50\% Inner Fit} &
\multicolumn{2}{c}{50\% Outer Fit} \\

\#
& $R$ (cm) & $\kappa$ 
& $R$ (cm) & $\kappa$  
& $R$ (cm) & $\kappa$  
& $R$ (cm) & $\kappa$  
& $R$ (cm) & $\kappa$  \\ 
\rowcolor[gray]{0.7}
1  & 118.397  &1.95  & 118.216 &  1.54 &  115.360 &  1.14 &  120.494  & 2.59 &  120.783  & 2.42 \\ 
2  & 113.124  &1.06  & 113.408 &  0.94 &  117.905 &  1.77 &  113.691  & 0.78 &  117.257  & 1.87 \\
3  & 117.110  &1.70  & 117.286 &  1.57 &  117.289 &  1.50 &  118.475  & 2.34 &  120.665  & 2.39 \\
\rowcolor[gray]{0.7}
4  & 122.392  &2.29  & 122.356 &  1.77 &  119.559 &  1.65 &  123.735  & 2.39 &  121.200  & 2.05 \\
5  & 113.331  &0.76  & 113.010 &  0.25 &  117.868 &  1.63 &  114.129  & 0.49 &  117.380  & 1.64 \\
\rowcolor[cmyk]{0,0.4,0,0}
6  & 112.906  &0.94  & 113.239 &  0.86 &  108.735 & -0.22 &  114.824  & 1.61 &  115.172  & 1.30 \\
7  & 113.538  &1.13  & 114.096 &  1.15 &  110.307 &  0.13 &  116.257  & 2.26 &  116.593  & 1.68 \\
8  & 114.325  &1.26  & 114.335 &  0.96 &  117.910 &  1.83 &  114.158  & 0.29 &  115.067  & 1.33 \\
9  & 114.372  &1.30  & 115.151 &  1.39 &  105.942 & -0.67 &  116.887  & 2.44 &  113.365  & 0.94 \\
\rowcolor[cmyk]{0,0.4,0,0}
10 & 112.035  &0.42  & 111.684 & -0.09 &  109.045 & -0.30 &  112.131  &-0.49 &  109.484  &-0.17 \\
\rowcolor[cmyk]{0,0.4,0,0}
11 & 111.766  &0.75  & 112.177 &  0.68 &  101.524 & -1.58 &  113.095  & 1.04 &  110.939  & 0.40 \\
\rowcolor[cmyk]{0,0.4,0,0}
12 & 112.117  &0.84  & 112.264 &  0.62 &  104.243 & -0.98 &  113.636  & 1.26 &  112.183  & 0.72 \\
\rowcolor[gray]{0.7}
13 & 122.464  &2.43  & 123.174 &  2.23 &  113.602 &  0.60 &  126.136  & 2.60 &  116.069  & 1.07 \\
\rowcolor[gray]{0.7}
14 & 117.964  &1.59  & 116.901 &  0.83 &  118.303 &  1.60 &  117.175  & 0.87 &  125.126  & 3.29 \\
15 & 113.674  &1.17  & 113.717 &  0.92 &  115.336 &  1.31 &  114.701  & 1.25 &  119.361  & 2.39 \\

\hline%
\hline%

\end{tabular}
\label{fit_result_tab}
\end{table}

The results for five different fits are presented in Table~\ref{fit_result_tab}. The ideal radius ($R$) and conic constant ($\kappa$) should be $R$=110~cm and $\kappa\approx-$1. However, the fitted $\kappa$ values range from -0.5 to 2.6, which indicates that most of the mirrors have oblate ellipsoid shape rather than paraboloid shape. This should increase optical aberrations, but simulation results described in Chapter \ref{chapter_geant4} indicates that the focusing result of oblate mirror is comparable to the spherical mirror (Mirror \# 6 fitted parameters were used for the study). The PMT positions need to be adjusted to optimize the focusing spot for the oblate mirror, as discussed in Chapter \ref{chapter_geant4}.

A perfect spherically curved mirror with $R$=110~cm should have $\kappa$=0, and its focal length ($f$) should be $f$=$R$/2=55~cm. For an oblate ellipsoid mirror, the focal length is shorter due to the oblateness (see Fig.~\ref{conic_plot}). All the detector mirror candidates have radii larger than 110~cm, and the positive $\kappa$ value will shorten the focal length to be around 55~cm. 

%

$R$ and $\kappa$ fitting results over different areas vs Mirror \# are plotted in Fig.~\ref{fit_result_graph}. The average value of $R$ or $\kappa$ over all mirrors is plotted as the solid line and $\pm\sigma$ limitations are plotted as the dashed lines, where $\sigma$ is the standard deviation. In addition, 6 sets of criteria were introduced to categorize the mirror quality based on the fitting results:

\begin{description}
\item[$R$ Best:] Select mirror when all 5 $R$ values are less than averages (under the solid line) for 5 area fits.
\item[$\kappa$ Best:] Select mirror when all 5 $\kappa$ values are less than averages for 5 area fits.
\item[Less Oblate:] Select mirror when all 5 $\kappa$ values are less than 1.
\item[Average Mirror:] Select mirror when all 5 $R$ and $\kappa$ values are less than (averages $+ \frac{\sigma}{2}$) for 5 area fits
\item[Fitting Sum:] Select mirror when the sum of 5 of $R$ and $\kappa$ values are less than $\sum$averages.
\item[Best Sum:] Select mirror when the sum of 5 of $R$ and $\kappa$ values are less than $\sum$(averages $-\sigma$).
\end{description}

Table~\ref{mirror_select_tab} shows the criteria and Mirror \# that passed the condition. Mirrors \#4, \#13 and \#14 did not pass any criterion, therefore they have the lowest quality; Mirror \#1 has large $R$ and $\kappa$ value which is also considered as low quality mirror. Mirrors \#2 and \#9 are on the boundary line for several criteria (see Fig.~\ref{fit_result_graph}), and it is necessary to check with the optical test results to further understand their quality, especially since Mirror \#9 passed the Best Sum criterion which indicates it is a high quality mirror. Mirrors \#6, \#7, \#10, \#11 and \#12 passed the most criteria; we conclude these mirrors have the highest quality in terms of fitted Radius and Conic Constant. Optical tests were required to confirm the high and low quality mirror results, and classify the average quality mirrors.

\newcommand{\mytoprule}{\specialrule{1pt}{0em}{0em}}

\begin{table}[t]%
\centering
\caption[Mirror Assessment Results]{Criteria and Mirror \# passed the condition.}
\renewcommand{\arraystretch}{1.2}

\begin{tabular}{l|c}


Criteria Name	      & Passed Mirror\# 			 			\\
\mytoprule
$R$ Best              & 6, 7, 8, 10, 11, 12		 	            \\
$\kappa$ Best   	  & 10, 11, 12					            \\
Less Oblate			  & 10							            \\
Average Mirror		  & 6, 10, 11, 12				            \\
Fitting Sum			  & 1, 2, 3, 5, 6, 7, 8, 9, 10, 11, 12, 15	\\
Best Sum			  & 6, 7, 9, 10, 11, 12						\\
Failed All			  & 4, 13, 14         						\\
\end{tabular}
\label{mirror_select_tab}
\end{table}

\begin{figure}[hp]
	\centering
	\includegraphics[width=0.9\textwidth]{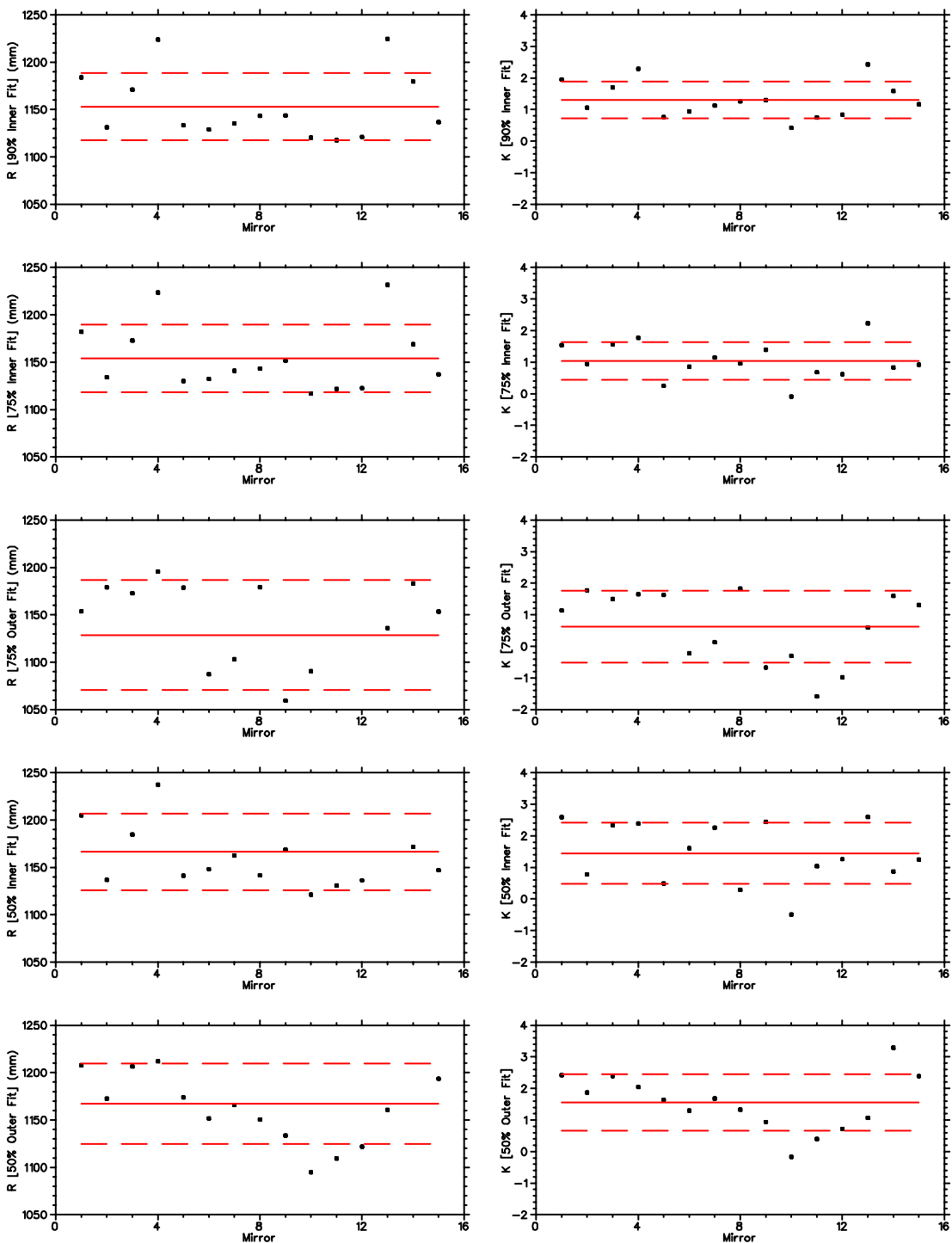}
	\caption[Mirror Quality Assessment Plot]{$R$ and $\kappa$ fitting result from coordinate measurements for all 15 mirrors. From top to bottom, it shows $R$ and $\kappa$ results for 90\%, 75\% Inner, 75\% Outer, 50\% Inner and 50\% Outer fit. Note that red solid lines give the average value of all mirrors; red dashed lines give 1 $\sigma$ over and below the averaged value. The expected radius and conic constant: $R_{expect}$=110~cm, and $\kappa_{expect}$=-1. \oic}
	\label{fit_result_graph}
\end{figure}

\subsection{Fit$-$Measurement Plots}

Ideally, the detector mirrors should have few imperfections at the reflecting surface. In order to check imperfections, we computed the difference plot between the conic constant fit and the experimental results. The fitted radius and $\kappa$ are used to generate a perfect oblate shape, then its $z$ coordinate is compared with the real measurement at each $x$, $y$ measuring point. The difference is defined as

\begin{equation}
	\Delta z= z_{measurement} - z_{generated}\,.
\end{equation}

In general, the central 50\% region has small $\Delta z$, whereas large $\Delta z$ only occur along the edges. The features of all fit$-$measurement plots can be categorized into four groups.

In the first group, mirrors have uniformly small $\Delta z$ for most areas, which indicates that the reflecting surface contains few imperfections. Difference plots for Mirrors \#10, \#11 and \#12 have this feature. Fig.~\ref{m10_f_m} shows the fit$-$measurement plot for Mirror \#10.

In the second group, mirrors have large $\Delta z$ ($|\Delta z|>$ 0.6~mm) along two opposite edges (corners), and small $\Delta z$($<$ 0.6~mm) along other edges (corners). Mirrors \#2, \#5, \#6, \#7, \#8 and \#9 have this feature. Fig.~\ref{m6_f_m} shows the fit$-$measurement plot for Mirror \#6.

In the third group, mirror have large $\Delta z$ ($|\Delta z|>$ 0.6~mm) along all four edges (corners). Plots for Mirrors \#3, \#14 and \#15 have this feature, and their imperfections are worse than the second group mirrors. Fig.~\ref{m14_f_m} shows the fit$-$measurement plot for Mirror \#14.

In the fourth group, $|\Delta z|$ is not flat near the central region. Plots for Mirrors \#13 and \#4 have this feature. Fig.~\ref{m13_f_m} shows the fit$-$measurement plot for Mirror \#13.

Based on the Monte Carlo Simulation result, the reflecting area near the edges are shown to be less important than the central area. Mirrors from the first and second groups have some imperfections mostly along the edges, while their $\Delta z$ are relatively small and one of these edges will be outside the beam envelope. On the other hand, group three mirrors have large $\Delta z$ along all four edges and group four mirrors have imperfections in central region, and as a result these mirrors have larger scale imperfections.

\begin{figure}[hp]
	\centering
	\subfloat[Mirror \#10, an example of a mirror with good uniformity (1st group).]{\includegraphics[width=0.75\textwidth]{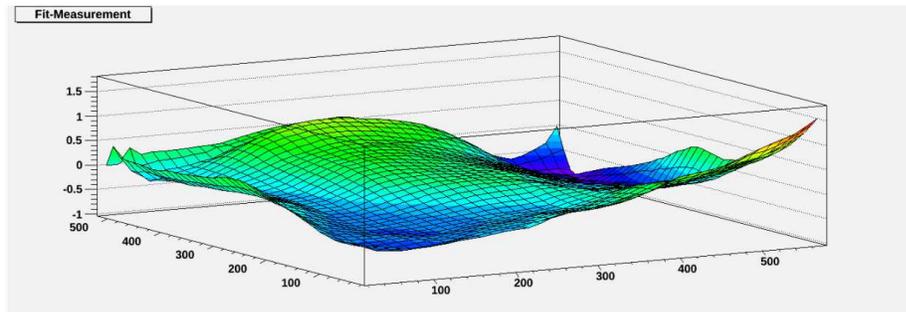}\label{m10_f_m} } \\
	\subfloat[Mirror \#6, an example of a mirror with large two edges (2nd group).]{\includegraphics[width=0.75\textwidth]{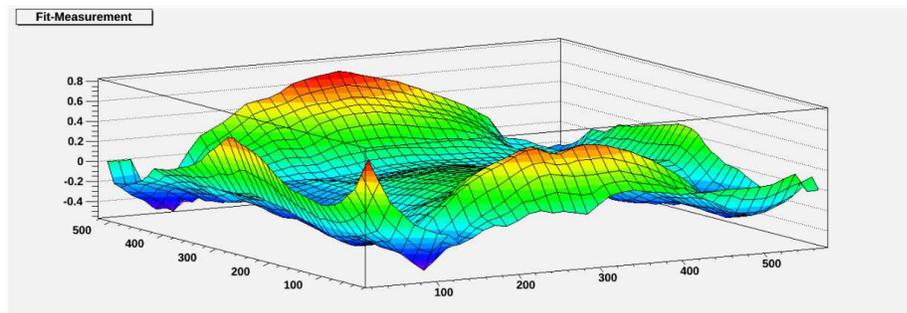}\label{m6_f_m}} \\
	\subfloat[Mirror \#13, an example of a mirror with large $\Delta z$ along 4 corners (3rd group).]{\includegraphics[width=0.75\textwidth]{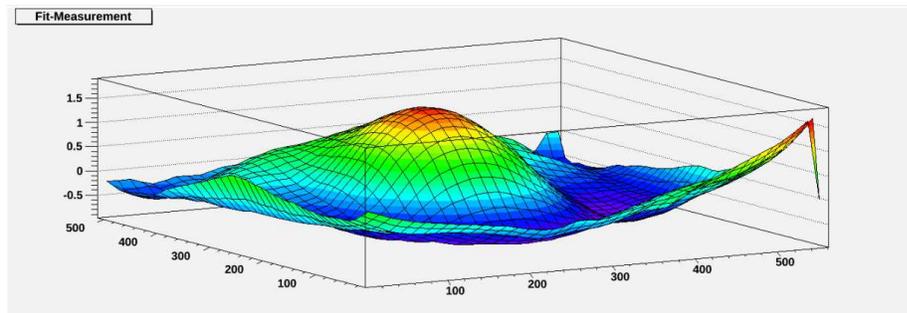}\label{m13_f_m}}\\
	\subfloat[Mirror \#14, an example of a mirror with poor central region (4th group).]{\includegraphics[width=0.75\textwidth]{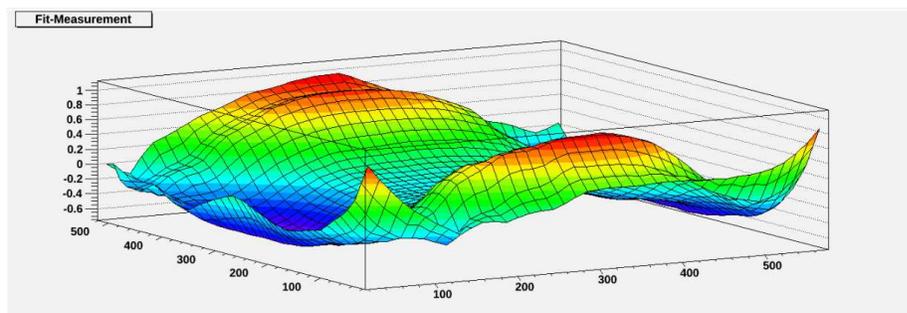}\label{m14_f_m}}
	\caption[Mirror $\Delta z$ Example Plots]{Mirror $\Delta z$ plots for Mirror \#10, \#6, \#13 and \#14. Units of the plot are~mm. \oic}
	\label{fig_mirrors_fit_measurement}
\end{figure}

%
%
%
%

\subsection{Uncertainty of Coordinate Measurements}

We requested that Dumur remeasure Mirrors \#10 and \#13. This allows the systematic uncertainty of a single coordinate measurement to be studied. For the 90\% area fit of Mirror \#10 the first measurement $a$ has fitted $R_{10a}$ = 112.04~cm and $\kappa_{10a}$ = 0.42, and the second measurement $b$ has fitted $R_{10b}$ = 110.8~cm and $\kappa_{10b}$ = 0.35. We used Eq. \ref{conic_equ} to reconstruct the $z$ value for 810 points across the mirror with both sets of fitting parameters, then compare the difference $\Delta z = z_a -z_b$. The averaged $\Delta z=$ 0.267~mm and $\sigma=$ 0.166~mm. Similarly, for 90\% fit of Mirror \#13 the first measurement $a$ has fitted $R_{13a}$ = 122.46~cm and $\kappa_{13a}$ = 2.43, the second measurement $b$ has fitted $R_{13b}$ = 122.56~cm and $\kappa_{13b}$ = 2.52. The averaged $\Delta z=$ 0.016~mm and $\sigma=$ 0.033~mm. We used the standard deviation of $\Delta z$ of Mirror \#10 (0.166~mm) to estimate the uncertainty of a single coordinate measurement. The Mirror \#10 measurements have larger $\Delta R = R_{10a}-R_{10b} = (112.04-110.8)~\textrm{cm} = 1.8$~cm and Mirror \#13 measurements have larger $\Delta\kappa= \kappa_{13b} - \kappa_{13a} =  0.11$. The uncertainties for $R$ and $\kappa$ are decided to be $\pm$2~cm and $\pm$0.15, which are very conservative estimations based on the $\Delta R$ and $\Delta \kappa$. Even though the mirror uncertainties are slightly overestimated, they are still not large enough to challenge the mirror quality conclusion.

\subsection{Coordinate Measurement Result Summary}

From the coordinate measurement fitting results, we conclude that all manufactured HGC mirrors have oblate elliptical curvature: $\kappa > $ 0, and their fitted radius of curvature are slightly larger than desired: $R >$ 110~cm. Mirrors \#1, \#4, \#13 and \#14 have the lowest quality in terms of fitted $R$ and $\kappa$. Mirrors \#6, \#7, \#10, \#11 and \#12 have the highest quality. Optical tests are needed to categorize the intermediate mirrors. The estimated systematic uncertainty for a single $z$ coordinate measurement is $\delta z$=$\pm$0.166~mm; the uncertainties in the fitted $R$ and $\kappa$ are $\pm$2~cm and $\pm$0.15 respectively. From the fit-measurement plots, we learned Mirrors \#10, \#11 and \#12 have the fewer imperfections; Mirrors \#2, \#5, \#6, \#7, \#8 and \#9 have some imperfections along two edges (corners).

\section{Optical Test}

\subsection{Methodology}
In this test a 1 mW laser beam is used as light source, and it is split to illuminate the whole mirror. A photograph of the reflected pattern is taken on a small screen at the optimal focusing distance. The equipment used to split the laser beam is shown in Fig.~\ref{optical_test}: a combination of laser splitter and concave lens. The distance between the concave lens and the mirrors $S_{O}= 624~cm$ and the mirror to focused image distance ($S_{i}$) is also recorded. Then, the focal length ($f$) can be calculated by the equation:

\begin{equation}
\frac{1}{f}= \frac{1}{S_{O}} + \frac{1}{S_{i}} \,.
\label{focal_equ}
\end{equation}

\begin{figure}[t]
	\centering
	\includegraphics[width=0.8\textwidth]{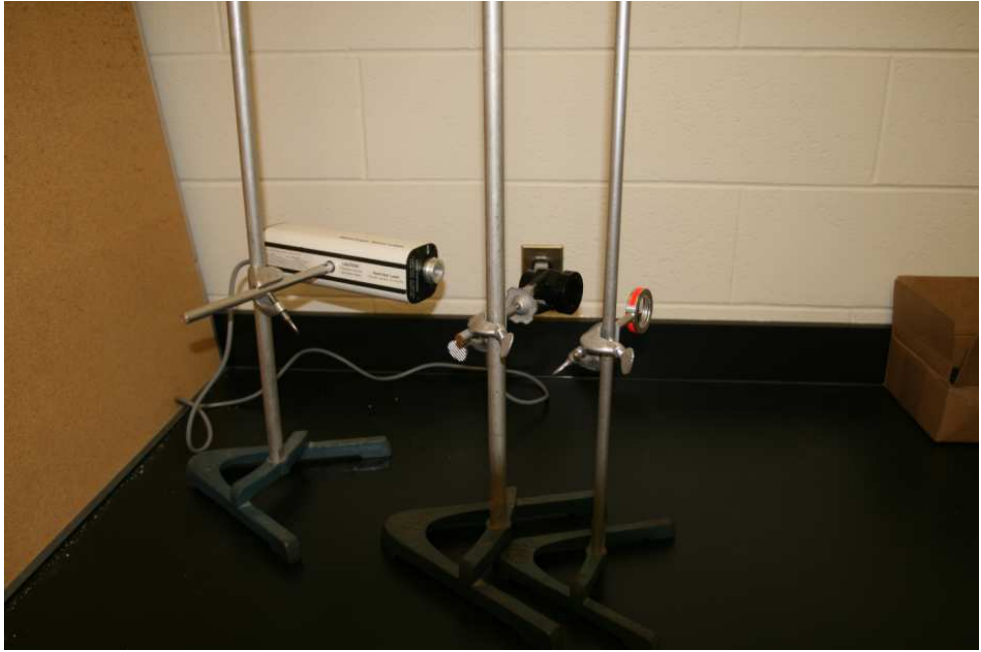}
	\caption[Optical Test Setup]{Optical test equipment: 1 mW lab laser, a laser splitter and a concave mirror. \oic}
	\label{optical_test}
\end{figure}

\subsection{Optical Test Results}
In this subsection we present the optical test results. For each mirror, photographs of the reflected laser beam pattern were taken at the optimal focus point and the horizontal distance ($S_{i}$) relative to the mirror was recorded. Due to the various optical aberrations and the fact that the mirror was not aluminized, reflections were obtained from both the front and back surfaces of the mirror. The spot shape was complex and not all components focused at the same distance. Thus, there is some systematic uncertainty in the image distance measurement, and aluminization will help to reduce this.

Table~\ref{optical_result_tab} shows $S_{i}$ and calculated focal length (FL) using Eq. \ref{focal_equ}. For a spherical mirror, the $f$ should be $f=R/2$. For an oblate ellipsoid mirror, $f<R/2$, and the $\kappa$ value determines the degree of shortness. We divide the 90\% Fitted Radius (FR) from the coordinate measurement by 2, and obtain the focal length for spherical shaped mirror, then compute the difference between it and the calculated focal length (FL). The results are listed in Table~\ref{optical_result_tab}. Fig.~\ref{k_vs_diff} shows the plot of 90\% fit $\kappa$ vs (FL$-$FR/2). From this plot, we can conclude a strong correlation between $\kappa$ and (FR/2$-$FL): the more oblate the mirror the shorter the focal length as determined from the optical test.

Except for Mirrors \#1 and \#13, all other mirrors have focal lengths between 54 and 57~cm, with the measurement uncertainty being 0.5~cm for $S_{i}$, and 2~cm for $S_{o}$. Although positive $\kappa$ values do not help to focus light, they bring the focal point closer to the optimised focal point despite the mirrors having a larger radius.

\begin{figure}[t]
	\centering
	\includegraphics[width=0.45\textwidth,angle=90]{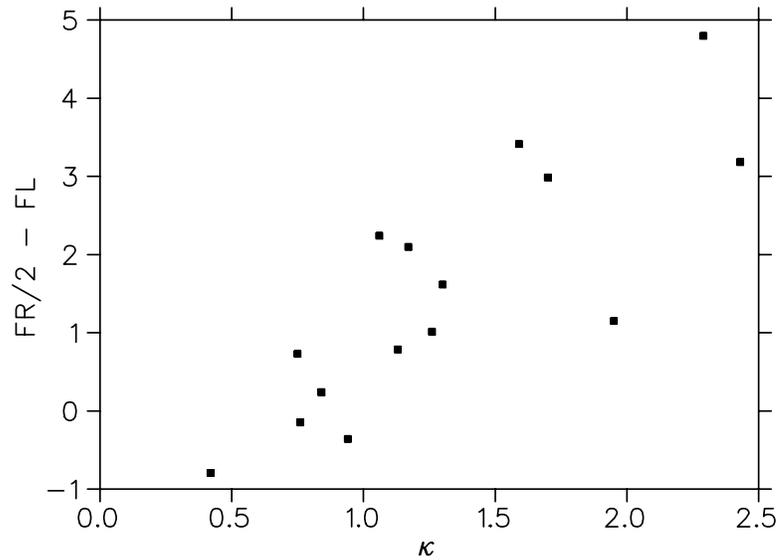}
	\caption[(Focal length$-$Fitted Radius/2) vs Conic Constant]{90\% fitted Conic Constant $\kappa$ is plotted against the difference between Focal Length and half of Fitted Radius (FR/2$-$FL) for 15 mirrors.}
	\label{k_vs_diff}
\end{figure}

The images of the reflected patterns are shown in Fig.~\ref{red}, the processed version is in Fig.~\ref{processed}. GIMP \cite{gimp} color tools were used to process the original image: the contrast was increased to +50, brightness was reduced to -50, red color was filtered out, and the brightest area was turned into blue. From the original image, it is very difficult to determine the size and shape of the brightest area, whereas the processed image shows the brightest area very clearly in blue. Direct observation confirms Mirrors \#1, \#4 and \#13 have the poorest quality: there are huge tails in Mirror \#1 image, and the light does not focus for Mirrors \#4 and \#13.

The reflected images include the reflection of the back surface where the imperfections are much worse due to the slumping process, thus it is impossible to categorize mirrors based only on their front surface quality. Some quantitative criteria are introduced to help determine the entire mirror quality using the processed reflection image. The conditions include the intensity, size and shape of the blue region, also the size and shape of the green region.

%
%

The size and shape of the brightest (blue) area are the most important factors to determine the mirror quality. Table~\ref{pixel_info} shows the number of green pixels, the number of blue pixels and the blue-green ratio for each mirror from Fig.~\ref{processed}, where the higher pixel number corresponds to the larger image area. Mirror \#1, \#4 and \#13 have the largest green pixel number and smallest blue pixel number, while their blue-green ratio is less than 10\%. They are confirmed to be the low quality mirrors. On the other hand, Mirrors \#2 and \#3 have the smallest green region and the highest blue-green ratio (larger than 30\%). It is difficult to conclude the optical quality of this type of mirrors, thus other test results are needed.  All other mirrors have green and blue pixel numbers around 40000 and 5000, and the blue-green ratio is around 0.13. In terms of shape, Mirror \#15 has two separate blue regions; Mirror \#5 has a long strip for the blue region. The best blue spots are produced by Mirrors \#6, \#7, \#9 and \#11, they have small circular blue spots with few spikes around them. Surprisingly, with the best coordinate measurement result, Mirror \#10 has a blue triangular spot at low intensity, the spot size is slightly larger than other high quality mirrors.

Mirrors \#1, \#4, \#13 have the largest green region. Mirrors \#5 and \#14's green regions are much larger than their blue. Mirror \#15 has two focused regions on the image. Other mirrors have reasonable good green regions, especially for Mirrors \#2 and \#10, which contain no large tail.  

Based on the criteria for the image results, we conclude that Mirrors \#6, \#7, \#9 and \#11 have the best optical quality; Mirrors \#2, \#8, \#10 and \#12 have reasonably good optical quality; Mirrors \#3, \#5, \#14 and \#15 have average quality, Mirrors \#1, \#4 and \#13 have the worst quality.

The shape of the blue spots for high and average quality mirrors can be classified into two groups: small blue spot and large blue spot. The small blue spot mirrors include \#6, \#7, \#9, \#8, \#11 and \#12, their characteristics are a small circular bright blue spot with few spikes and tails in the green region. The large blue spot mirrors include \#2 and \#10, their characteristics are a relatively larger blue spot and no tails in the green region. It is difficult to determine which group represents the better focusing ability, therefore, we decided to send one mirror from each group for aluminization. The optical test will be repeated and the reflection of the back surface should be completely eliminated. The comparison of the reflected pattern before and after the aluminization for the testing mirror will give us some intuitive hints for the front surface quality of other mirrors. Mirrors \#2 and \#8 are the perfect candidates for the initial aluminization test, as they are average quality mirrors which share the same characteristics with the high quality mirrors, they will contribute to our understanding of the front surface optical quality for both groups, as well as the ECI aluminization technique. 

\subsection{Optical Test Summary}
From the optical test results, we extracted focal lengths corresponding to full mirror illumination and photographed the reflected pattern of the split laser beam. All of the high and average quality mirrors from the coordinate measurements have focal lengths from 54~cm to 57~cm, with uncertainty $\pm$2~cm, which are close to the optimal value. Mirror \#6, \#7, \#9 and \#11 have the highest quality reflected spot, and \#2, \#8, \#10, \#12 have reasonably good reflected spot. Mirror \#2 and \#8 are selected to be aluminized, and optical test will be repeated after the aluminization.

\begin{table}[t]%
\centering
\caption[Optical Test Results]{ $S_{i}$ column shows the measurement for mirror to image distance; Focal Length column gives the calculated focal length with $S_{i}$ using Eq. \ref{focal_equ}; Fitted Radius is the 90\% fitted radius from coordinate measurement; FR/2 is the Fitted Radius/2; FR/2$-$FL is the difference between half of the Fitted Radius and Focal Length; $\kappa$ is the 90\% fitted conic constant from coordinate measurement. Note: gray highlighted mirrors have the worst optical testing results.}

\begin{tabular}{cccccccccccc}

\toprule%
Mirror & $S_{i}$ & Focal Length  &  Fitted Radius &   FR/2      &    FR/2$-$FL  &    $\kappa$    \\        
\#	   &  (cm) 	 &  (cm)	 	 &     (cm)       &   (cm)      &    (cm)      &     		    \\
\toprule%

\rowcolor[gray]{0.7}
1      &  64.0   &   58.05       & 118.40         & 59.20       &    1.15      &   1.95         \\
2      &  59.5   &   54.32       & 113.12         & 56.56       &    2.24      &   1.06         \\
3      &  61.0   &   55.57       & 117.11         & 58.56       &    2.99      &   1.70         \\ \rowcolor[gray]{0.7}
4      &  62.0   &   56.40       & 122.39         & 61.19       &    4.80      &   2.29         \\
5      &  62.5   &   56.81       & 113.33         & 56.67       &   -0.14      &   0.76         \\
6      &  62.5   &   56.81       & 112.91         & 56.45       &   -0.36      &   0.94         \\
7      &  61.5   &   55.98       & 113.54         & 56.77       &    0.79      &   1.13         \\
8      &  61.7   &   56.15       & 114.33         & 57.16       &    1.01      &   1.26         \\
9      &  61.0   &   55.57       & 114.37         & 57.19       &    1.62      &   1.30         \\
10     &  62.5   &   56.81       & 112.04         & 56.02       &   -0.79      &   0.42         \\
11     &  60.5   &   55.15       & 111.77         & 55.88       &    0.73      &   0.75         \\
12     &  61.3   &   55.82       & 112.12         & 56.06       &    0.24      &   0.84         \\ \rowcolor[gray]{0.7}
13     &  64.0   &   58.05       & 122.46         & 61.23       &    3.19      &   2.43         \\ \rowcolor[gray]{0.7}
14     &  61.0   &   55.57       & 117.96         & 58.98       &    3.41      &   1.59         \\ 
15     &  60.0   &   54.74       & 113.67         & 56.83       &    2.10      &   1.17         \\
\bottomrule
\end{tabular}

\label{optical_result_tab}
\end{table}

\begin{table}[t]%
\centering

\caption[Pixel Counting Results]{Pixel information of Fig.~\ref{processed}. The gray highlighted mirrors have the lowest blue-green ratio.}

\begin{tabular}{cccc}
\toprule%
Mirror  &   Green Pixel & Blue Pixel  &   Blue-Green  \\        
\#	    &   \#      	& \#      	  &     Ratio  	     \\ \toprule%
\rowcolor[gray]{0.7}
1		&	58072		&	 5144	  &	   0.09			 \\
2       &   30201       &    11896    &    0.39          \\
3       &   20590       &    14494    &    0.70          \\
\rowcolor[gray]{0.7}
4       &   84718       &    6800     &    0.08          \\
5       &   38860       &    5607     &    0.14          \\
6       &   36418       &    5427     &    0.15          \\
7       &   40475       &    5050     &    0.12          \\
8       &   39362       &    6050     &    0.15          \\
9       &   42580       &    5321     &    0.12          \\
10      &   35126       &    5026     &    0.14          \\
11      &   48764       &    5900     &    0.12          \\
12      &   40120       &    4806     &    0.12          \\
\rowcolor[gray]{0.7}
13      &   87806       &    1021     &    0.01          \\
14      &   36801       &    5098     &    0.14          \\
15      &   43228       &    7805     &    0.18          \\
\bottomrule
\end{tabular}
\label{pixel_info}
\end{table}

\begin{figure}[hp]
	\centering
	\includegraphics[width=0.9\textwidth]{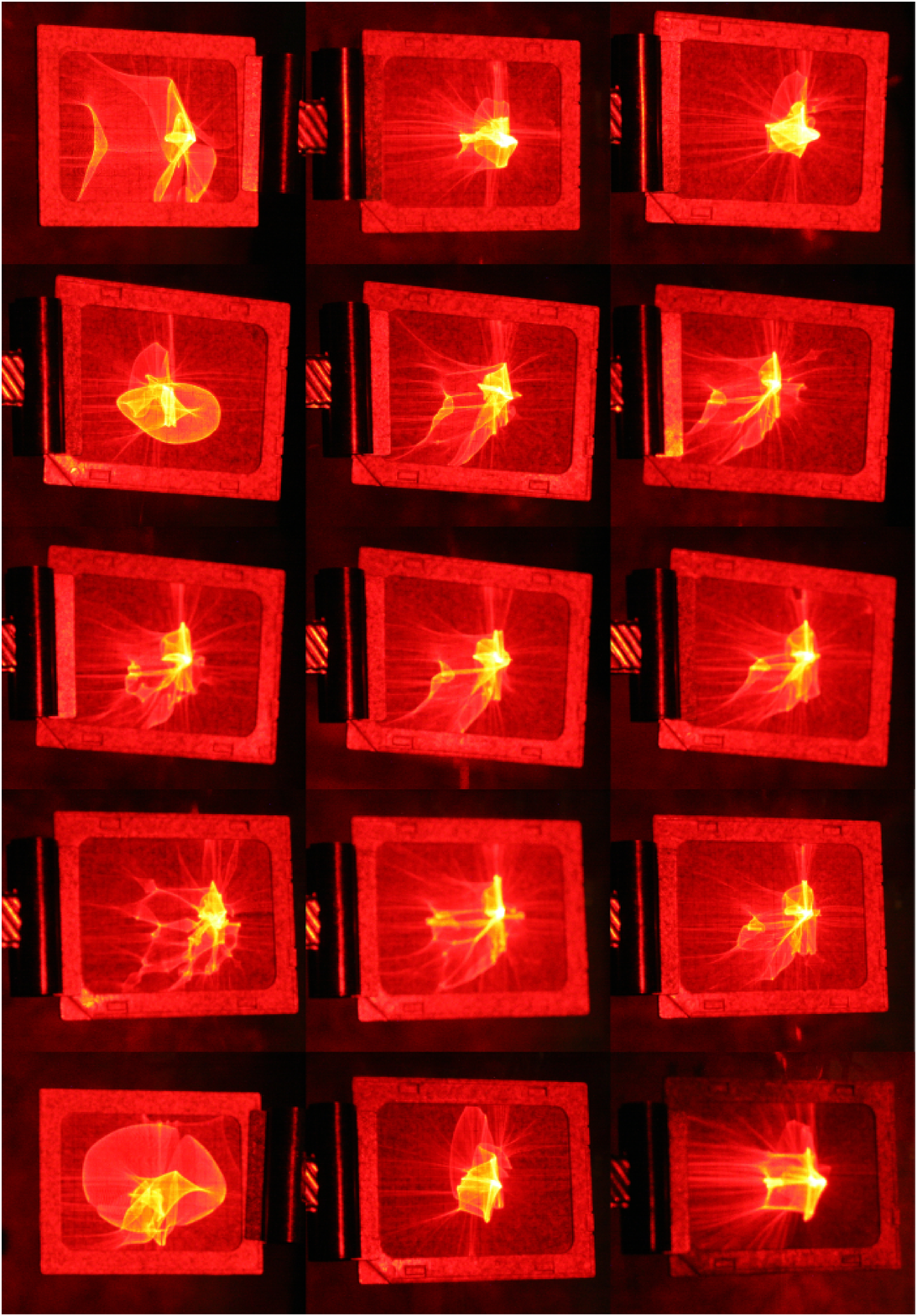}
	\caption[Return Raw Spot Images]{Return (raw) spot image for all mirrors. The mirror number is incremented horizontally, top left corner is Mirror \#1 and bottom right corner is Mirror \#15. \oic}
	\label{red}
\end{figure}

\begin{figure}[hp]
	\centering
	\includegraphics[width=0.9\textwidth]{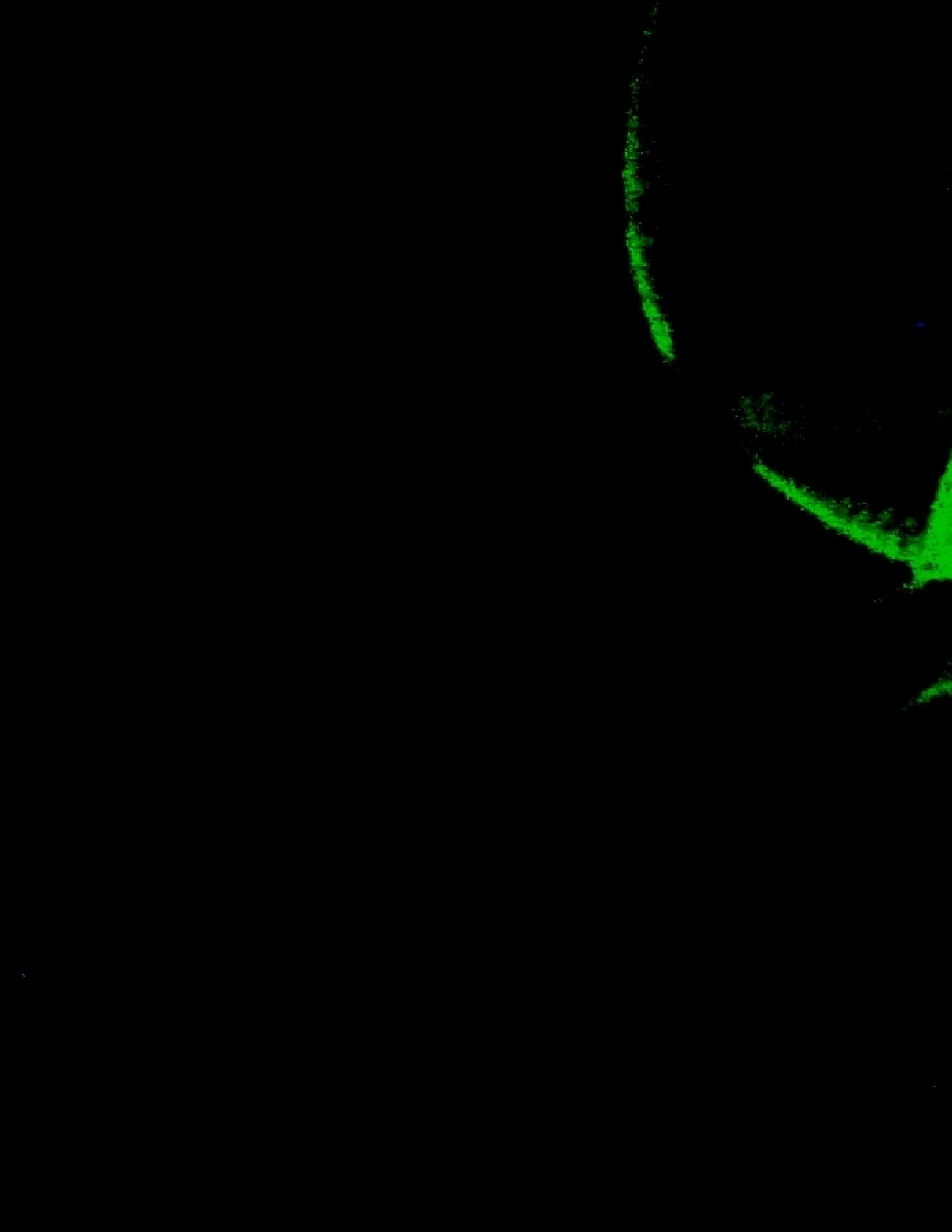}
	\caption[Processed Return Spot Images]{Processed return spot image for all mirrors. The mirror number is incremented horizontally, top left corner is Mirror \#1 and bottom right corner is Mirror \#15. \oic}
	\label{processed}
\end{figure}

\section{Combined Result}
\paragraph{Mirror \#1: Bad} Mirror \#1 failed nearly all criteria in coordinate measurement test, it has large fitted $R$ (118.4~cm) and $\kappa$ (1.95). The actual focal length is too large (58~cm). Spot image shows large tails in the green region, and blue region is not focused to a circular spot.
 
\paragraph{Mirror \#2: Average} According to coordinate measurement result, Mirror \#2 failed most of the criteria,  because its fitted parameters are large for 75\% outer and 50\% outer fit. On the other hand, it has reasonable fitted $R$ (113.1~cm) and $\kappa$ (1.06) in central region. The optical test for Mirror \#2 is better than expectation, there is no large green tail in the image, and blue spot is focused to a relatively large region (similar to Mirror \#10). Mirror \#2 is the perfect candidate to be sent for the aluminization, to help understand characteristics of Mirror \#10.

\paragraph{Mirror \#3: Average} Mirror \# 3 failed most of the criteria for the coordinate measurement test, it has large fitted $R$ (117~cm) and $\kappa$ (1.7). From the optical test, Mirror \#3 has long focal length (58.56~cm). From the image, the blue region failed to focus to a spot.

\paragraph{Mirror \#4: Bad} Coordinate measurement and optical tests confirmed Mirror\# 4 is one of the worst mirrors: large fitted $R$ and $\kappa$, large imperfection in the center, the return spot image failed to focus.

\paragraph{Mirror \#5: Average} Although Mirror \#5 failed some strict criteria, its fitted $R$ (113.3~cm) and $\kappa$ (0.76) look quite reasonable. The optical result is not as good as the expectation, blue region failed to focus to a spot. All these results may suggest this mirror has better quality in the center than the edge.

\paragraph{Mirror \#6: Good} Mirror \#6 passed nearly all criteria for the coordinate measurement test, and its optical test result looks satisfying. It is considered to be one of the best mirrors. 

\paragraph{Mirror \#7: Reserve} Mirror \#7 passed most of the criteria, including the Best Sum criterion, which indicates it is a high quality mirror from the coordinate measurement test. The optical focal length is reasonable (56~cm) and the reflected spot image has high quality. 

\paragraph{Mirror \#8: Average} Mirror \#8 has very similar fitted $R$ and $\kappa$ to \#7 and \#9, however it did not pass the Best Sum criterion. Mirror \#8 has the same spot characteristics to \#6, \#7, \#8, \#9, \#11 and \#12, and it is the most average mirror of the group. Mirror \#8 will contribute to understand the spot image of other mirrors. 

\paragraph{Mirror \#9: Reserve} Mirror \#9 passed most of the criteria, including the Best Sum criterion, which indicates it is a high quality mirror from the coordinate measurement test. The optical focal length is reasonable (55.6~cm) and the reflected spot image has high quality. 

\paragraph{Mirror \#10: Good}  Mirror \#10 passed all criteria for the coordinate measurement test, it has the best fitting result and has fewer imperfections. However, its spot image looked different from other good mirrors, Mirror\#2 is sent for aluminization to help understand this type of reflected light spot.

\paragraph{Mirror \#11: Good}  Mirror \#11 passed nearly all criteria for the coordinate measurement test, has few imperfections and its optical test result looks satisfying. It is considered to be one of the best mirrors.

\paragraph{Mirror \#12: Good}  Mirror \#12 passed nearly all criteria for the coordinate measurement test, has few imperfections and its optical test result looks satisfying. It is considered to be one of the best mirrors.

\paragraph{Mirror \#13: Bad} Coordinate measurement and optical test both confirmed Mirror \#13 is one of the worst mirrors: large fitted $R$ and $\kappa$, large imperfections in the center, return spot image does not focus.
 
\paragraph{Mirror \#14: Low Average} Despite reasonably good spot image, Mirror \#14 has huge fitted $R$ (118~cm) and $\kappa$ (1.59). Also the focal length from the optical test is very long (59~cm).

\paragraph{Mirror \#15: Average} Mirror \#15 has average quality according to coordinate measurement test, the optical image test shows two separate blue regions which indicate certain parts of the mirror have bad imperfection.

Table~\ref{table_master} gives an overall summary for the coordinate measurement and optical test result, and ranking of all 15 HGC mirrors.

\begin{table}[t]
\centering
\caption[Overall Mirror Quality Summary]{Overall mirror quality summary from coordinate measurement and optical test results. 90\% Fit $R$ and $\kappa$ are form Table \ref{fit_result_tab}; \# of Failed Criterion is deduced from Table \ref{mirror_select_tab}, FR/2$-$FL results are from Table \ref{optical_result_tab}, Blue-Green ratio results are from Table \ref{pixel_info}. The worst quality mirrors are highlighted in red; the average quality mirrors are highlighted in yellow; the best quality mirrors are highlighted in green. \oic}
\begin{tabular}{ccccccc}
\toprule%
Mirror  & 90\% Fit& 90\% Fit & \# of Failed &  FR/2$-$FL &   Blue-Green  &  Overall \\        
\#	    & $R$ (cm)& $\kappa$ & Criterion    &   (cm)    &    Ratio 	    &  Ranking \\    \toprule  \rowcolor[rgb]{1,0.3,0.3}
1		& 118.397 &  1.95    & 5            &   1.15    &    0.09	    &  Bad     \\    \rowcolor[cmyk]{0,0,0.5,0}
2       & 113.124 &  1.06    & 5            &   2.24    &    0.39       &  Average \\    \rowcolor[cmyk]{0,0,0.5,0}
3       & 117.110 &  1.70    & 5            &   2.99    &    0.70       &  Average \\    \rowcolor[rgb]{1,0.3,0.3}                           
4       & 122.392 &  2.29    & 6            &   4.80    &    0.08       &  Bad     \\    \rowcolor[cmyk]{0,0,0.5,0}
5       & 113.331 &  0.76    & 5            &  -0.14    &    0.14       &  Average \\    \rowcolor[rgb]{0.3,0.8,0.3}
6       & 112.906 &  0.94    & 2            &  -0.36    &    0.15       &  Good    \\    \rowcolor[cmyk]{0,0,0.5,0} 
7       & 113.538 &  1.13    & 3            &   0.79    &    0.12       &  Reserve \\    \rowcolor[cmyk]{0,0,0.5,0}
8       & 114.325 &  1.26    & 4            &   1.01    &    0.15       &  Average \\    \rowcolor[cmyk]{0,0,0.5,0}
9       & 114.372 &  1.30    & 4            &   1.62    &    0.12       &  Reserve \\    \rowcolor[rgb]{0.3,0.8,0.3} 
10      & 112.035 &  0.42    & 0            &  -0.79    &    0.14       &  Good    \\    \rowcolor[rgb]{0.3,0.8,0.3} 
11      & 111.766 &  0.75    & 1            &   0.73    &    0.12       &  Good    \\    \rowcolor[rgb]{0.3,0.8,0.3} 
12      & 112.117 &  0.84    & 1            &   0.24    &    0.12       &  Good    \\    \rowcolor[rgb]{1,0.3,0.3}                           
13      & 122.464 &  2.43    & 6            &   3.19    &    0.01       &  Bad     \\    \rowcolor[rgb]{1,0.3,0.3}
14      & 117.964 &  1.59    & 6            &   3.41    &    0.14       &  Bad     \\    \rowcolor[cmyk]{0,0,0.5,0}
15      & 113.674 &  1.17    & 5            &   2.10    &    0.18       &  Average \\    
\bottomrule
\end{tabular}
\label{table_master}
\end{table}

\section{Conclusion}
\label{mirror_selection_con}

By combining the results of both tests, we conclude: Mirror \#6, \#10, \#11 and \#12 have excellent quality and are recommended to be the detector mirrors; Mirror \#7 and \#9 can be reserved as backups; Mirror \#2 and \#8 are sent for test aluminization. The optical test will be repeated after aluminization of \#2 and \#8, so that reflected spot can be compared with our first result. The test aluminization result for Mirror \#8 will be discussed in Chapter \ref{reflectivity}.

}

%% file: truck/reflectivity.tex
{

\label{reflectivity}

\section{Introduction}



In November 2011, HGC Mirrors \#2 and \#8 were shipped to Evaporated Coatings Inc. (ECI) \cite{eci} for an aluminization test based on the conclusion from the mirror selection study (see Section \ref{mirror_selection_con}). The aluminized mirrors were then sent to Jefferson Lab. The University of Regina and Jefferson Lab (Hall C, Detector Group and Free Electron Laser Facility) are collaborating to construct a permanent setup to measure mirror reflectivity between 165-400nm, in order to verify the mirror aluminization technique by ECI. In this chapter, we will describe the permanent reflectivity setup at Jefferson Lab and present the reflectivity results of the aluminized Mirror \#8.

The reflectivity quality control by ECI was not performed on the test mirrors. A number of 1" diameter test samples were placed at various positions in the vacuum deposition tank and aluminized the same way as the test mirrors (see Fig \ref{sample_pos}). Their reflectivities were later measured by ECI to indicate the quality of the aluminized test mirrors. At the end of the chapter, the ECI sample reflectivity results are compared to our measurement.

\section{Methodology}

\subsection{Equipment list}

The list of important components used for the setup construction is given below:

\begin{itemize}
	\item 2 $\times$ AXUV-100G Photo-Diode UV Detectors (\#02, \#18) with Ceramic Shoulders manufactured by International Radiation Detectors, Inc. (IRD) \cite{irdinc}
	\item Thorlab MC100 Optical Chopper System with MC1F2 Chopping Blade
	\item SR530 Lock-in Amplifier 
	\item The McPherson \cite{mcphersion} Model 218 Vacuum Ultraviolet (VUV) Monochromator with Holographic 200~nm Blaze (1200 Grids/mm) Grating
	\item 1mW Melles Griot \cite{melles} Alignment Laser
	\item 3 Watt Hamamastu \cite{hamamastu} Deuterium UV Light Source 
	\item Deep Ultraviolet (DUV) Flipper Mirror: Melles Griots DUVA-PM-5010M-UV
	\item F1.5 Focusing Lens: Edmund Optics UV Plano-Convex 50mm Dia.x75mm FL Uncoated Lens
	\item F4 Focusing Lens: Edmund Optics CAF2 PCX50.8 Lens 
	\item 1 $\times$ Translation Stage, 1 $\times$ Rotation Stage and 8 $\times$ Stepper Motors 
\end{itemize}

The mirror reflectivity measurements used HGC Mirror \#8 and IRD UV detector \#02. The HGC mirror was clamped along the left and bottom edges with metal clamps (two along each side). The clamped mirror edges were sandwiched between rubber gaskets to isolate them from the metal pieces.

\subsection{Monochromator}
The monochromator is an optical device that transmits a mechanically selectable, narrow wavelength band of light. It employs a diffraction grating to spatially separate the light at different wavelengths, and subsequently a mechanical device is used to select a specific wavelength and guide the light out with a reflecting mirror. The basic components of the monochromator include: two reflecting mirrors, a diffraction grating and a wavelength adjustment mechanism. However, the inside structure of a monochromator varies significantly, depending on the purpose and manufacturer.

\begin{figure}[t]
	\centering
	\includegraphics[width=0.75\textwidth]{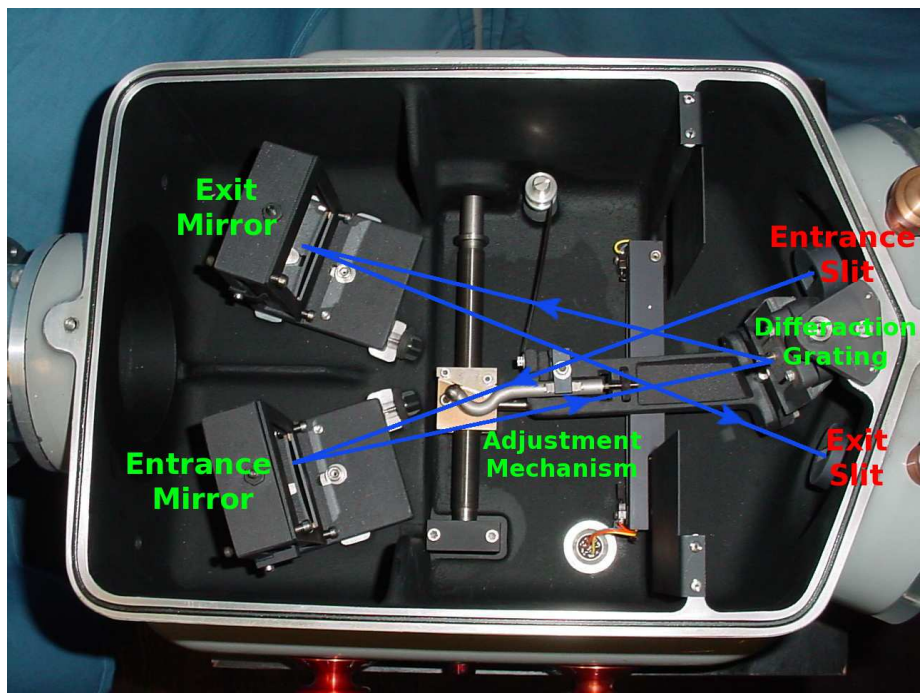}
	\caption[McPherson 218 VUV Monochromator]{McPherson 218 VUV monochromator. The light path inside the monochromator is in blue \cite{monochrom}. \oic}
	\label{pic_monochrom}
\end{figure}

The monochromator used for our reflectivity measurement is the McPherson 218 vacuum ultraviolet (VUV) model as shown in Fig.~\ref{pic_monochrom}. It has adjustable entrance and exit slits, two reflecting mirrors and a diffraction grating. During the operation, the incoming light passes though the entrance slit, and is reflected by the entrance reflecting mirror toward the diffraction grating. Through the process of light diffraction and interference, a spectrum is created and projected onto the exit reflecting mirror, then a narrow band of light is let through by the exit slit. Both the entrance and exit reflecting mirrors are at fixed positions; the incident angle of the light to the diffraction grating is adjusted by the wavelength adjusting mechanism, this will select which part of the light spectrum is reflected by exit mirror and passes through the exit slit. The complete light path inside the monochromator is indicated in Fig.~\ref{pic_monochrom}.

The VUV model monochromator is capable of operating under a vacuum of (10$^{-6}$ torr) or under a purged condition. The wavelength adjustment is achieved using a remote electronic controller which makes data acquisition much more efficient. The optimal operating wavelength range is dictated by the diffraction grating used. To optimize the light signal around 200~nm, a holographic 200~nm blaze (1200 G/mm) grating is used, so that the coherent light spectra has a peak intensity around 200~nm. The system must be carefully calibrated before taking any measurement.

\subsection{Setup}

%
\begin{figure}[p]
	\centering
	\subfloat{\includegraphics[width=0.75\textwidth]{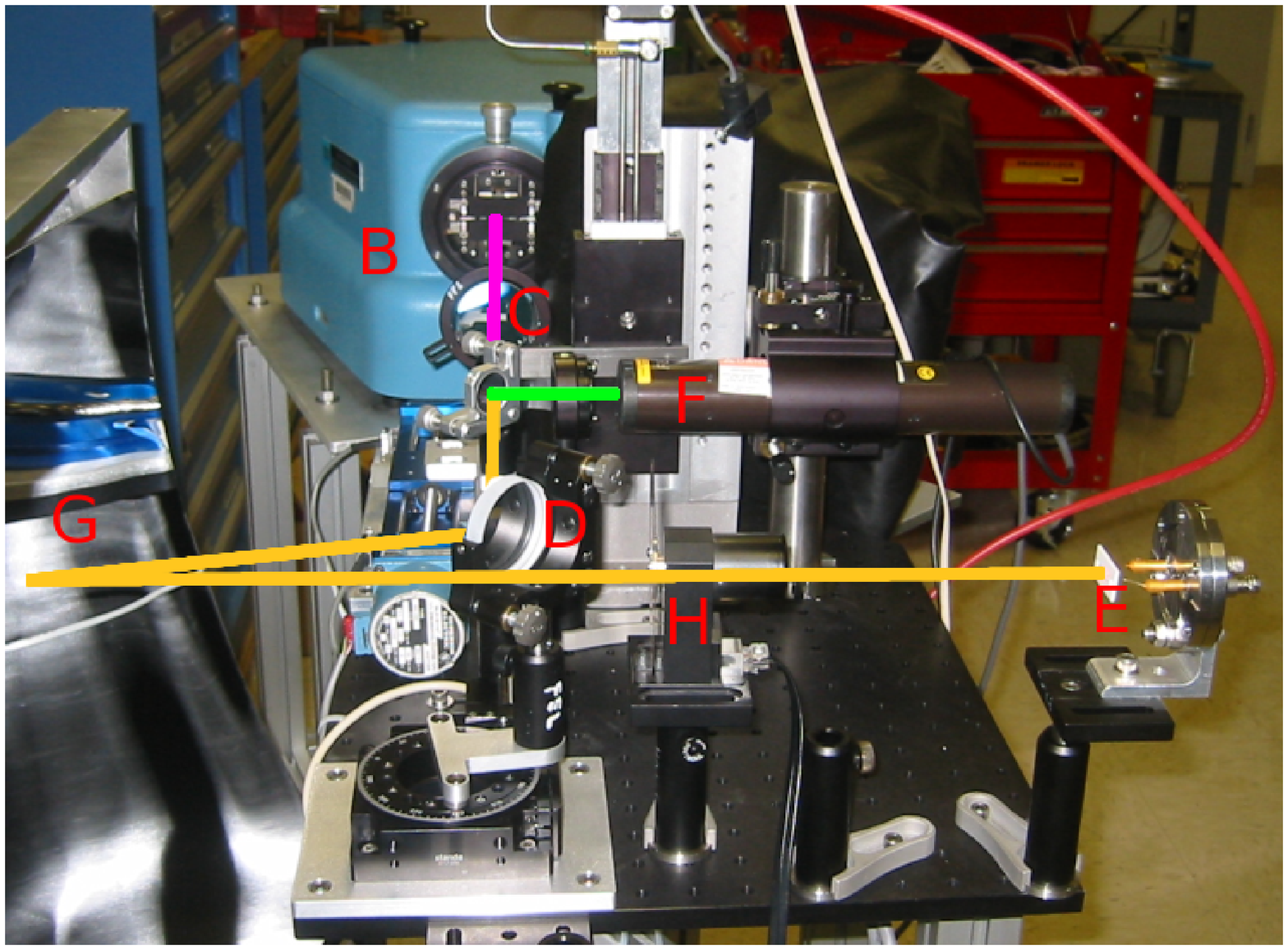}} \\
	\subfloat{\includegraphics[width=0.75\textwidth]{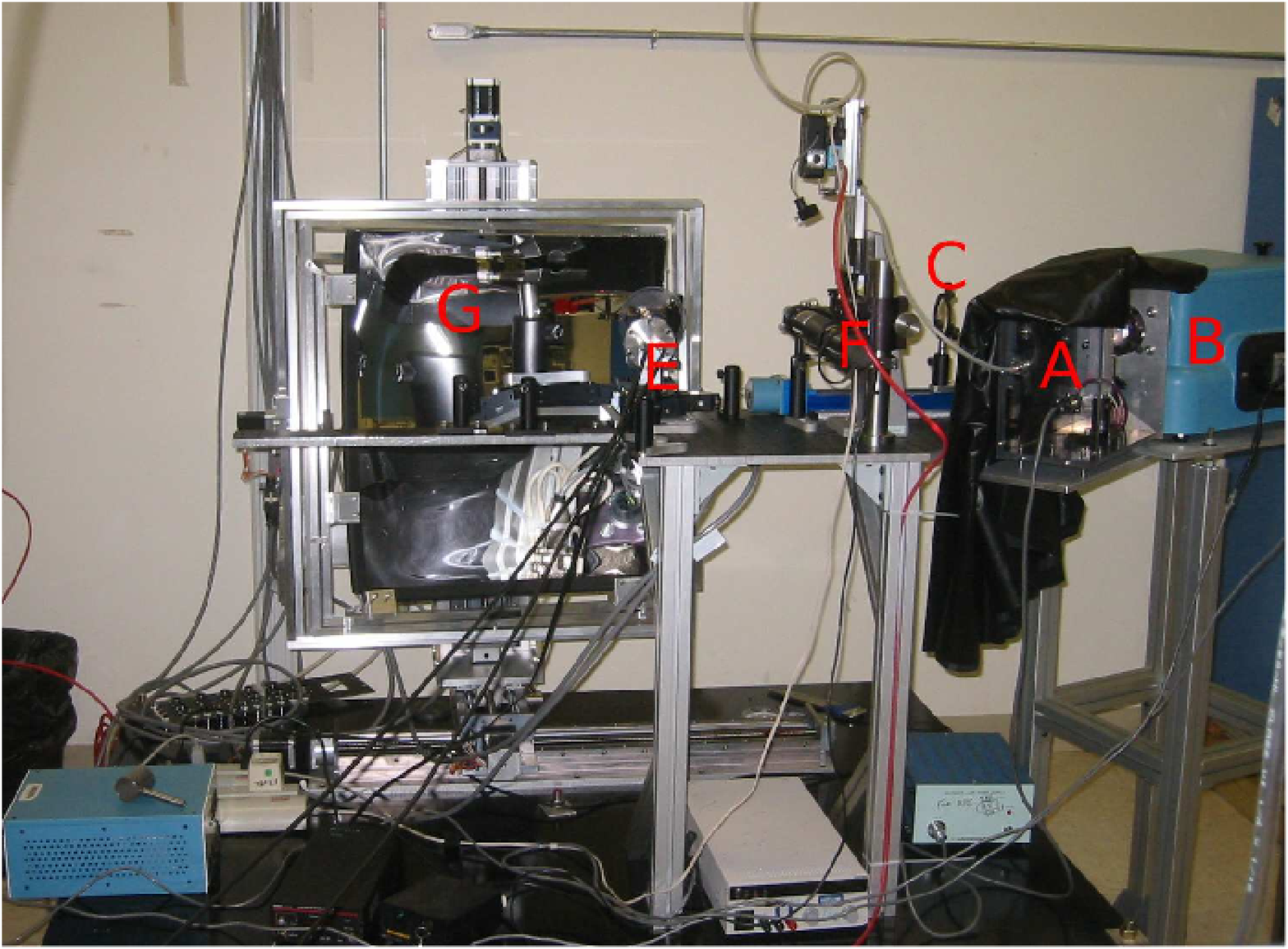}}
	\caption[Mirror Reflectivity Setup Images]{Images of the mirror reflectivity setup.  A: Light source box; B: Monochromator; C: Focusing lens; D: DUV Flipper Mirror; F: Alignment Laser Stage; E: Detector; G: Test Mirror; H: Optical Chopper. The light path is shown on the top image, the monochromator light path is in magenta; the alignment laser light path is in green; the shared (common) light path is in yellow. \oic}
	\label{setup}
\end{figure}

The main body of the setup is constructed in three parts, as shown in Figs.~\ref{setup} and \ref{schematic}: the light source box, the monochromator and the detector hutch. In Fig.~\ref{setup} the hutch was not yet installed.

The light box houses the Hamamatsu deuterium lamp and F1.5 focussing lens. The distance between the lamp and the lens is 4~cm, and the distance between the lens and the light entering the slit of the monochromator is also 4~cm.

Inside of the detector hutch, the test mirror is clamped onto the hosting bracket and then mounted to the angle adjusting frame. The stepper motors attached to the frame are able to change the pitch and yaw angle with an accuracy of 0.1$^\circ$. The whole frame sits upon a three axis translation stage, which allows the source light from the monochromator to reach any spot on the sample mirror.

During the reflectivity measurement, the detector position is fixed at a constant point. This means the sample mirror's position and angle need to be adjusted to reflect the source light towards the detector. Since the human eye does not respond to Medium Ultraviolet (MUV) or Far Ultraviolet (FUV) light, an alignment laser is introduced to match the monochromator light path, in order to help guide the mirror adjustment. The laser is installed perpendicular to the monochromator light path, and a pneumatically controlled stage device inserts a small reflecting mirror to intercept the light path if the laser is needed for alignment (see Fig.~\ref{setup}).

The air starts to block UV light around 190~nm, so for the reflectivity measurement at wavelengths below this value, an oxygen free environment is required. A high vacuum environment would be ideal, but the cost of a large vacuum chamber is beyond the budget. Instead, a pure N$_2$ environment is used, because the leak tight requirements for a purged system are much less demanding. All three parts of the setup can be separately fed with cool N$_2$ gas. The volume of the hutch needs several hours of purging before reaching the condition necessary for reflectivity measurements. The purging process continues during the data taking period to maintain a low oxygen level and also to cool the UV lamp. An oxygen monitor is placed inside of the hutch to read out the oxygen level. For the HGC mirror reflectivity measurement, the wavelength range of interest is 190-400~nm, therefore a pure N$_2$ environment is not required.

\subsection{Methodology}

The list below gives the symbols we used to represent the distances between the important components of the setup throughout the chapter: 
\begin{description} 
\item $d_o$: Distance between beam out slit of monochromator to focusing lens.
\item $d_{l-f}$: Distance between focusing lens and flipper mirror. 
\item $d_{f-d}$: Distance between flipper mirror and detector.
\item $d_{l-d}$: Distance between focusing lens and detector. 
\item $d_{f-m}$: Distance between flipper mirror and Mirror \#8.
\item $d_{m-d}$: Distance between Mirror \#8 and detector.
\item $d_{i}$: Total length of light path from the lens and detector (The sum of $d_{l-f}$, $d_{f-d}$, $d_{l-d}$, $d_{f-m}$ and $d_{m-d}$ for each mode). 
\item $A$: Focused beam spot at the detector at 225~nm wavelength.
\item $T_{flipper}$ Flipper mirror angle according to the rotation stage angle indicator. 
\end{description}

We performed wavelength scans between 190-400~nm at three different modes to compute the final reflectivity results and their descriptions are listed as follows: 
\begin{description}
\item[No Reflection (NR) Mode: ] Light from the monochromator slit was focused by the focusing lens and the image was directly projected onto the detector. The slit-lens distance ($d_o$) was 31~cm, and the lens-detector distance ($d_{l-d}$) was 69~cm. The wavelength scan at this mode was used as the reference measurement for determination of the flipper mirror reflectivity. Only one measurement was taken.

\item[Flipper Mirror Reflection (FMR) Mode:] A schematic diagram is shown in Fig.~\ref{fmrmode}. The $d_o$ was the same as for the NR mode. However, the light was then reflected off the Deep Ultra Violet (DUV) flipper mirror and projected onto the detector. The total light path of this mode was equivalent to the NR mode. The flipper mirror was rotated to 10.75$^\circ$ according to the rotation stage indicator and the light incident angle to the flipper mirror was 47.5$^\circ$. Other key parameters are listed in Table \ref{key_para}. The wavelength scans at this mode were used to determine the flipper mirror reflectivity, as well as to provide a reference measurement to compute the test mirror reflectivity. A total of eight wavelength scans were taken in the 8 hour testing period, which includes: FMR 1 and FMR 2, were measured immediately after the NR measurement; FMR, Right FMR, Center FMR, Top FMR, Bottom FMR and Corner FMR, were measured before the each of the Mirror~\#8 reflection measurements. The measurement names were given based on the Mirror~\#8 reflectivity measurement location as shown in Fig.~\ref{mirror_pos}.  

\item[Mirror \#8 Reflection (M8R) Mode:] A schematic digram of this mode is shown in Fig.~\ref{rmrmode}. With $d_o$ reduced to 22.2~cm, the flipper mirror was rotated by 95$^\circ$, where the rotating stage indicator should record 105.75$^\circ$. After the focusing lens, the light was first reflected by the flipper mirror towards the test mirror instead of the detector, then the test mirror reflected the light towards the detector. The light incident angles to the flipper mirror and test mirror were 47.5$^\circ$ and 5.5$^\circ$, accordingly. From Table \ref{key_para}, the total light path of this mode was two times longer than the NR and FMR modes. The focused spot size was also larger. A wavelength scan in this mode was used to compute the test mirror reflectivity. There were six measurements in total in this mode at the locations indicated in Fig.~\ref{mirror_pos}. Each Mirror~\#8 reflection measurement was taken after the corresponding FMR measurement with the same naming convention.
\end{description}

The flipper mirror reflectivity is computed as
\begin{equation}
R_{flipper} = \frac{I_{FMR}}{I_{NR}}
\end{equation}
where $I_{FMR}$ is the signal strength at arbitrary wavelength in Flipper Mirror Reflection Mode, and $I_{NR}$ is the signal strength at the same arbitrary wavelength in the NR Mode. 

The Mirror \#8 reflectivity is determined as
\begin{equation}
R = \frac{I_{M8R}}{I_{FMR}}
\end{equation}
where $I_{M8R}$ is the signal strength at arbitrary wavelength in M8R Mode, and $I_{FMR}$ is the signal strength at the same arbitrary wavelength in the FMR Mode.

Table \ref{key_para} summarizes the distances between the important components in different measurement modes. The focal length of a lens is dependent on the wavelength of the incoming light. At all three measurement modes, the focal length for the optics are optimized for 225~nm where the signal is the strongest, so a sharp image could be projected on the detector. The focused image sizes at 225~nm are listed in Table \ref{key_para}. A wavelength scan from 190-400~nm would imply for most of the measurements (wavelengths), the projected image on the detector would be slightly out of focus (blurry).  However, the active area of the detector is 1~cm$ \times$1~cm, which is sufficient to detect the slightly blurry image.

\begin{figure}[p]
	\centering
	\subfloat[No Reflection Mode]{\includegraphics[width=0.49\textwidth]{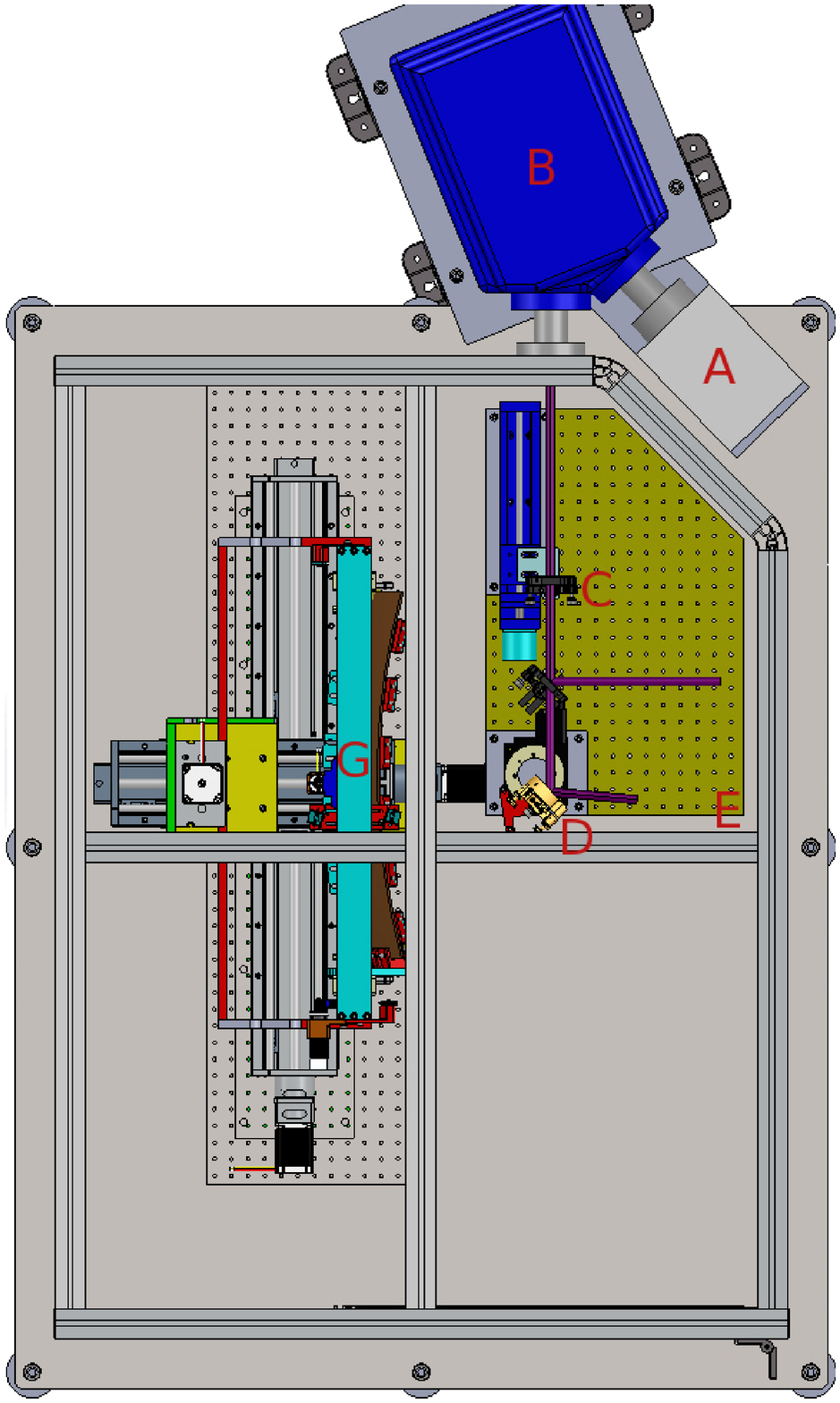} \label{fmrmode}}
	~
	\subfloat[Flipper Mirror Reflection Mode]{\includegraphics[width=0.49\textwidth]{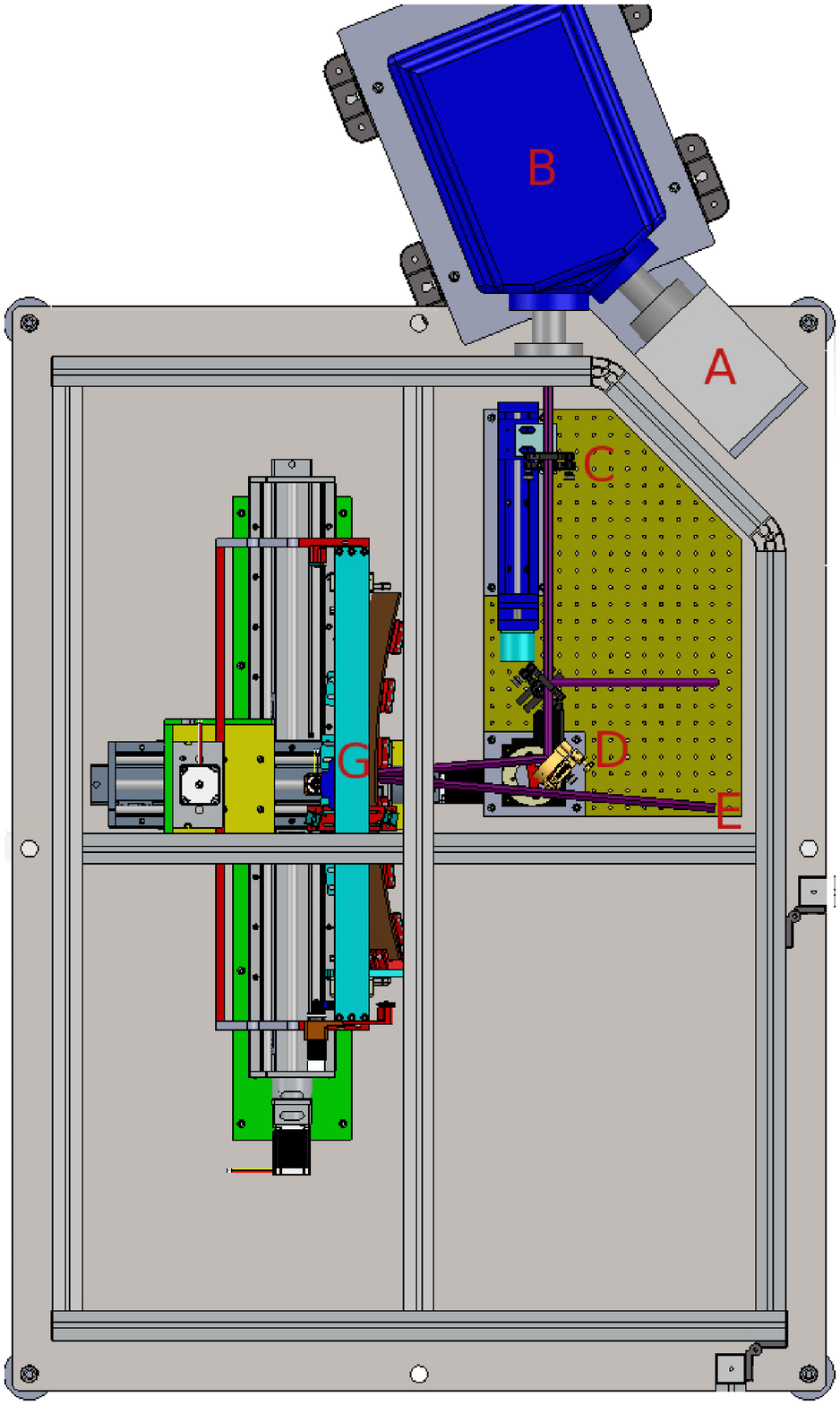} \label{rmrmode} }
	\caption[Schematic Diagram of the Reflectivity Measurement Modes]{Schematic diagram for different measurement modes. All important components are labelled with a capital letter. A: Light source box; B: Monochromator; C: Focusing lens; D: DUV Flipper Mirror; E: Detector; G: Mirror Holding Frame. Position of the focusing lens (C) and flipper mirror (D) are different between the two modes \cite{gould}. \oic }
	\label{schematic}
\end{figure}

\begin{figure}[t]
\centering
	\subfloat[Witness sample placement]{\includegraphics[width=0.5\textwidth]{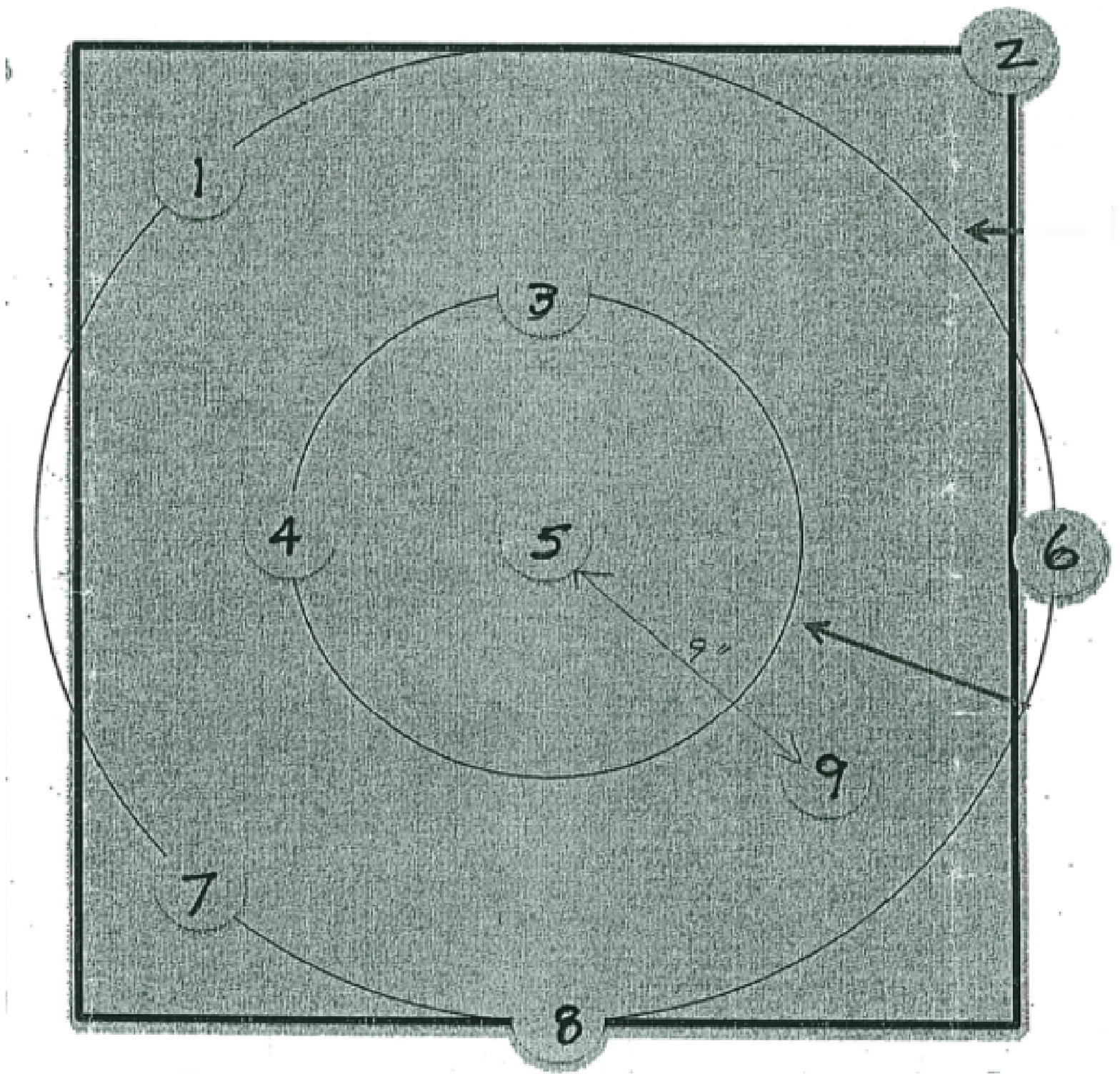}\label{sample_pos}}
	\subfloat[Mirror Measurement Position]{\includegraphics[width=0.5\textwidth]{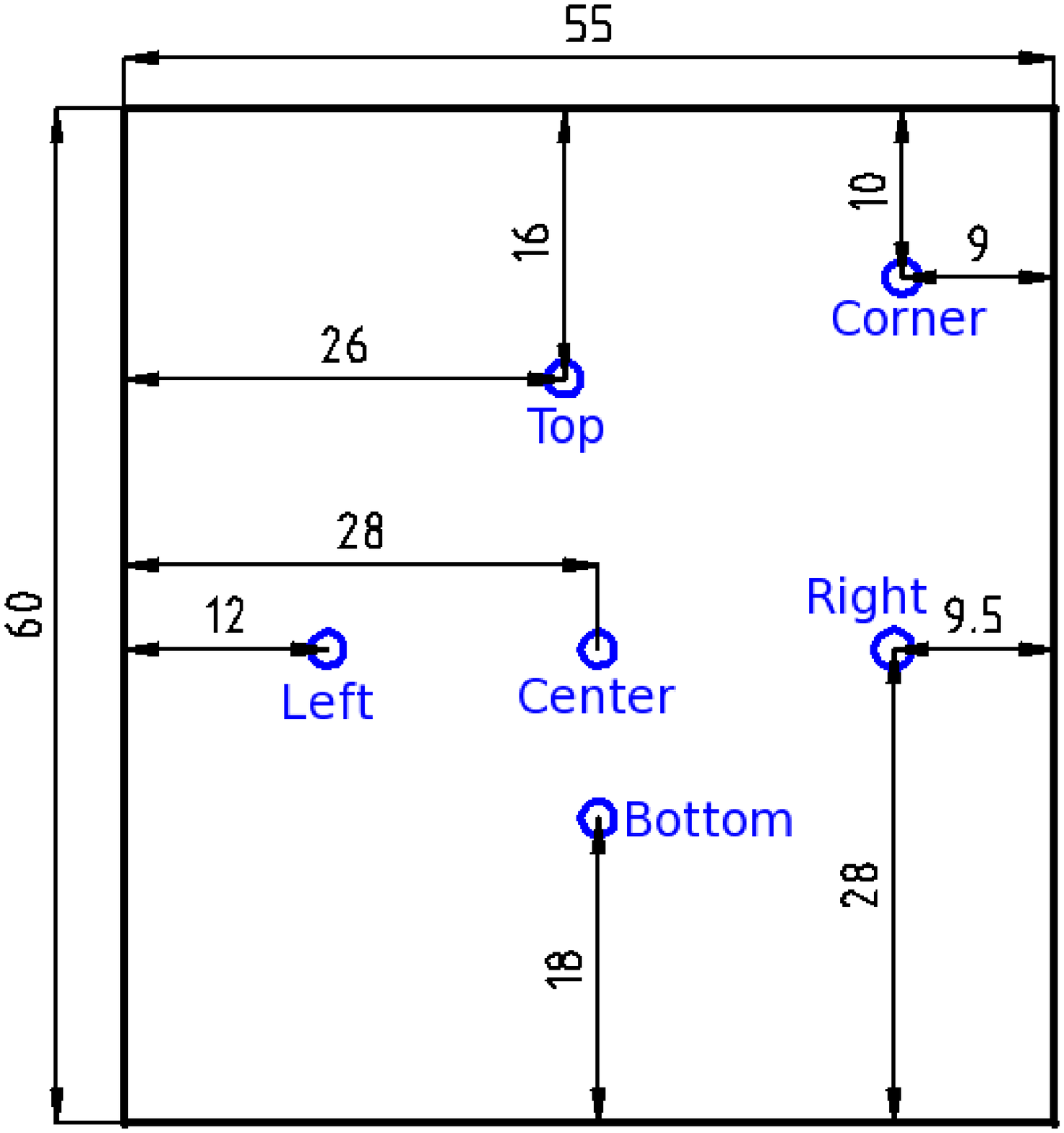}\label{mirror_pos}}
	\caption[Reflectivity Measurement Location]{(a) Witness sample locations with respect to the main mirror position in the ECI aluminization. (b) Reflectivity measurement locations on Mirror \#8. \oic}
\end{figure}

\begin{table}[t]%
\centering
\caption[Key Parameters for the Reflectivity Measurement]{The key parameters for all measurement modes.}
\begin{tabular}{cccc}
\toprule
                &   No Reflection     & Flipper Mirror     & Real Mirror          \\
                &   Mode  			  & Reflection Mode    & Reflection Mode      \\
\toprule                              
$d_o$           &  31~cm               &  31~cm              & 22.2~cm               \\
$d_{l-f}$       &  -                  &  42~cm              & 43.3~cm               \\
$d_{f-d}$       &  -                  &  27~cm              & -                    \\
$d_{l-d}$       &  69~cm               &  -                 & -                    \\
$d_{f-m}$       &  -                  &  -                 & 34~cm                 \\
$d_{m-d}$       &  -                  &  -                 & 60~cm                 \\
$d_i$	        &  69~cm               &  69~cm              & 137.3                \\
$A$             &  6~mm$ \times$5~mm     &  6~mm$ \times$5~mm    & 8~mm$ \times$7~mm       \\
$T_{flipper}$   &  -                  &  105.75$^\circ$    & 10.75$^{\circ}$      \\

\bottomrule
\end{tabular}
\label{key_para}
\end{table}

\subsection{Detector and Signal Schematic Diagram}

The IRD AXUV-100G photo-diode detector has a different signal response depending on the incoming photon wavelength, with the response curve shown in Fig.~\ref{pic_axuv-100G_response_curve}.

For all measurement modes, the continuous light signal was converted to a pulsed light signal by a Thorlab MC100 optical chopper before reaching the detector. The chopper had a two-slot chopping blade operated at a frequency of 14~Hz. The chopper generated a reference frequency, which was fed to the reference channel of the lock-in amplifier. The detector generated a pulsed DC differential signal through two connecting points, and the two signals were fed into the input channels A and B of the lock-in amplifier. Then, the lock-in amplifier would subtract signal B from A, and perform the filtering and amplifying algorithm. The schematic diagram of the signal treatment is shown in Fig.~\ref{signal}.

\begin{figure}[t]
	\centering
	\includegraphics[width=0.75\textwidth]{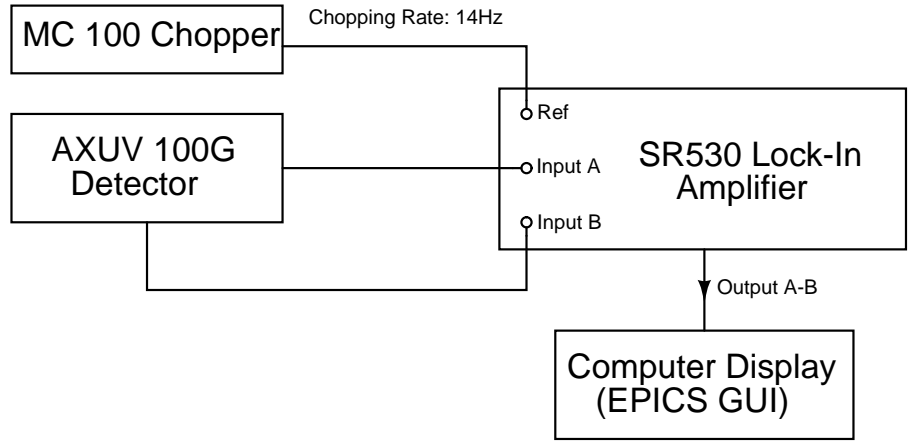}
	\caption[Signal Output Diagram]{Systematic diagram for the signal output.}
	\label{signal}
\end{figure}

\subsection{Lock-in Method and Monochromator Scanning Setting}

The lock-in technique is used to detect and measure very small AC signals. A lock-in amplifier can make an accurate measurement of small signals, even when the signals are obscured by background noise thousands of times larger. Essentially, a lock-in filter is a filter with an arbitrarily narrow bandwidth which is tuned to the frequency of the signal. Such a filter rejects most unwanted noise to allow the signal to be measured.

All lock-in measurements share a few basic principles. The technique requires that the experiment be excited at a fixed frequency in a relatively quiet part of the noise spectrum. Then, the lock-in can detect the response from the experiment in a very narrow bandwidth at the excitation frequency.

In this measurement, the SR530 lock-in amplifier was operated in $R,\theta$ display mode, where $R$ stands for the signal magnitude and $\theta$ is the phase between measured signal and the reference. The reference signal was generated by the optical chopper, which chopped at the rate of 14 Hz. 
%

There was no special option needed to filter the signal from the detector for our measurement. During the data taking, the sensitivity and time constant were the most important parameters that could dictate the data quality. The sensitivity must be set to not overload the system. The time constant dictated how fast the RC circuit of the output channel discharges, which would control signal registration. A shorter time constant would allow a faster response to the signal change but larger fluctuation in signal strength, and a longer time constant would register stable signal strength but take longer to react to any change in signal. The preferred time constant settings during the measurement were 1~s and 3~s. 

For the data taking, the monochromator was configured to perform a wavelength scan from 190-400~nm at 5~nm steps. The dwell time at each wavelength was 20~s, and the associated time constant was 1~s or 3~s to generate a flat signal output. If the dwell time was 60~s, the recommended time constant would be 3~s or 10~s (the time constant does not need to be larger than 10~s). 

A baseline (background) measurement should be taken before any measurement. The monochromator light was diverted away from the detector so that only the background light falling into the chopping frequency window could be registered. The baseline (background) signal was measured to be 1.6$\times10^{-5}$ V consistently, and $R_{offset}$ was set for this value during all reflectivity measurements.

The signal magnitude $R$ can be can be calculated as \cite{sr530}
\begin{equation}
R=\sqrt{(x+x_{offset})^2+(y+y_{offset})^2} + R_{offset}
\label{R_mode}
\end{equation}
where $x$ and $y$ are the horizontal and vertical components of the signal magnitude $R$.


\section{Results}

\subsection{Deuterium UV Lamp Thermal Test Results}
The 3 Watt Hamamatsu deuterium UV source's stability has been questioned from the beginning. It was purchased 20 years ago by the FFL Facility at Jefferson Lab. A thermal effect test was carried out to investigate the stability of the UV lamp for long operation at room temperature. The monochromator was set for 225~nm wavelength, corresponding to a maximum signal output of 0.25 mV. The No Reflection (NR) mode was used for this test, where the light from the monochromator was directly focused onto the detector by the focusing lens. The $x$ and $y$ outputs from the lock-in amplifier were monitored separately, as shown in Fig.~\ref{thermal}.

\begin{figure}[t]
	\centering
	\includegraphics[width=0.75\textwidth]{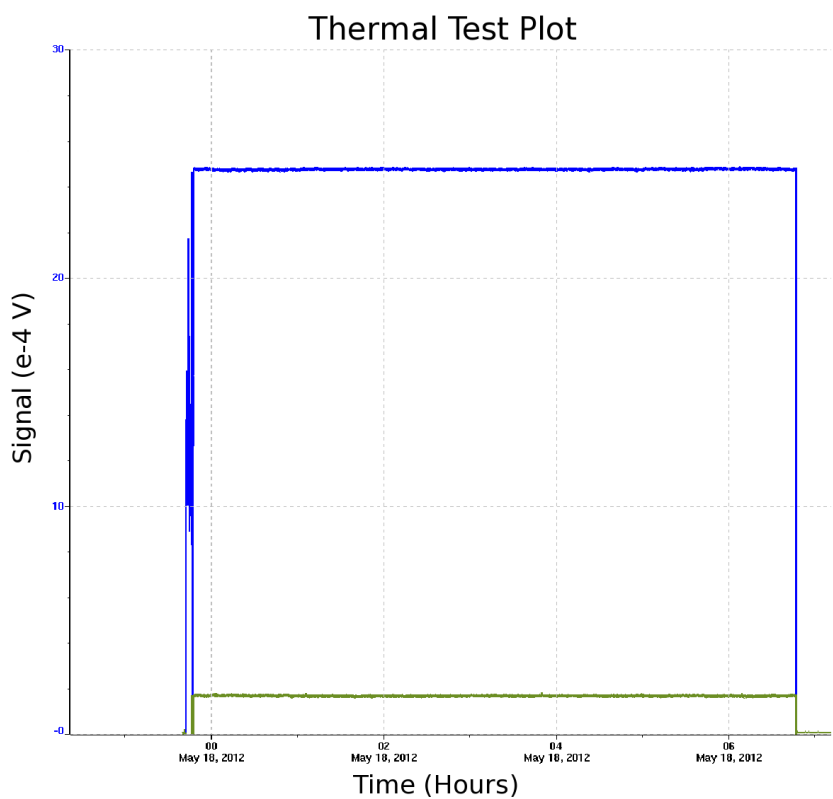}
	\caption[Thermal Test Result]{Lock-in amplifier $x$, $y$ strength monitoring of the Deuterium lamp for 6 hour testing period. The blue curve is the $x$ output and green curve is the $y$ output. $x$ and $y$ are the horizontal and vertical component of the signal magnitude $R$. \oic} 
	\label{thermal}
\end{figure}

After 6 hours of operation, we observed no fluctuation in either $x$ or $y$ signal output, therefore the signal magnitude $R$ was also constant according to Eq. \ref{R_mode}. From this result, we can safely conclude the UV source performance did not change due to heat and long operation, even without any method of cooling and ozone removal. For measurement after the facility is fully commissioned, the light source box will be constantly purged with cold N$_2$ vapour, thus the UV source intensity should be stable during the long data taking process.

%

\begin{figure}[p]
	\begin{centering}
	\subfloat[Detector Response]{\includegraphics[width=0.75\textwidth]{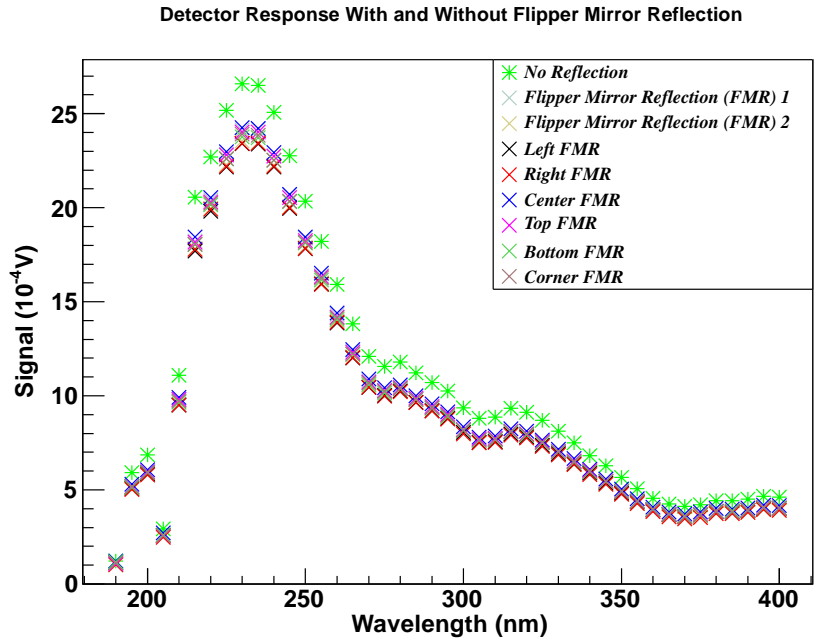}\label{fmr_response}} \\
	\subfloat[Flipper Mirror Reflectivity]{\includegraphics[width=0.75\textwidth]{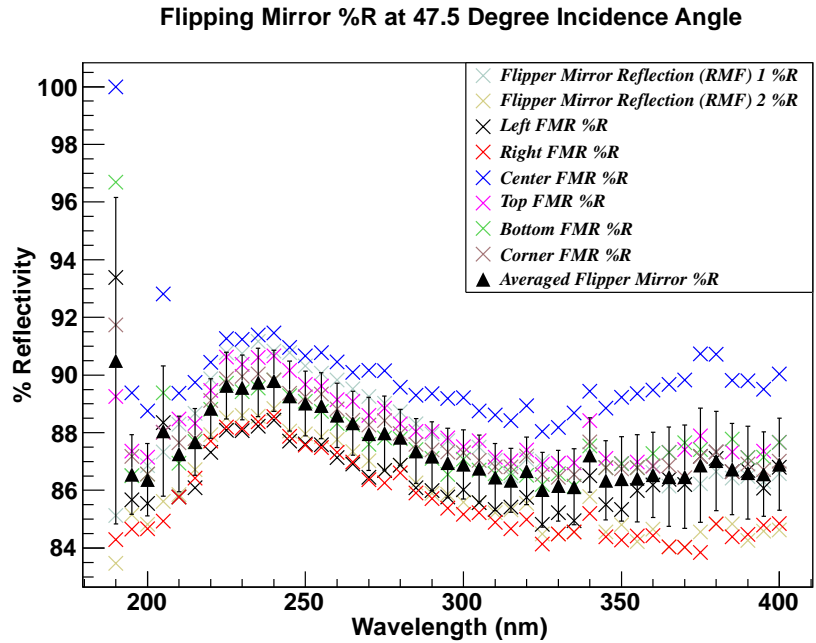}\label{fmr_reflectivity}}
	\caption[Flipper Mirror Reflectivity]{Flipper mirror reflectivity results. (a) shows the detector response; (b) shows the percentage reflectivity results. The standard deviations (error bars) of flipper mirror reflectivity curve are used as the systematic errors ($\delta_{measurement}$) for the Mirror \#8 reflectivity curve. \oic}
	\label{fmr_results}
	\end{centering}
\end{figure}

\subsection{UV Detector Response Uniformity}
The UV detector response uniformity was tested with a focused light beam at 225~nm wavelength over an area of 6 mm $\times$4 mm. Five measurements were taken with the light beam shining at different locations: one at the center and four at the corners with 1 mm clearance to the edges. The response signal differences were around 1-2\%. During the reflectivity measurement, after changing the measurement location on the mirror, the focused light beam was not guaranteed to be returned to the same location on the detector. The detector uniformity would contribute an uncertainty in the mirror reflectivity result. However, we think it was already included in the total error estimated by data reproducibility from the eight repetition measurements at FMR Mode.

\subsection{DUV Flipper Mirror Reflectivity Result}

In Fig.~\ref{fmr_response}, one measurement of NR mode and eight measurements of FMR mode wavelength scan curves are plotted. Measurements FMR1 and FMR2 were taken immediately after the NR measurement. The other six FMR measurements were taken separately within a 6 hour period. The ratios between each of the FMR curves and the corresponding NR curves are plotted in Fig.~\ref{fmr_reflectivity}. The averaged reflectivity curve and the standard deviations (error bars) are also shown. The standard deviation of the eight reflectivity measurements is between 1-2\% at most of measured wavelength, excepting at 190~nm, where the signal is too close to the baseline to be trusted. The standard deviations (error bars) of flipper mirror reflectivity curve in Fig.~\ref{fmr_reflectivity} represent the reproducibility error of a measurement, they are used as the systematic uncertainties ($\delta_{measurement}$) for Mirror \#8 reflectivity curve in Fig.~\ref{real_reflectivity}.

Between 220-400~nm, the averaged reflectivity gradually increases as the wavelength goes deeper into the UV region, peaks around 235~nm at 89.5\%, and then starts to decrease. At 200~nm the reflectivity is 86.5\%. The measured curve is different from the typical reflectivity curve for the Melles Griots DUV flipper mirror, see App. \ref{duv_spec}. The typical curve peaks around 197~nm at 94.5\%, and gradually decreases as the wavelength increases. At 250~nm the reflectivity is 91\%. The difference between the measured and typical reflectivity curves is around 3\% between 220-250~nm; this difference keeps increasing as the wavelength gets shorter; at 200~nm, the difference is 8\%.

The possible explanations for such discrepancy are stated as following: the typical DUV mirror reflectivity curve is a theoretical prediction under the ideal condition, whereas the quality of individual DUV mirror could deviate; also the mirror surface was scratched during the alignment process and the scratches would result in some scattered light.


\subsection{HGC Mirror \#8 Reflectivity Results}

Fig.~\ref{real_response} shows six sets of FMR and M8R detector response curves. Each set of FMR and M8R measurements were taken after the mirror was adjusted to a new position for testing. For the FMR measurements, the signal output was unrelated to the mirror position because the light was only reflected from the flipper mirror. For the M8R measurements, the lens and flipper mirror positions needed to be adjusted, so the light could reflect of the flipper mirror and Mirror \#8. The FMR and M8R measurements were taken back-to-back, and the lens and flipper mirror positions were adjusted remotely, so that the testing environment was not disturbed. The orientation of the measured position (Left FMR, Right FMR and so on) on the mirror are demonstrated in Fig.~\ref{mirror_pos}.


%

Fig.~\ref{real_reflectivity} shows the reflectivity curves of Mirror \#8, which are the ratios between the M8R and FMR measurements of the same label, for example (Left M8R curve)/(Left FMR curve). From the results it is seen that between 200-300~nm the reflectivity at the center is worse than at the corners; between 300-400~nm the reflectivity at the center is marginally better than at the corners. The systematic uncertainties for the Mirror \#8 reflectivity curve in Fig \ref{real_reflectivity} are taken as the same as that for the flipper mirror reflectivity curve, which is shown in Fig.~\ref{fmr_reflectivity}.


\begin{figure}[p]
	\begin{centering}
	\subfloat[Detector Response]{\includegraphics[width=0.75\textwidth]{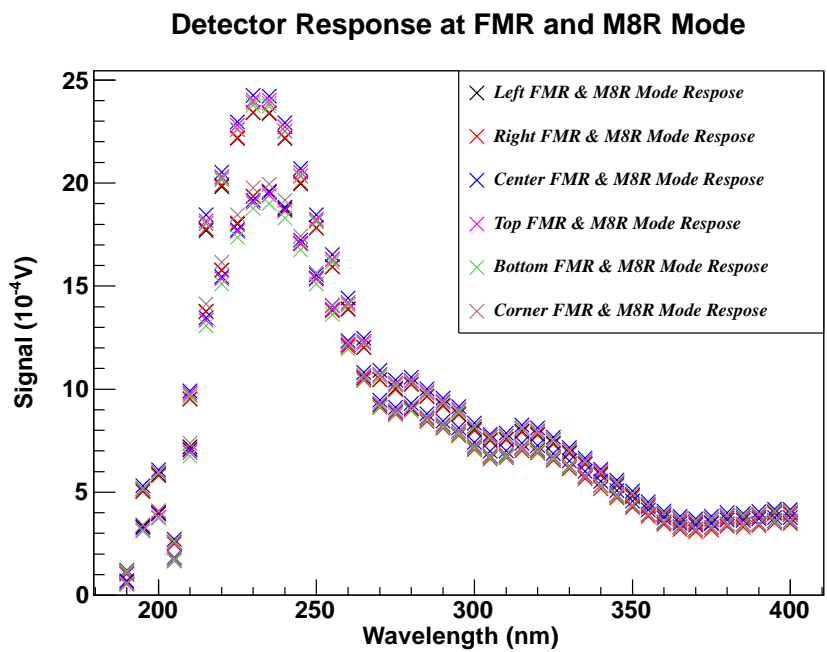}\label{real_response}}\\
	\subfloat[Mirror \#8 Reflectivity Results]{\includegraphics[width=0.75\textwidth]{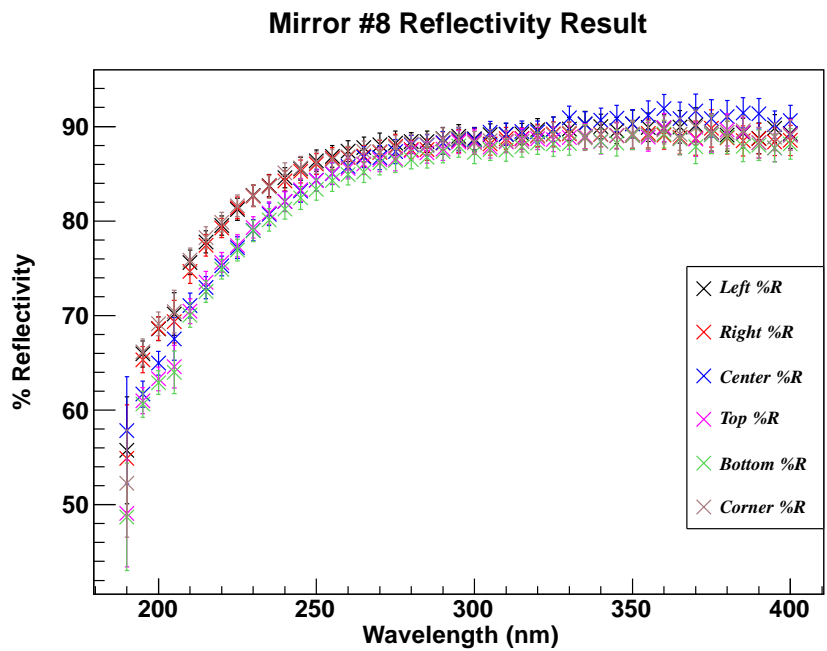}\label{real_reflectivity}}
	\caption[Mirror \#8 Reflectivity Results]{Mirror \#8 reflectivity results. In (a), the top set of measurements are in FMR mode and bottom set of measurements are in M8R mode. \oic}
     \label{pic_mirror8reflectivity}
	\end{centering}
\end{figure}

During the aluminization process, ECI placed nine 1$\inchsign$ witness samples at different locations across the test mirror as shown in Fig.~\ref{sample_pos}. The samples' reflectivities were tested for product quality control, and these measurements are attached in App.~\ref{sample_results}. We did not measure the reflectivity of the witness samples because the dispersed UV light beam from the monochromator is too large to be completely reflected by them. 

The ECI sample results have significant jitter, as shown in Fig.~\ref{ECI_sample_result_200-250}. The systematic uncertainty can only be estimated depending on the variation level of the jitter. The estimated error below 230~nm is $\delta_{ECI}$ = $\pm$1\% and above 230~nm is $\delta_{ECI}$ = $\pm$0.5\%.

Witness sample reflectivity curves \#4, \#5 and \#7 were chosen to compare with the measured reflectivity curves at positions Left, Center and Corner between 200-250~nm wavelength. The mirror testing positions of \#4 and Left, \#5 and Center, \#7 and Corner are similar. The witness sample curves were subtracted from the measured curves, and the difference curve is plotted in Fig.~\ref{difference}. For most of the points, the difference between the ECI witness sample and Mirror \#8 reflectivity is between 0-3\%, which means the measured Mirror \#8 reflectivity is higher than the witness sample reflectivity provided by ECI. At the corner, the differences are slightly below 0, but 0 is still within the error bar. In the ECI witness sample results there is a sharp peak at 205~nm which does not exist in our measurement. 


The error bar in Fig.~\ref{difference} is calculated by
\begin{equation}
 \delta_{Difference}= \sqrt{\delta_{ECI}^2 + \delta_{Measurement}^2}
\end{equation}
where $\delta_{ECI}$ is the estimated ECI witness sample curve error and $\delta_{Measurement}$ is our measurement error.

\begin{figure}[t]
	\centering
	\includegraphics[width=0.75\textwidth]{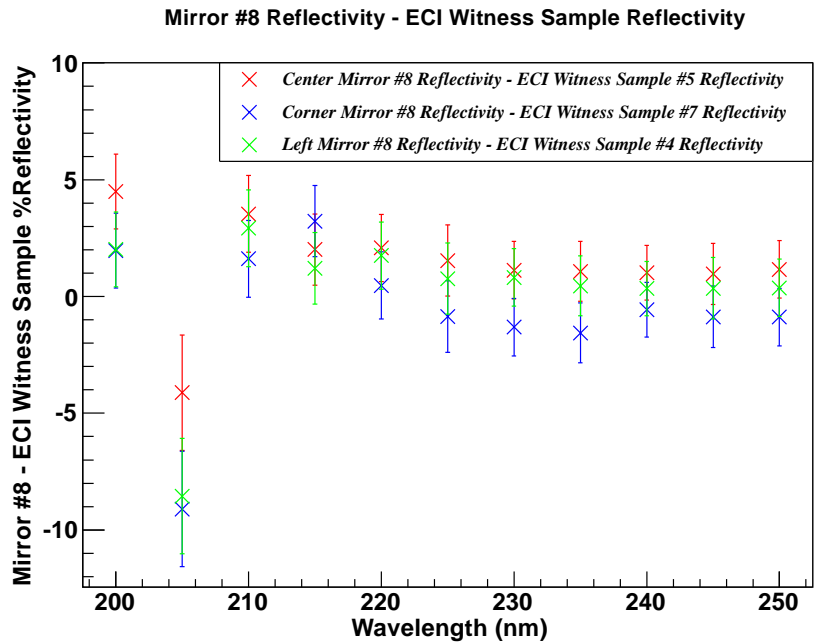}
	\caption[Difference Between ECI and Measured Result]{The difference between the witness samples and the Mirror \#8 measurement reflectivity results. \oic}
	\label{difference}
\end{figure}

\section{Conclusion}
Based upon our HGC Mirror \#8 reflectivity measurements, we confirm that the ECI witness sample measurement represents the reflectivity performance of the aluminized mirror surface between 200-400~nm. At 200~nm wavelength, the reflectivity is 10\% lower than the original ECI claimed value of 76\% with the worst witness sample results. However, this difference decreases rapidly as the wavelength increases. At 250~nm, the difference between the sample curves and theoretical curves ranges between 0-5\%. Based on the Geant4 simulation results described in Chapter \ref{chapter_geant4}, the difference between the theoretical curve and worst sample curve corresponds to a decrease of 2.5\% in terms of the detected photo-electrons and 0.13\% in terms of detected pion events. This effect is insignificant compared to other significant contributions to the UV absorption in the deep UV region, i.e. the C$_4$F$_8$O gas absorption etc. We believe the ECI aluminization quality meets our performance specifications and is sufficient for the aluminization of mirrors to be actually used in the HGC. 

We are also delighted to report that the reflectivity setup at Jefferson Lab is capable of measuring the reflectivity of large size optics between 200-400~nm wavelength at normal atmospheric condition and is potentially able to measure the reflectivity down to 165~nm wavelength under the N$_{2}$ purged environment.

\section{Suggestions for Future Measurement}
It is not obvious if the current UV source is suitable for UV measurements below 190~nm. Further testing is required to confirm that the signal strength is sufficiently strong and stable at such low wavelength. One of the possible options is to replace the 3 Watt Hamamastu deuterium lamp with a Newport Oriel 30 Watt deuterium lamp, which has much higher intensity at lower wavelength.

At the moment, the setup is in a semi-automatic state, where some manual fine adjustments are required. Also, the cameras which can help with the alignment and status check are yet to be installed. For the Noble Gas \v{C}erenkov mirror reflectivity measurements, the detector hutch enclosure with pure N$_2$ environment will forbid any manual adjustment and alignment. Some level of automation is needed. However, it is not obvious at the moment how accurate automatic alignment can be achieved. Manpower needs to be committed to resolve this issue.

}

%% file: truck/g4_simulation.tex
{

\label{chapter_geant4}

\section{Introduction}
In order to better understand the optical alignment and expected performance of the HGC, a computer simulation model was constructed using the standard software toolkit for high energy physics known as Geant4 \cite{geant4-1, geant4-2}. The original detector simulation platform was developed by the University of Virginia and many customizations have been implemented to study the HGC detector.    

In this chapter, we will describe the basic setup of the HGC simulation model and important results for the simulated performances.

There are three main detector performance criteria that we are studying: the photon focusing efficiency, the photon detection efficiency and the charged particle identification efficiency.

The photon focusing efficiency detection is defined as
\begin{equation}
\eta_{\gamma} = \frac{\textrm{Number of photons within 12~cm diameter of 4 PMTs}}{\textrm{Total Number of Photons}}\,,
\label{eqn_photon_gamma}
\end{equation}
this is the ratio between the number of photons focused onto the 12~cm diameter active area for all four PMTs and the total number of \v{C}erenkov photons. $\eta_{\gamma}$ is used to study the alignment of the optical components.

The photon detection efficiency is defined as
\begin{equation}
\eta_{pe} = \frac{\textrm{Number of Photo Electrons in all 4 PMTs}}{\textrm{Total Number of Photons}}\,,
\label{eqn_photon_pe}
\end{equation}
this is the ratio between the number of photo-electrons output by all four PMTs and the total number of \v{C}erenkov photons. $\eta_{pe}$ is used to study the effect of threshold cuts to optimize the $\pi/K$ separation in the analysis of the simulation data. The quantities $\eta_{\gamma}$ and $\eta_{pe}$ are closely related, and they will be explained in detail in Section \ref{g4_study_pmt} where the PMT optimization results are presented.

The charged particle identification efficiency is defined as
\begin{equation}
\eta_{particle} = \frac{\textrm{Number Identified Particles}}{\textrm{Total Number of Particles}}\,,
\label{eqn_particle_efficiency}
\end{equation}
this is the ratio between the number of identified charged particles passing the simulated HGC analysis threshold and the total number of charged particles passing through the SHMS. $\eta_{particle}$ is used to study the particle identification efficiency, which is of primary concern to the experiments that will make use of the HGC. It is explained in more detail in Section \ref{g4_expected_perfomance}, where the overall projected performance results are presented.

\begin{figure}[t]
	\centering
	\includegraphics[width=0.90\textwidth]{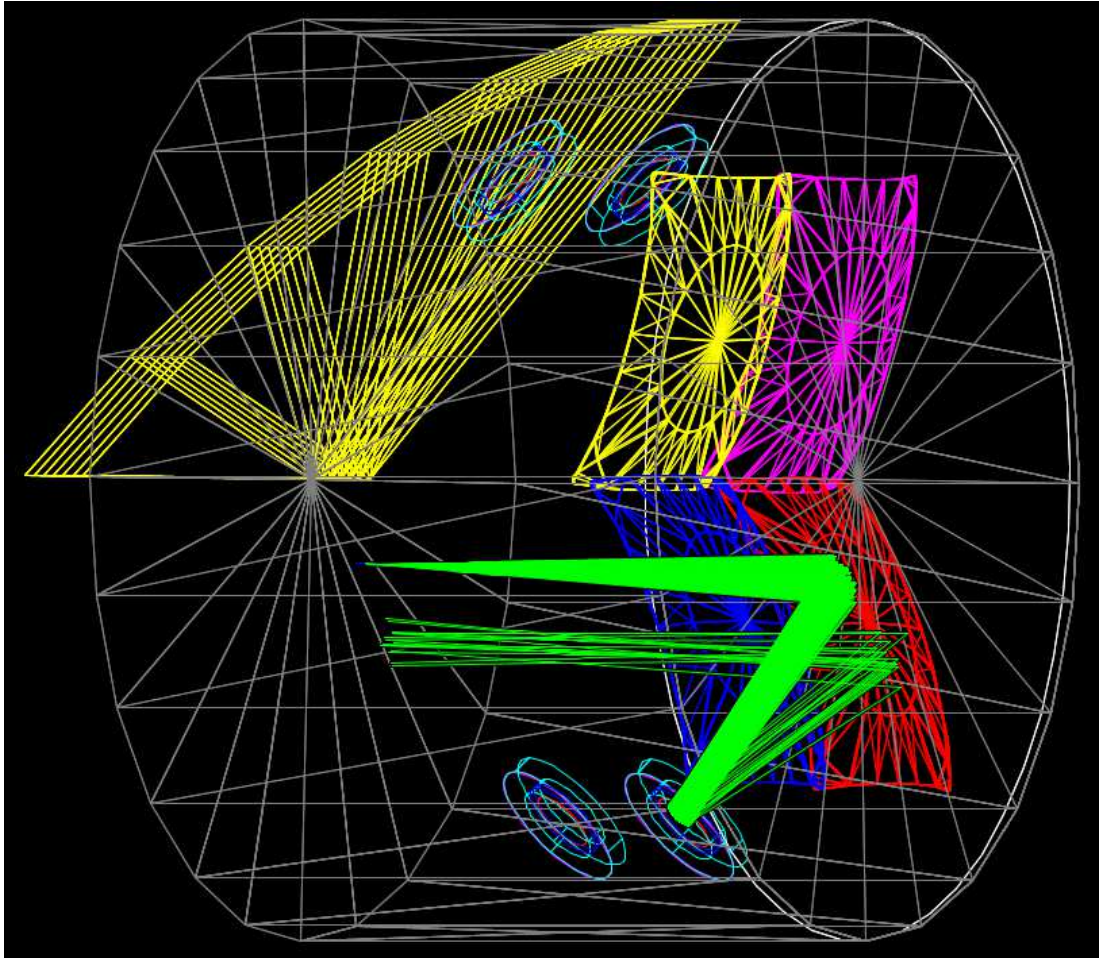}
	\caption[Geant4 HGC Detector Model]{HGC Geant4 model. The planes for Mirrors \#1, \#2 and \#4 are switched off to give better view to the photon trajectories in green. See the text for further details. \oic}
	\label{pic_g4_detector}
\end{figure}

\section{Simulation Setup}


In the simulation, the HGC detector dimension and optics arrangement corresponds to the real specification described in Chapter \ref{sec_hgc}, where the inner diameter of HGC detector is 170~cm, and length 117.1~cm. The center of the detector is offset 5~cm in the $+x$ direction. The detector is configured to contain C$_4$F$_8$O gas at 0.95 atm pressure, which has refractive index of 1.00137. 

The center corner of the four mirrors are interleaved in the order of \#: 4, 3, 2, 1, where mirror \#4 is in $-x$, $-y$ quadrant (magenta), \#3 is in $-x$, $+y$ quadrant (yellow), \#2 is in $+x$, $-y$ quadrant (red) and \#1 is in $+x$, $+y$ quadrant (blue). The closest mirror to mirror approach is between 7-10~mm. There is a 5~cm overlap between Mirrors \#1 and \#2 and the same for mirrors \#3 and \#4~in the $y$ direction. The coordinate convention used in this chapter is the same one as described in Section \ref{coordination}. In the HGC detector frame, the $z$ axis is parallel to trajectory of the charged particle, the x axis points down to represent increasing momentum in particle distribution, and the $y$ axis points to the left of the detector. In this chapter, the bottom two quadrant mirrors are referred as $+x$ mirrors and top two as $-x$ mirrors. The $+x$ mirrors are tilted at 20$^\circ$ and the $-x$ mirrors are at -16.2$^\circ$ with respective to the $x$ axis.

The simulation can run with three different types of mirror curvature: parabolic ($\kappa<0$), spherical ($\kappa=0$) or oblate elliptical ($\kappa>0$), where $\kappa$ is the conic constant defined in Section \ref{mirror_selection_methodology}. From Chapter \ref{mirror_selection}, we concluded that all of the manufactured HGC mirrors are oblate ellipsoid. Thus, the fitted radius R$=$112.96~cm and conic constant $\kappa=$0.94 for Mirror \#6 were chosen to create the oblate elliptical mirrors to provide more realistic simulation to study the detector performance. The Mirror \#6 quality is regarded as the average among the high quality mirrors based on the mirror selection study results (see Section \ref{mirror_selection_con}). All optical alignment and projected performance results presented in this chapter are with the oblate mirrors and the difference between the oblate and spherical mirror performance is also presented herein.

Each of the HGC mirrors reflect the \v{C}erenkov radiation towards the corresponding 5" PMT. In the simulation, the PMTs are tilted at $\pm$42$^\circ$ with respect to the x-axis. In the Geant4 model, the PMT assemblies consist of four parts: cathode, quartz window and flange. Quartz window properties such as the chemical composition and absorption length were configured according to the properties of Corning 7980 quartz \cite{corning}, where the material absorption length are converted from the transmissivity curve from the manufacturer, see App. \ref{app_quartz}. The 10\% photon loss was included in addition to the reflection due to the Fresnel reflection at the quartz surface as a safety factor in the simulation. The position of the cathode in the model corresponds to the real PMT cathode position, and the wavelength dependent quantum efficiency and radial dependent efficiency provided by the manufacturer \cite{hamamastu} were implemented to predict the PMT performance and particle identification efficiency. 

In Fig.~\ref{pic_g4_detector}, a single \v{C}erenkov event for a $\pi^+$ with 7~GeV/c momentum is shown. The pion trajectory is in blue and enters the detector at the tip of the green \v{C}erenkov cone. The pion trajectory is mostly overlapped by the \v{C}erenkov photon trajectories shown in green. The generated \v{C}erenkov photons are reflected by the curved mirror towards the PMT assembly, and a small portion of the photons are reflected back towards the mirror by the quartz window; eventually these photons will be absorbed by the detector wall. The position and energy of all photons are recorded and later plotted.

By default, there are 7 tracking planes to record the photon trajectory near the PMT location: 3 planes before and 3 planes after cathode plane, as shown in Fig.~\ref{pic_g4_detector}. The planes for Mirrors \#1, \#2 and \#4 are switched off in this figure to give a better view of the photon trajectories in green. The planes are separated by 2~cm in the $z$ direction and are parallel to each other. The PMTs can be turned on or off in the simulation; the PMT-on mode is used to study the particle identification efficiency, and the photons will stop at the cathode, therefore three tracking planes after the cathode plane are not used; the PMT-off mode is used to study the optical alignment, therefore all 7 planes are used.

The HGC simulation is capable of studying the particle performances of $\pi^{\pm}$, $K^{\pm}$ and protons, and the maximum sample size for a single simulation is 0.5 M events. In order to make sure the particle distribution at SHMS focal plane is accurate, 0.5 M charged particle events were generated for our study by a collaborator \cite{jones} using the single arm SHMS Monte-Carlo (MC) of Hall~C. This is a detailed MC of the SHMS magnetic optics, and the trajectories of 0.5 M charged particles are accurately tracked through the magnetic field elements. The generated particle distribution file records for every event: position and momentum in the $x$, $y$ directions in front of the HGC detector, and the momentum ratio $\delta$ with range of $-13<\delta<22$\%, where $\delta = \frac{\Delta P}{P}$ is the percentage value used to reconstruct the momentum acceptance based upon the SHMS central momentum. The incident particle momentum for each event is computed as
$$P_{Real}=P_{Central}(1+\delta/100)$$
where $P_{Central}$ is the specified central momentum of the SHMS and it is set for 3 and 7~GeV/c in our study. 


\begin{figure}[p]
	\centering
	\subfloat[Particle Distribution]{\includegraphics[width=0.5\textwidth]{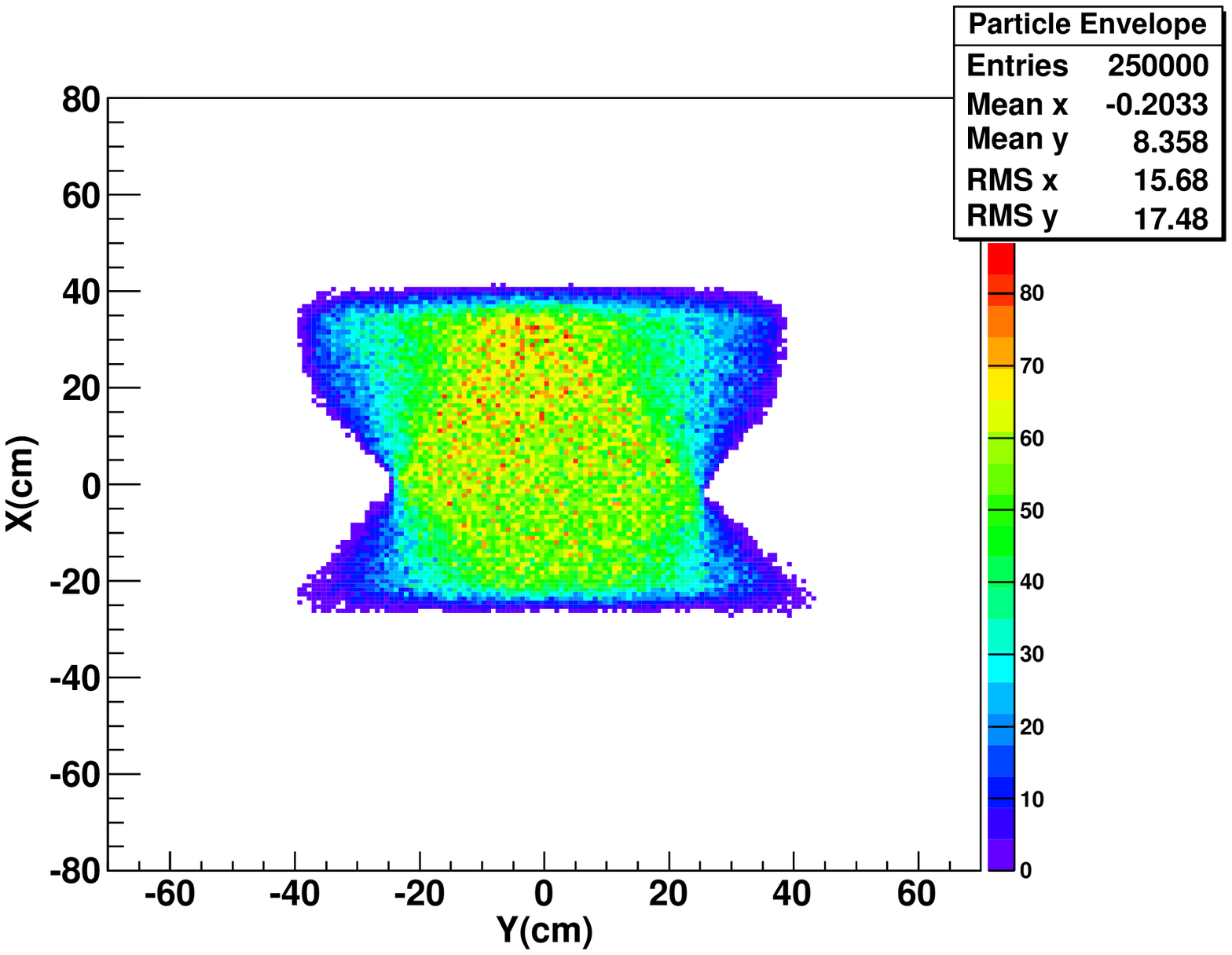}\label{particle_distribution}}
	\subfloat[Total \v{C}erenkov Photon Envelope]{\includegraphics[width=0.5\textwidth]{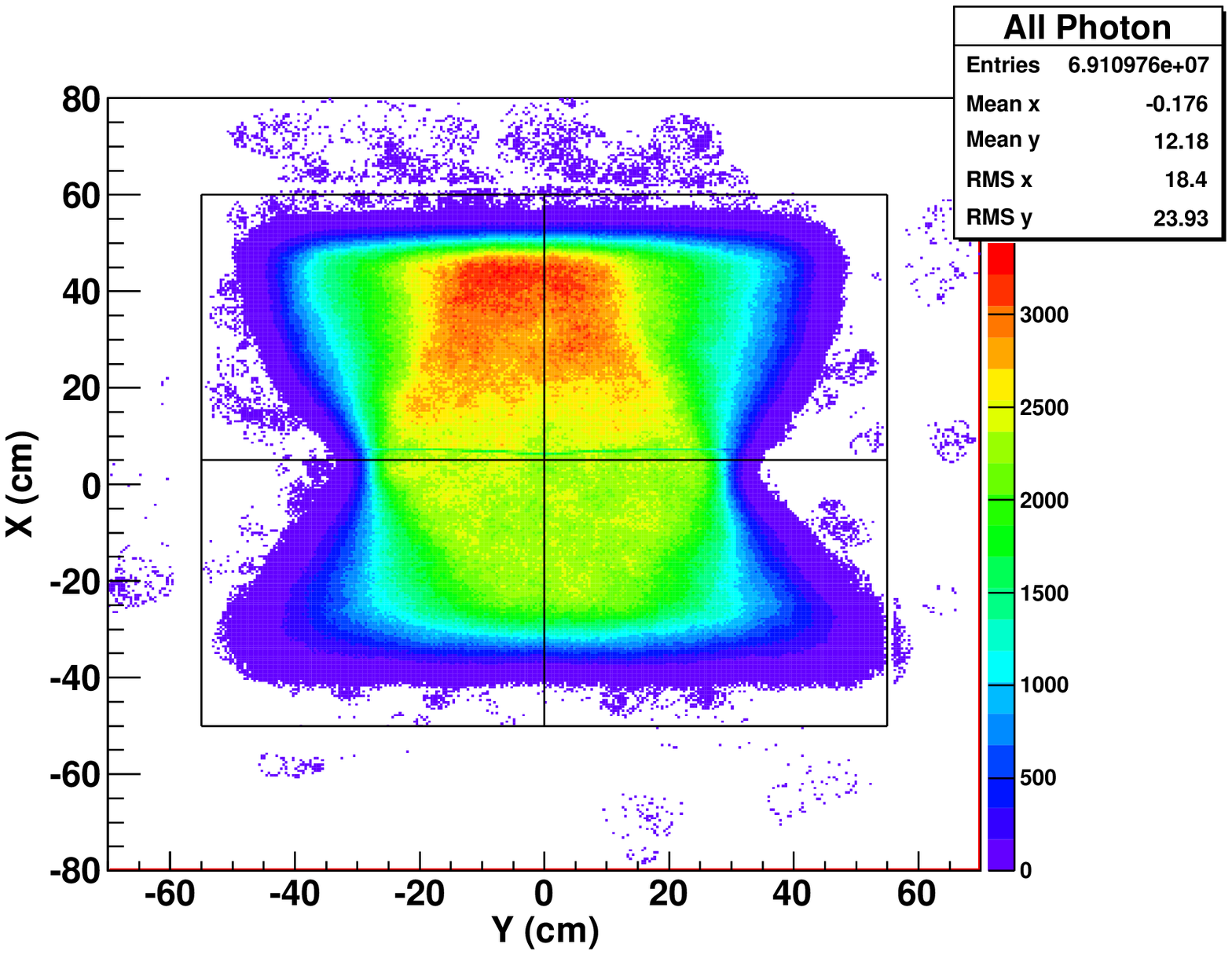}\label{cerenkov_envelop}}\\
	\subfloat[Captured \v{C}erenkov Photon Envelope]{\includegraphics[width=0.5\textwidth]{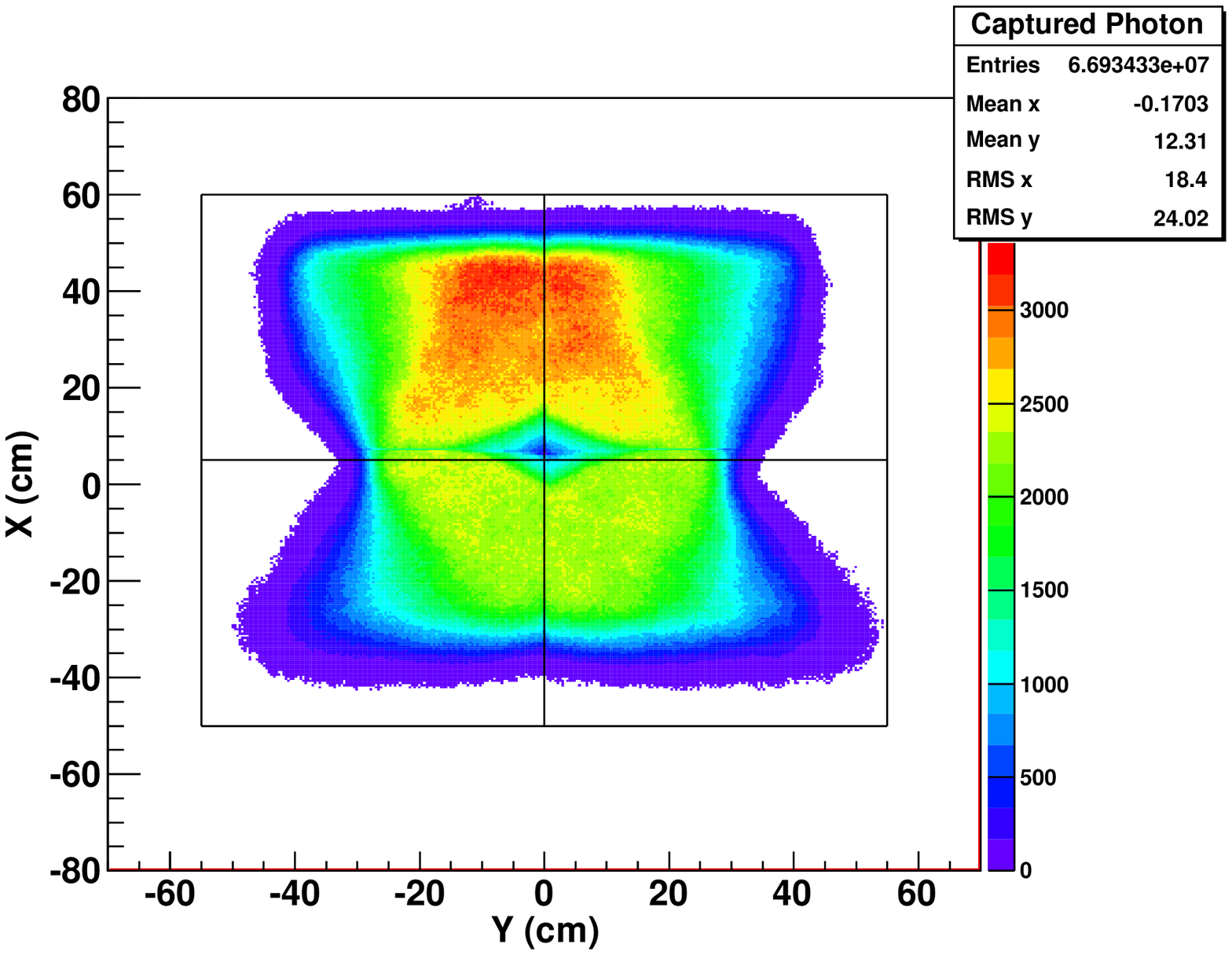}\label{pic_capture}}
	\subfloat[Missed \v{C}erenkov Photon Envelope]{\includegraphics[width=0.5\textwidth]{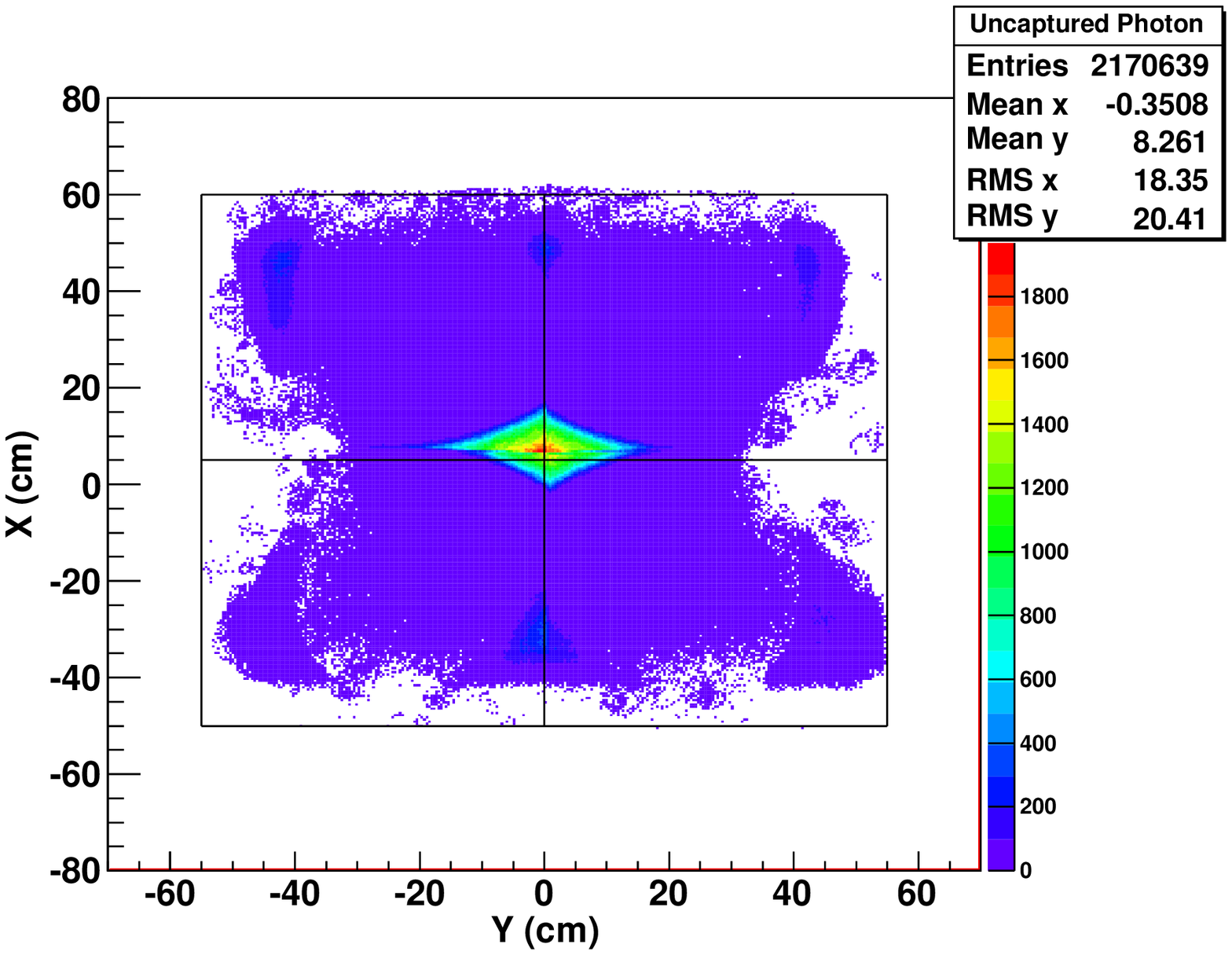}\label{pic_miss}}
	\caption[Charged Particle and Photon Envelope]{All plots are generated with 7~GeV/c central momentum $\pi^+$. (a) the $\pi$ envelope in front of the HGC detector; (b) the \v{C}erenkov photon envelope at the mirror plane; (c) the focused \v{C}erenkov photon envelope at the mirror plane; (d) missed focused photon envelope at the mirror plane. Note that all plots show the events distribution in x-y plane. The photons in (c) are reflected by the mirrors and optically focused onto the PMT active area when all PMT efficiencies are switched off in the simulation. Thus, photons which are missed due to the PMT inefficiency are not included in this plot, only those due to the imperfect optics. The photons in (d) are also reflected the mirror but not focused onto the PMT active area. Approximate mirror boundaries are shown in (b), (c) and (d). \oic}
\end{figure}

\section{Particle and \v{C}erenkov Envelopes}

The generated particle distribution by the single arm SHMS MC at the HGC detector front surface is plotted in Fig.~\ref{particle_distribution}, where the detector front surface is 18.80~m and the SHMS focal plane is $z$=18.10~m from the scattering chamber along the $z$ direction. The resultant \v{C}erenkov radiation envelope projected onto the mirror surfaces in the x-y plane at 7~GeV/c is shown in Fig.~\ref{cerenkov_envelop}. Note that the \v{C}erenkov angle $\theta_C$ at 7~GeV/c is 2.84$^{\circ}$, so the \v{C}erenkov radiation distribution is taller and wider than of the charged particles which generate it. Most of the photons are projected in a 90~cm$\times$80~cm region in the $x$-$y$ plane; the four black squared brackets indicate the approximate mirror boundaries. Thus, this leaves at least 5~cm clearance along the mirror edges for the mirror mounting scheme onto the detector.  Fig.~\ref{pic_capture} shows the distribution of reflected photons which are then focused onto the active region of the PMTs, which is indicated as the inner circle in Fig.~\ref{pic_xy_projection}. Note that PMT efficiencies are switched off in the simulation. Thus, photons which are missed due to the PMT inefficiency are not included in this plot, only those due to the imperfect optics. Fig.~\ref{pic_miss} shows the reflected photons which are not focused onto the active region of the PMTs. From the plotted distribution, the mis-focused photons are mainly distributed at the interleaved mirror corners near the detector center. This will lead to a localized region of lower efficiency near the center.


\section{PMT Optimization}
\label{g4_study_pmt}

\begin{figure}[p]
	\centering
	\subfloat[PMT \#1 $x$-$y$ Focused Spot Size]{\includegraphics[width=0.5\textwidth]{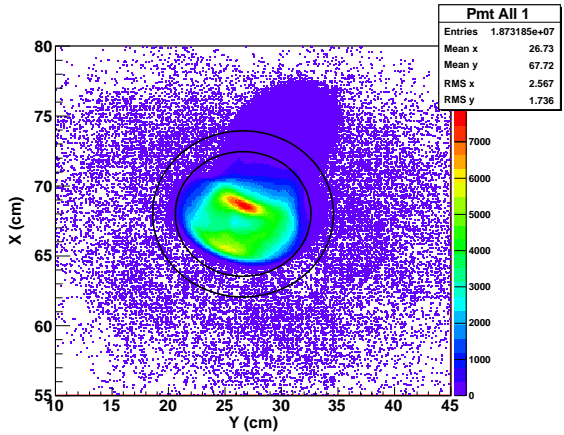}}
	\subfloat[PMT \#2 $x$-$y$ Focused Spot Size]{\includegraphics[width=0.5\textwidth]{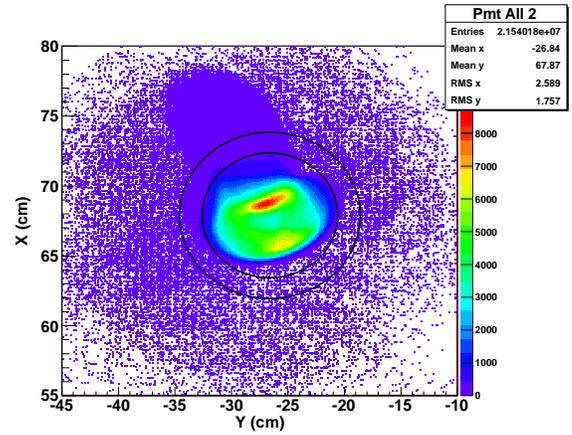}}\\
	\subfloat[PMT \#3 $x$-$y$ Focused Spot Size]{\includegraphics[width=0.5\textwidth]{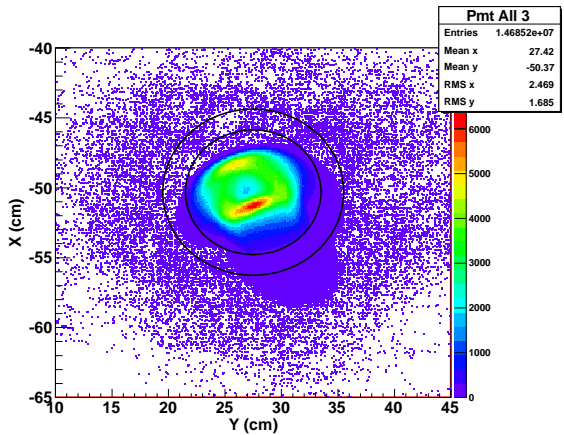}}
	\subfloat[PMT \#4 $x$-$y$ Focused Spot Size]{\includegraphics[width=0.5\textwidth]{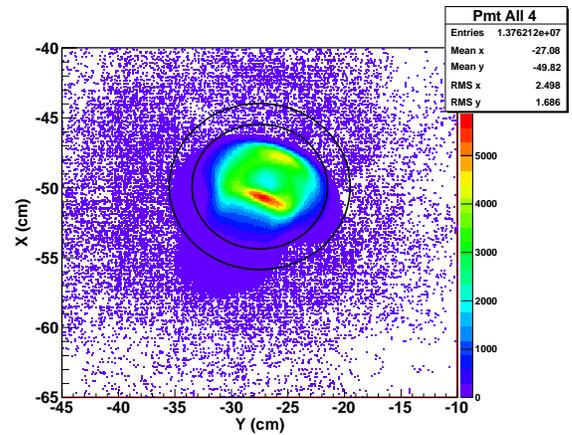}}
	\caption[Focused \v{C}erenkov Radiation Spot onto PMTs ]{The $x$-$y$ plane projection of the photon hit position at the PMT cathode plane for 7~GeV/c $\pi^+$. The inner circle indicates the PMT active area which has radius of 6cm; the outer circle gives the boundary of the quartz window which has radius of 8cm. Notice that the $x$ axis is vertical and $y$ axis is horizontal in HGC coordinate. However, the mean and RMS x, y values refer to the plotting program convention of x = horizontal (HGC $y$ axis) and y = vertical (HGC $x$ axis). \oic}
	\label{pic_xy_projection}
\end{figure}

\begin{figure}[p]
	\centering
	\subfloat[PMT \#1 x-z Photon Path Distribution]{\includegraphics[width=0.5\textwidth]{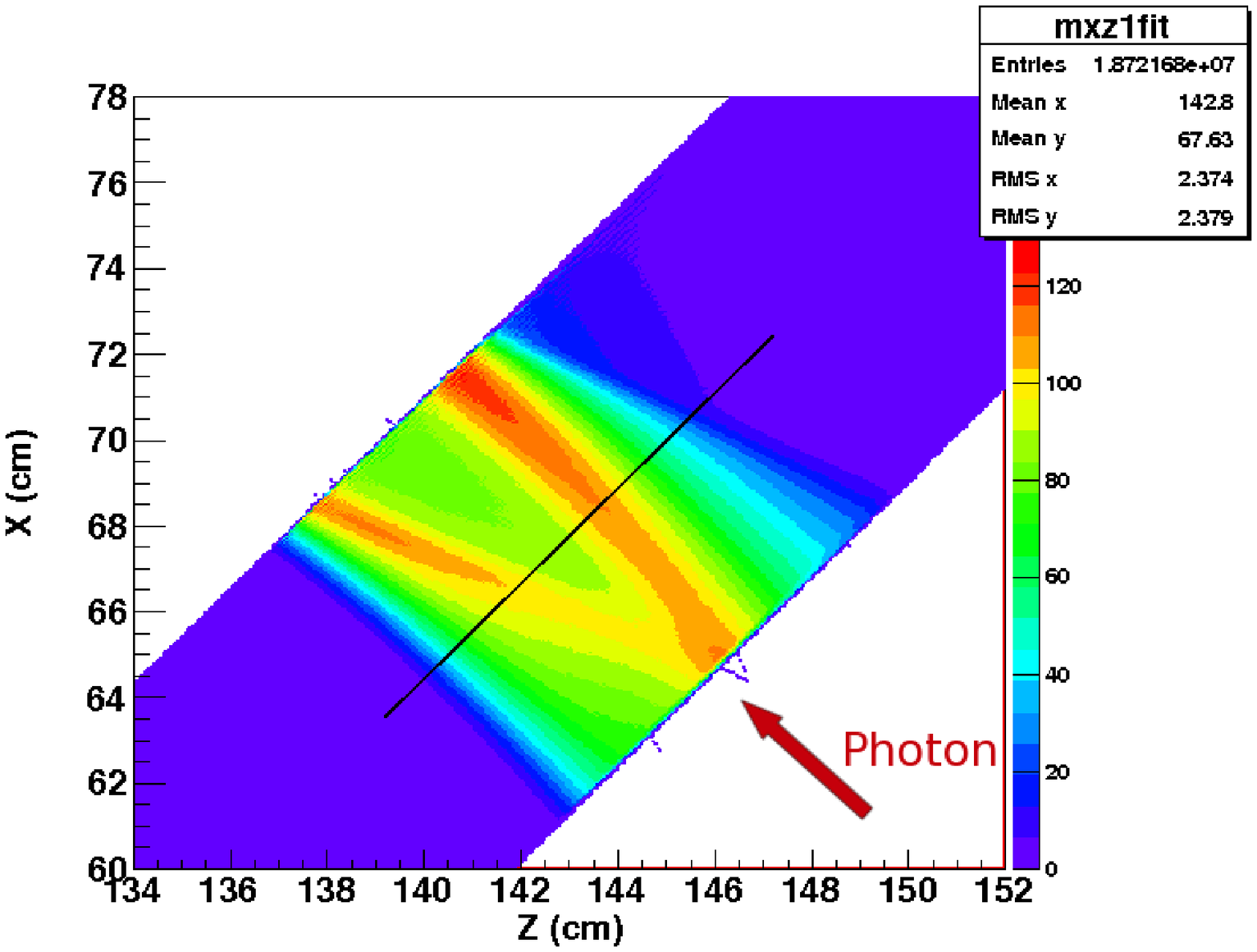}}
	\subfloat[PMT \#2 x-z Photon Path Distribution]{\includegraphics[width=0.5\textwidth]{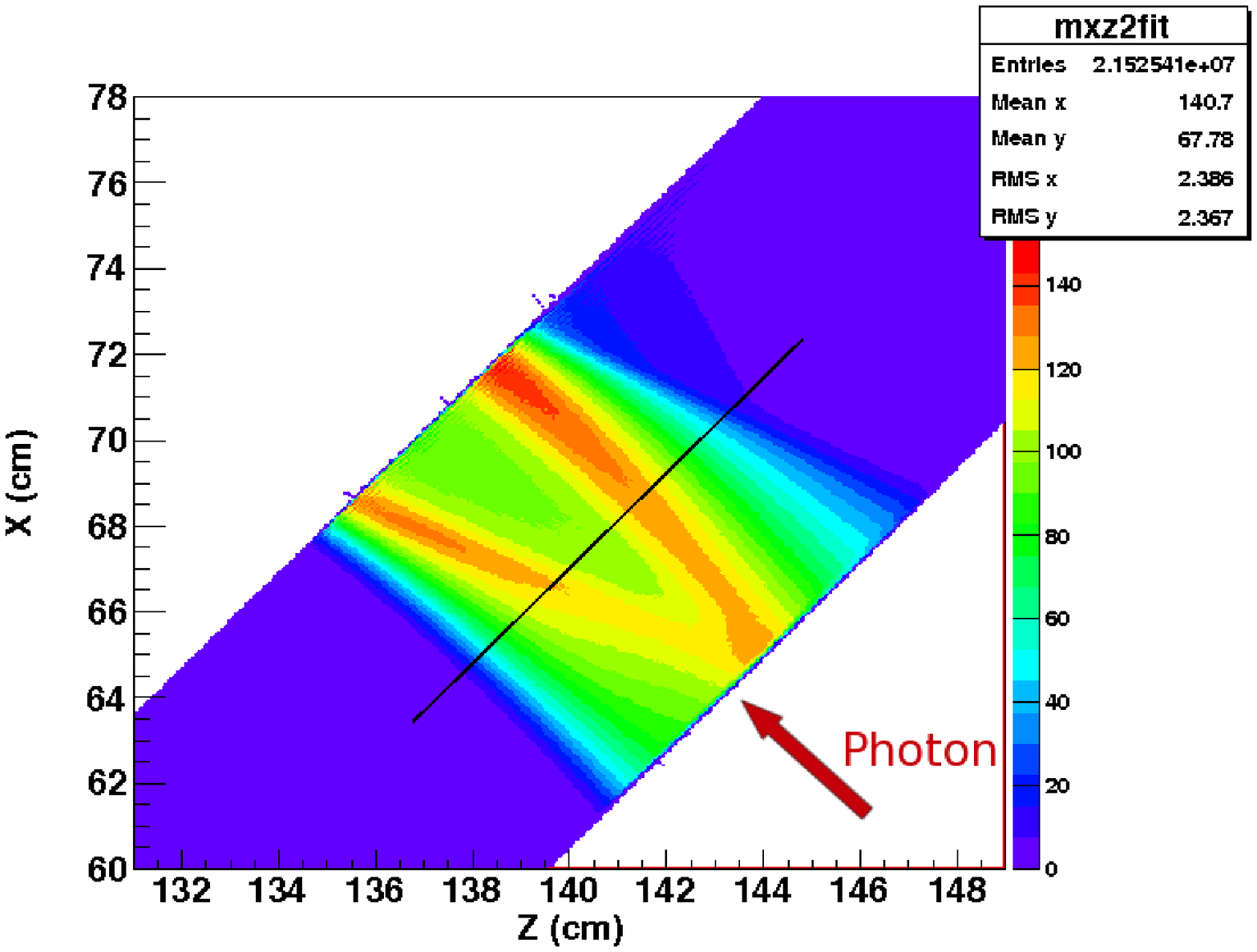}}\\
	\subfloat[PMT \#3 x-z Photon Path Distribution]{\includegraphics[width=0.5\textwidth]{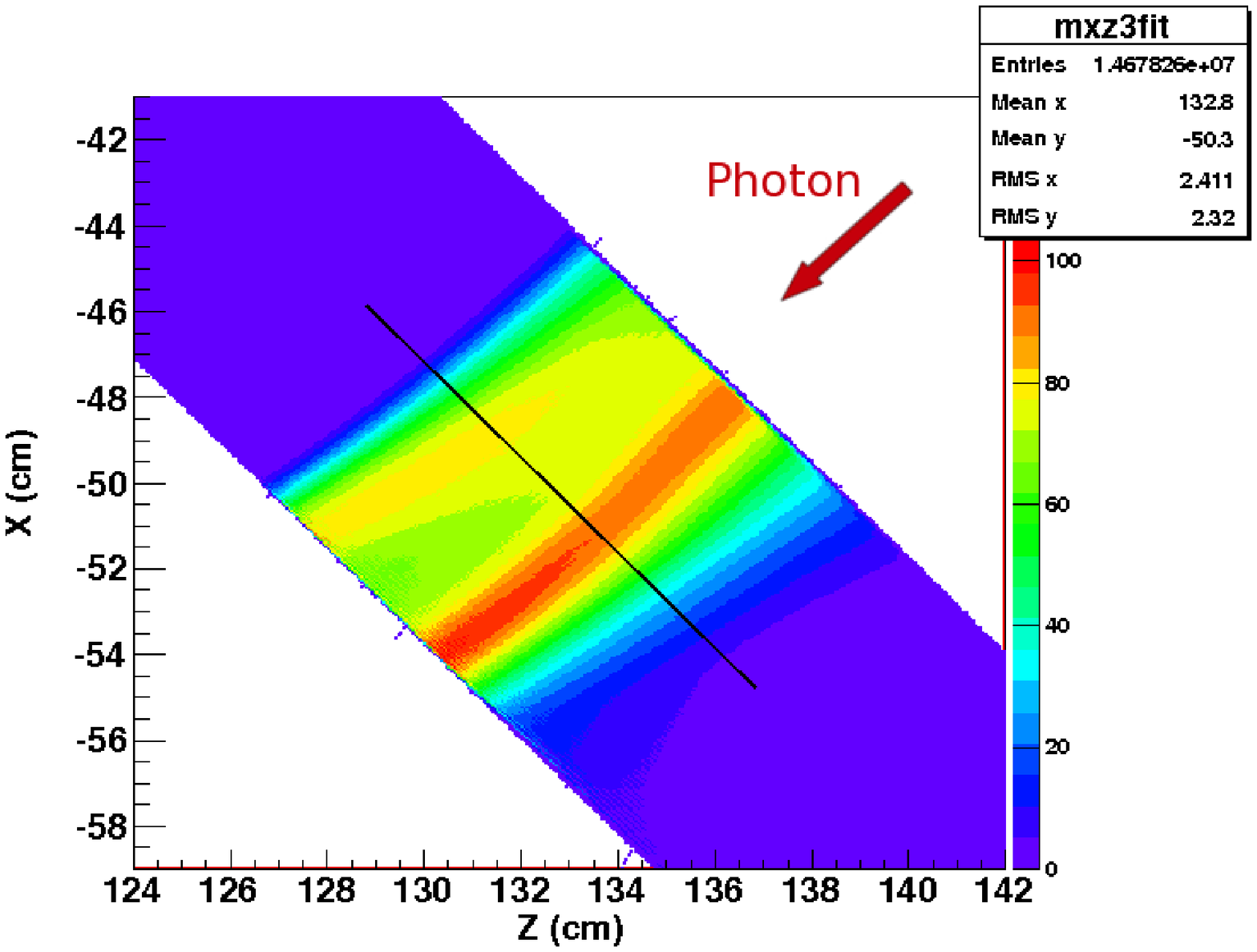}}
	\subfloat[PMT \#4 x-z Photon Path Distribution]{\includegraphics[width=0.5\textwidth]{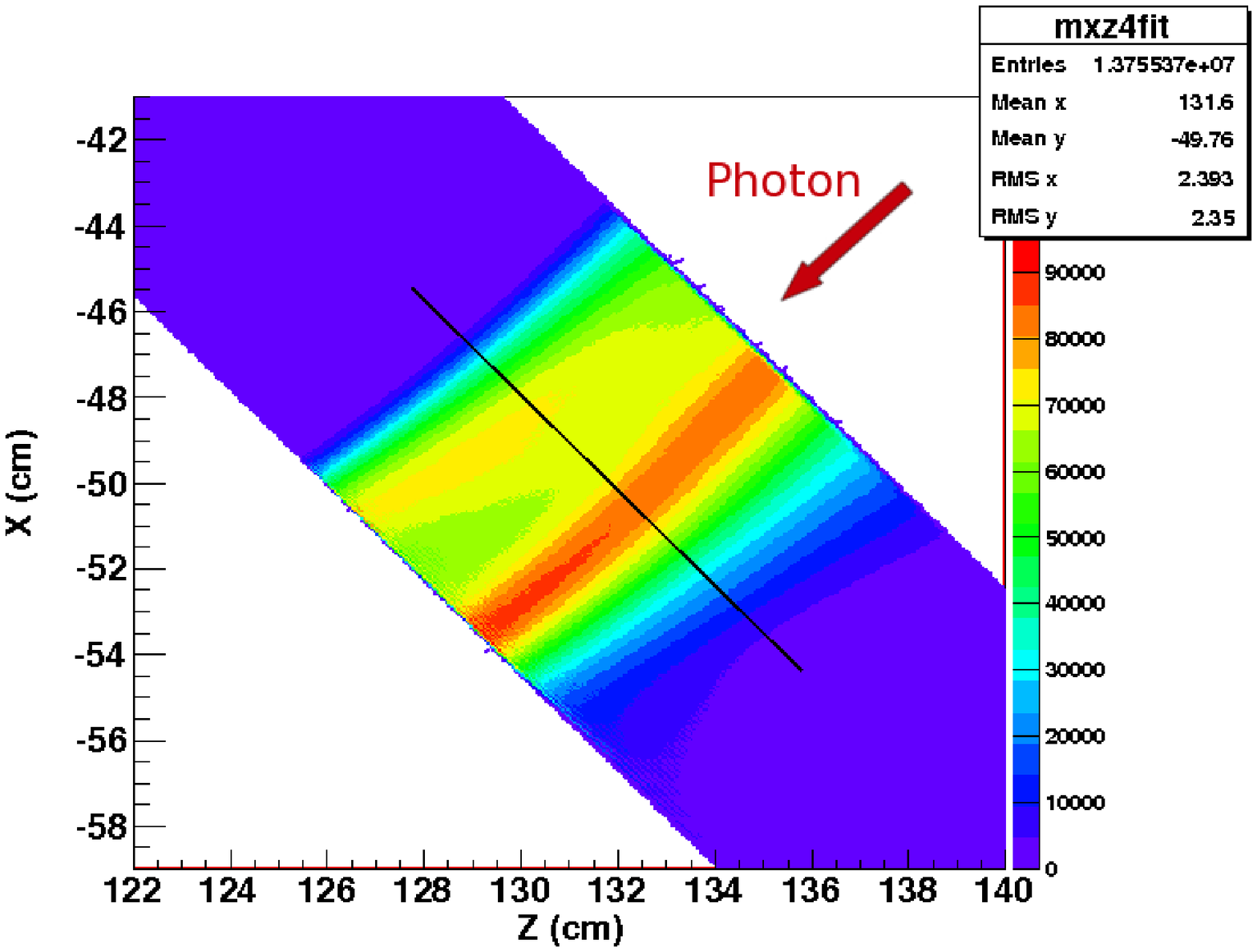}}
	\caption[Side view of the Reconstructed Photon Path]{The reconstructed photon path through 7 tracking planes in the x-z plane. The black line indicates the position of the cathode plane. \oic}
	\label{pic_xz_projection}
\end{figure}

Efficiencies $\eta_{\gamma}$ and $\eta_{pe}$ are defined in Equation (\ref{eqn_photon_gamma}) and (\ref{eqn_photon_pe}). The quantities $\eta_{\gamma}$ and $\eta_{pe}$ are very closely related: $\eta_{\gamma}$ is for the PMT-off mode and $\eta_{pe}$ is for the PMT-on mode. In the PMT-on mode, focused photons have a certain probability to be converted into photo electrons, and this probability is dictated by the various of efficiencies of the PMT assembly. The simulation takes into account efficiencies such as the quartz window and adaptor transmissivity, reflection  at the front face of the quartz window and PMT quantum and positional efficiency. For PMT-on mode, a $\pi^{+}$ at 7~GeV/c central momentum has $\eta_{pe}$=15.78\%. In the PMT-off mode, the PMT assembly efficiencies are assumed to be 100\%, which means all photons are focused onto the active area of the PMT are converted into photo electrons, where the missing photons are due to the imperfect optics. For PMT-off mode, $\pi^{+}$ at 7~GeV/c central momentum has $\eta_{\gamma}$=96.8\%. The PMT optimization studies in this chapter were using the PMT-off mode.

Fig.~\ref{pic_xy_projection} shows the $x$-$y$ projection of the photon hits at the PMT cathode plane. The inner black circle indicates the boundary of the PMT active area (12~cm in diameter); the outer black circle is the boundary of the quartz window. From the plotted distribution, most of the photons are focused onto the center of the PMT active area, and form a donut shaped spot. The missing photons are distributed in the large purple tail which extends outside of the PMT active region. To maximize the number of detected photons, ideally one needs to adjust the mirror angle to bring the purple tail into the PMT active region without losing too much light from the donut spot. Note also that at 3~GeV/c the donut shaped spot will be more compact, due to the smaller \v{C}erenkov cone angle. 

%
%
%
%

Fig.~\ref{pic_xz_projection} shows the reconstructed \v{C}erenkov photon trajectories in the $x$-$z$ plane using the position information from the 7 tracking planes. The black line is the cathode plane. The tracking planes are tilted by 42$^{\circ}$ with respect to the x-axis. From the graphs, we can conclude that the PMT optical performance is insensitive to small changes in the PMT tilt angle (1$^{\circ}$ or 2$^{\circ}$), and most of the photons pass through the cathode plane in the compacted central region. This is the justification for all PMTs to have the same tilt angle, but the mirrors have the different tilt angles with respect to the $x$ axis.

Fig.~\ref{pic_area_size} shows the Root Mean Square (RMS) area of the spot size at each tracking plane for each PMT. The RMS area is computed using the RMS x and y values inside of the PMT active region as shown in Fig.~\ref{pic_xy_projection}. Since the focused spots are similar to elliptical shapes, therefore the area can be calculated as 
\begin{equation} 
\textrm{RMS Area} = \frac{\textrm{RMS y}}{\sin 42^\circ} \times \textrm{RMS x} \times \pi\,,
\label{eqn_rms_area}
\end{equation} 
where the first term is corrected for the PMT tilt angle. The RMS x and RMS y in Equation (\ref{eqn_rms_area}) correspond to the RMS values for the x and y axes on the plot, not according to the HGC coordinate. A smaller RMS area would suggest the photons are more compactly distributed on the plane, thus easier to focus onto the active area of the PMT. 

The PMT cathode plane is labeled \#50 among all tracking planes. Planes \#53, \#52, \#51 are in front and \#49, \#48, \#47 are behind the cathode plane, planes are separated by 2~cm along the $z$ axis. Fig.~\ref{pic_area_size_34} shows that PMTs \#3 and \#4 have the smallest RMS spot size at plane \#50. However, PMTs \#1 and \#2 have the smallest RMS spot size between plane \#50 and \#49; this suggests that PMT \#1 and \#2 are 1~cm too close to the mirror. This small imperfection is due to the recent angle change for PMT \#1 and \#2 from 45$^\circ$ to 42$^\circ$ for the manufacture convenience, since Fig.~\ref{pic_xz_projection} shows the small PMT angle change results in little difference in performance. Further efforts will be committed to re-optimize the optics arrangement as well as other new studies following the release of on updated SHMS particle distribution from Hall~C.

\begin{figure}[t]
	\centering
	\subfloat[PMT 1 \& 2]{\includegraphics[width=0.5\textwidth]{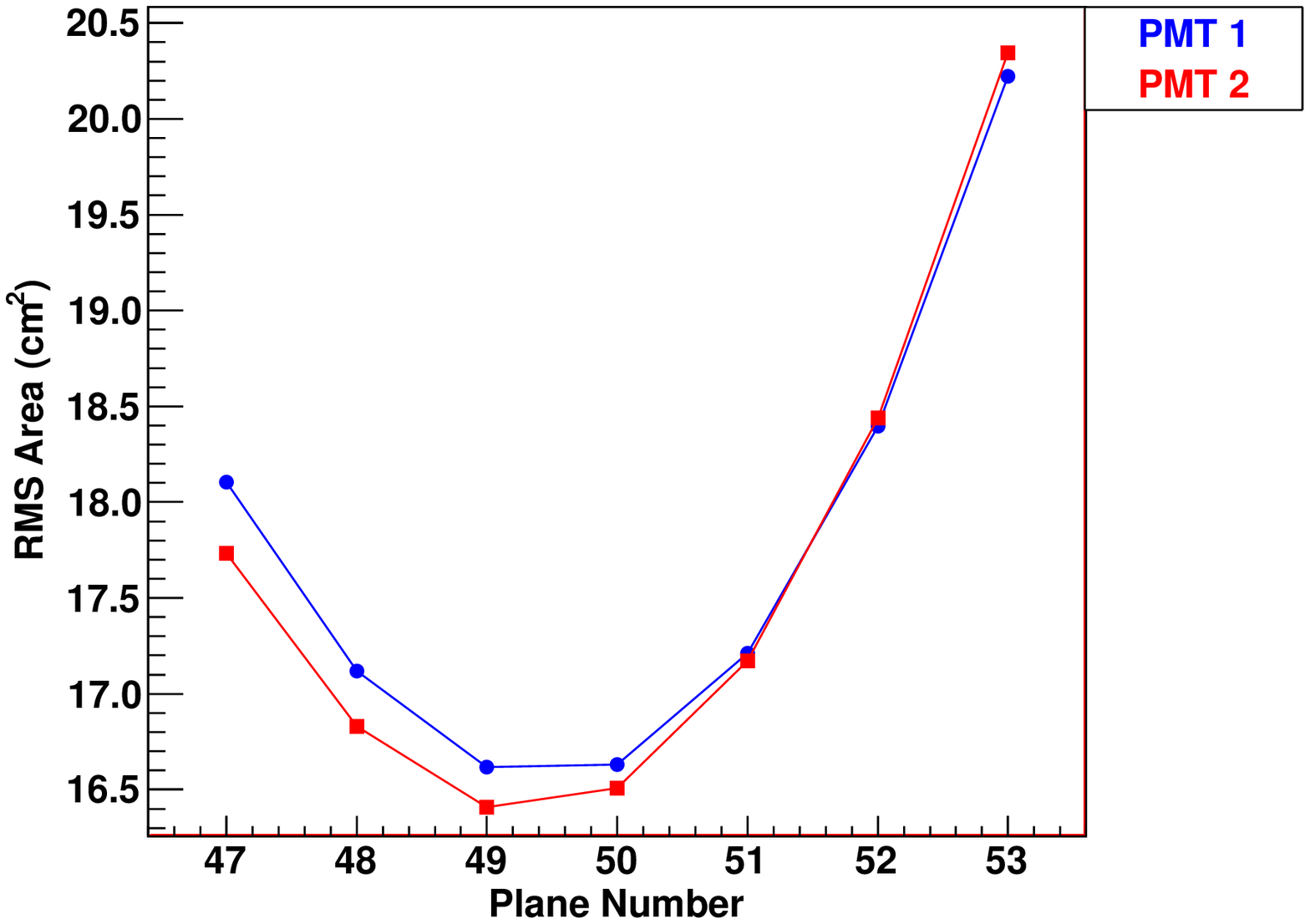}\label{pic_area_size_12}}
	\subfloat[PMT 3 \& 4]{\includegraphics[width=0.5\textwidth]{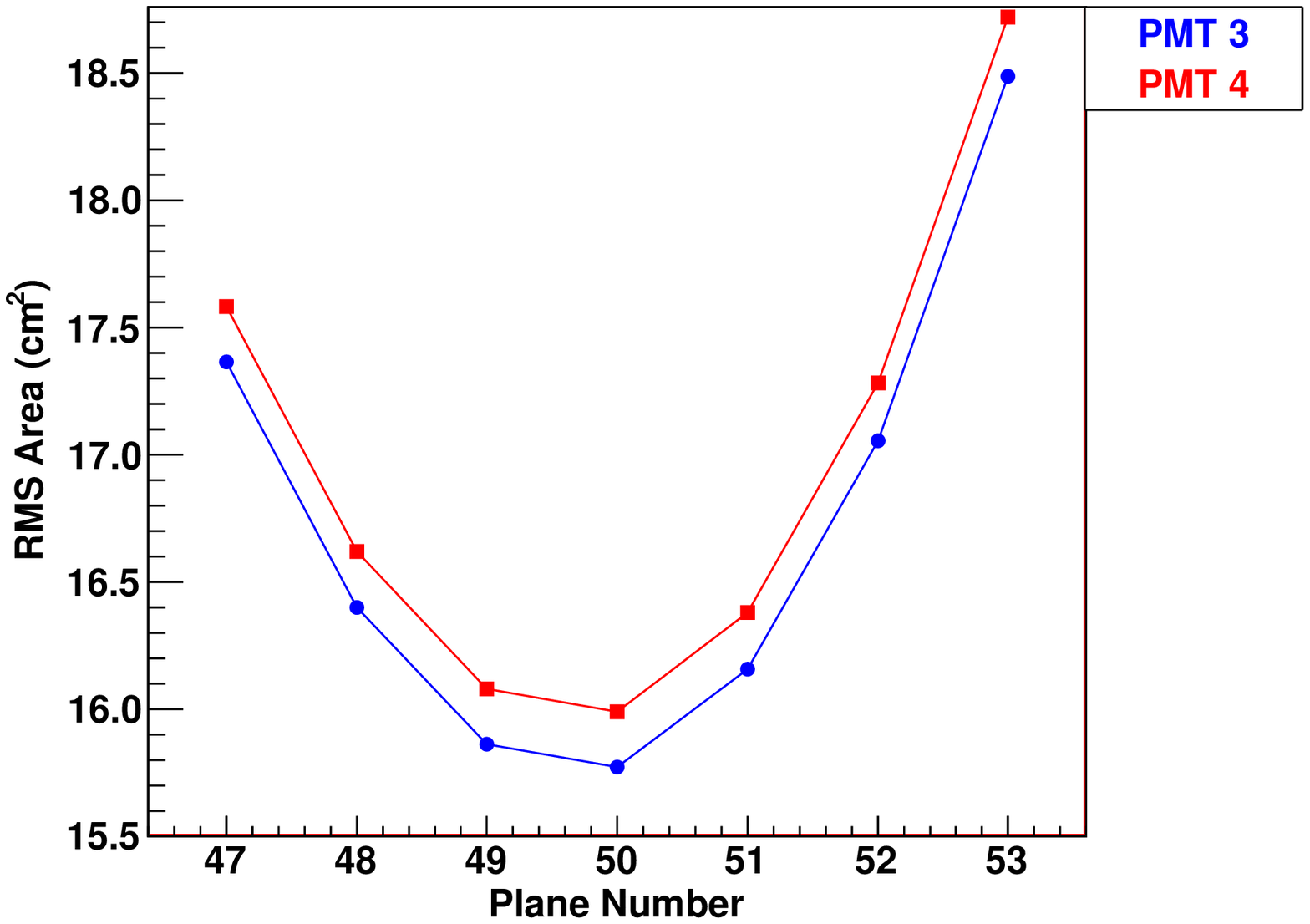}\label{pic_area_size_34}}
	\caption[Focused Spot RMS Size vs. Plane Number]{Focused spot root mean square size vs. plane number. \oic}
	\label{pic_area_size}
\end{figure}


\section{Projected Performance}
\label{g4_expected_perfomance}
The particle identification efficiency, $\eta_{particle}$, is defined in Equation (\ref{eqn_particle_efficiency}); it describes the overall performance of the detector. It is only studied using the PMT-on mode. The particle identification efficiency was studied with 3 and 7~GeV/c central momentum $\pi^{+}$, $K^{+}$ and proton. Based on the previous HMS experiments, the $\pi^+$-$K$-proton ratio is estimated to be 5:2:2 at 3 GeV and 5:3:2 at 7 GeV momentum in the simulation. The $\pi^+$ sample size for 3 and 7 GeV simulations were 250000. The simulated photo-electrons spectra were analyzed in the same manner that actual HGC data will be analyzed in SHMS experiments. The particle identification efficiency is highly dependent upon the signal threshold, which is applied in the analysis to reduce the noise signal. The optimization of the threshold signal level is one of the major goals of our study.

\begin{figure}[t]
	\centering
	\subfloat[3~GeV/c]{\includegraphics[width=0.5\textwidth]{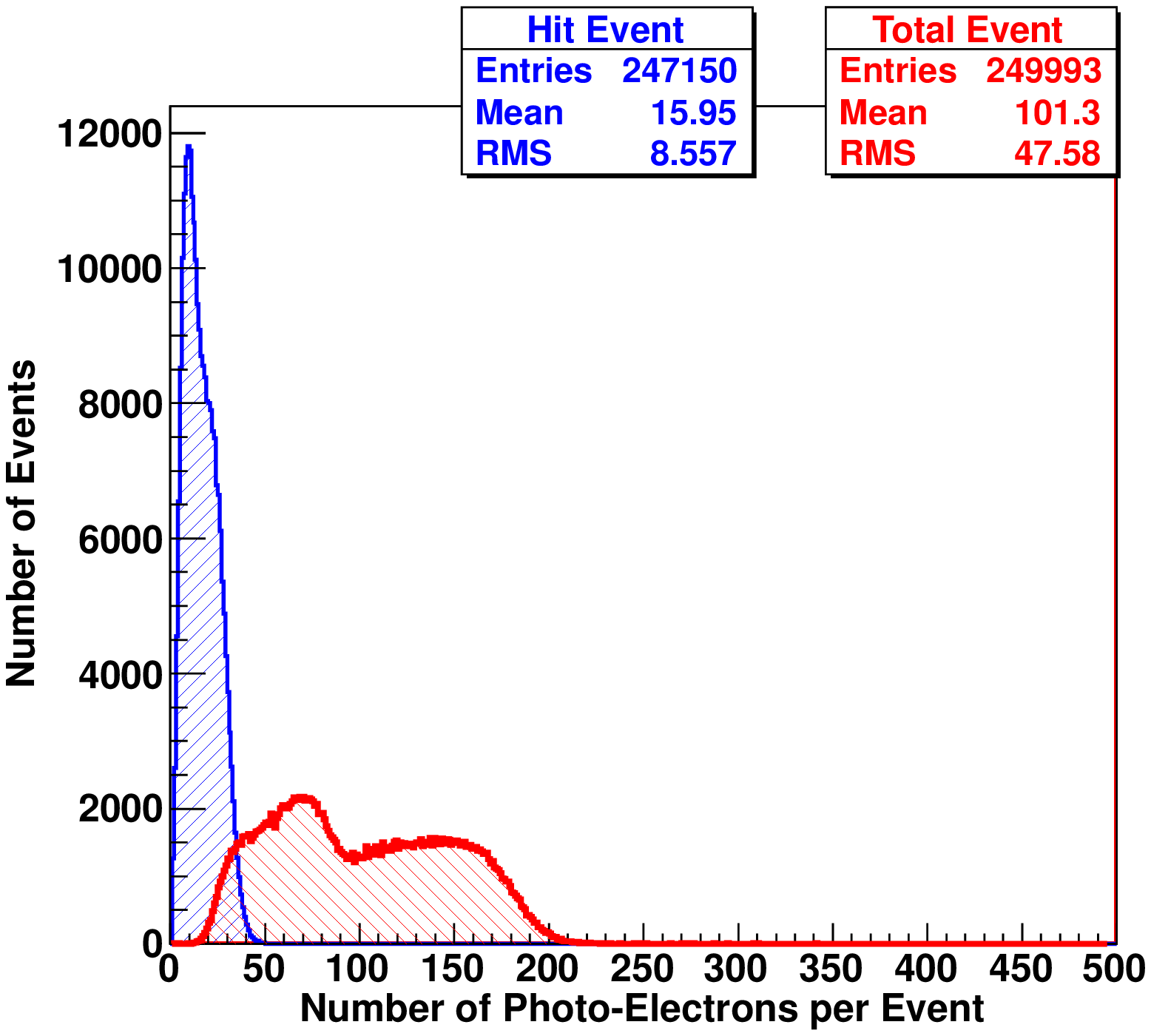}\label{pic_pion_photon_3gev}}
	\subfloat[7~GeV/c]{\includegraphics[width=0.5\textwidth]{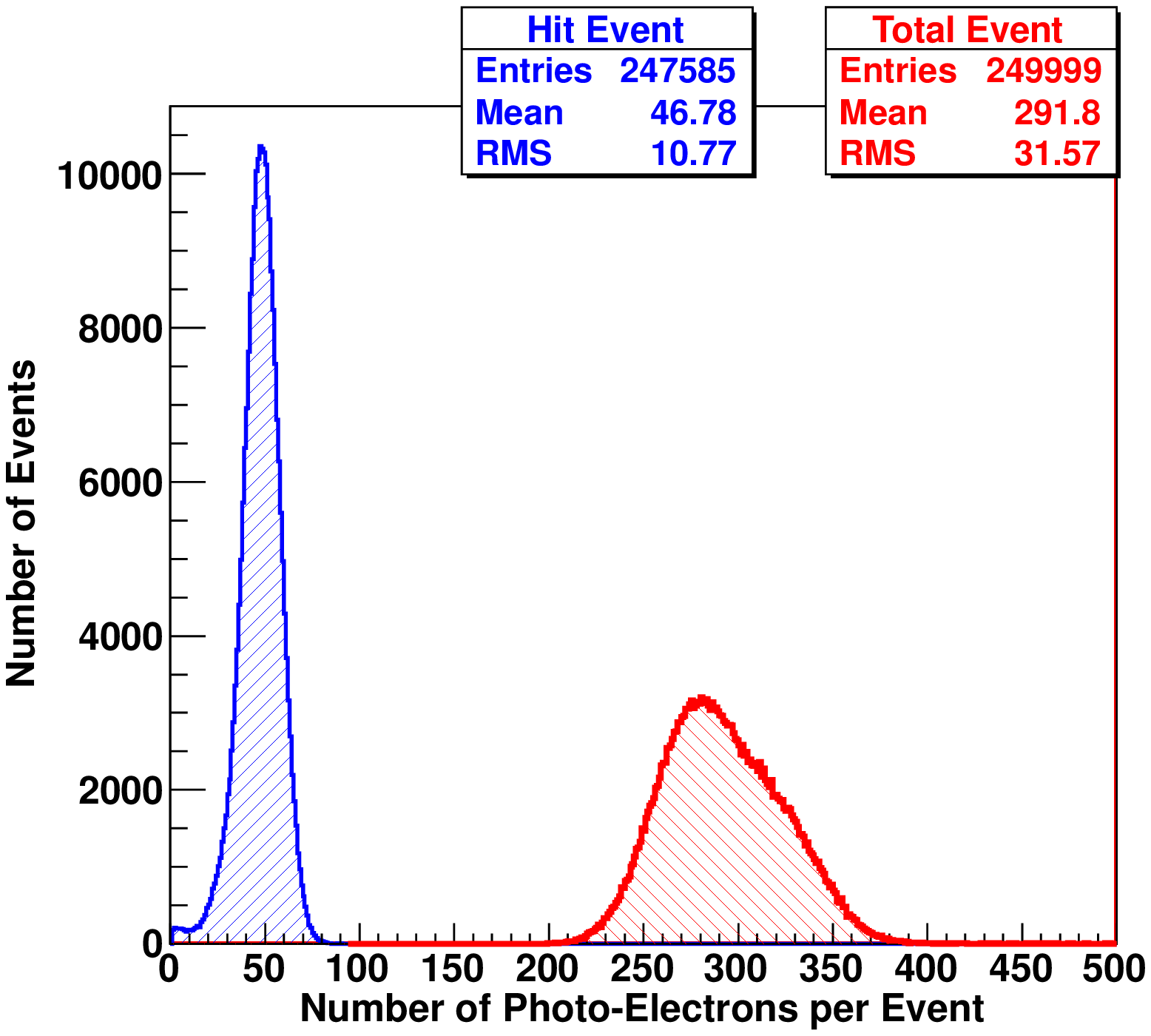}\label{pic_pion_photon_7gev}}
	\caption[Photons and Photo-e Distribution]{The $\pi^+$ \v{C}erenkov photon production and detected photon electron distribution at 3 and 7~GeV/c central momentum.  The red curve is generated \v{C}erenkov photons per event distribution, while the blue curve is photo-electrons emitted by the PMT photo-cathode per event distribution. \oic}
	\label{pic_pion_photon}
\end{figure}

\subsection{\v{C}erenkov Photon Detection for $\pi^+$ Events}

Fig.~\ref{pic_pion_photon} shows the total generated \v{C}erenkov photons and photo-electrons per event by all four PMTs at 3 and 7~GeV/c central momentum $\pi^{+}$. The total generated photon curve for 3~GeV/c $\pi^{+}$ has wide double peaks, which is due to the momentum difference of the particles entering different parts the detectors along the x direction. For example, the particles travel through $+x$ mirror (\#1 and \#2) regions have higher momentum than those travel through $-x$ mirror (\#3 and \#4) regions; the particles with higher central momentum can generate more \v{C}erenkov photons. The range for the SHMS momentum acceptance is created by the magnetic focusing elements (both dipole and quadrupole) in front of the detector stack, and it generates a larger range in velocities for lower central momentum $\pi^+$, which results a wider total photon distribution at 3~GeV/c than at 7~GeV/c central momentum. The total \v{C}erenkov photon production for 7~GeV/c $\pi^{+}$ has a Gaussian distribution with the mean 291.8 and standard deviation 31.6. The photo-electron (blue) distributions have sharp peaks, and the mean are 16.0 and 46.8 for 3~GeV/c and 7~GeV/c  $\pi^{+}$. This concludes that at same refractive index (pressure) the particle with higher momentum generates more \v{C}erenkov photon. The photo-electron distribution is summed over all four PMTs. This is easy to do in the simulation. However, in the actual experimental analysis, care will have to be taken to properly match the PMT signal spectra, so they can be correctly added together.

%
%
%
%


%
%

\subsection{Signal Threshold}

Positrons ($e^+$) have a \v{C}erenkov refractive index threshold that is lower than the detector refractive index, so they are capable of generating \v{C}erenkov photons in the HGC detector. The detected $e^{+}$ events can be rejected by other particle identification detectors like the lead-glass detector or the Noble Gas \v{C}erenkov detector at high central momenta. When $K^+$ or protons pass through the HGC detector, they can knock electrons free from the C$_4$F$_8$O gas particles; these secondary $e^-s$ at sufficiently high momenta can generate \v{C}erenkov radiation. This process is known as delta radiation and it is one of the primary sources for background signals in the HGC. The delta radiation can also be modeled by the Geant4 simulation. An effective way to reduce this background is to introduce a signal threshold, however, the threshold should not exclude too many real $\pi^+$ events. 


Fig.~\ref{pic_performance} shows the number of events vs. the detected number of photo-electrons per event, and Fig.~\ref{pic_trigger_rate} shows the particle detection efficiency vs. the number of photo-electrons. Results are shown for both 3 and 7~GeV/c central momenta for $\pi^{+}$, $K^{+}$ and proton. In Fig.~\ref{pic_performance} it can be seen that the detected pion events generate many more photo-electrons (on average) than delta radiation from $K^+$ and proton at 7~GeV/c momentum. However, the generated photo-electron difference is much smaller at 3~GeV/c. Thus, different signal thresholds were used for the different particle central momenta to ensure $\eta_{particle}>$ 98\% for $\pi^+$. At 3~GeV/c, the analysis threshold was set for 2 photo-electrons, giving $\eta_{particle}$ = 98.4\%; at 7~GeV/c, the threshold is set for 10 photo-electrons where $\eta_{particle}$ = 98.4\%. 1.1\% and 0.88\% of $K^{+}$ delta radiation events at 3 and 7~GeV/c central momentum pass the applied threshold. However, these $K^+$ events will have high probability to be identified by the Aerogel \v{C}erenkov Detector and so we are confident that our goal of 1000:1 $\pi^+$/$K^+$ separation ratio needed for SHMS experiments will ultimately be achieved.

\begin{figure}[p]
	\centering
	\subfloat[3~GeV/c]{\includegraphics[width=0.75\textwidth]{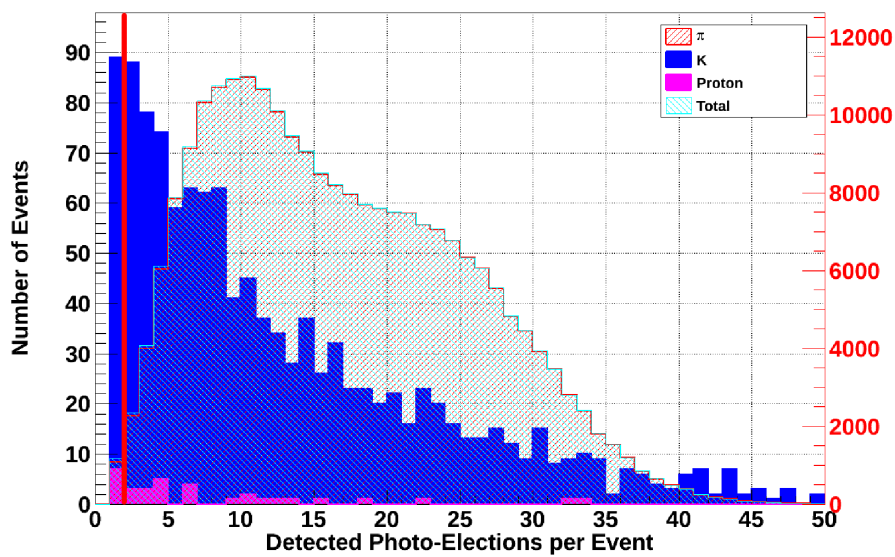}} \\
	\subfloat[7~GeV/c]{\includegraphics[width=0.75\textwidth]{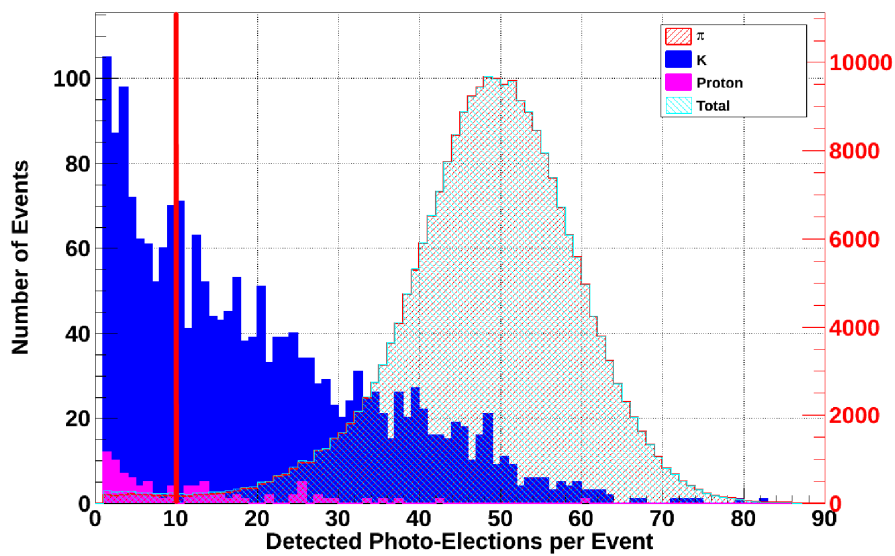}}
	\caption[Photo-e Distribution for Different Particles]{Particle events vs. the detected photo-electron per event plots at 3 and 7~GeV/c central momentum. $K^{+}$ distribution is in blue, proton distribution is in magenta, $\pi^+$ distribution is shaded in red and total distribution is shaded in cyan. $K^{+}$ and proton events are plotted against the left y axis in black; $\pi ^{+}$ and total events are plotted against the right y axis in red. Note that the $\pi ^{+}$ and total events distribution are overlapping. The red solid lines indicate the signal threshold. \oic}
	\label{pic_performance}
\end{figure}

\begin{figure}[p]
	\centering
	\subfloat[3~GeV/c]{\includegraphics[width=0.75\textwidth]{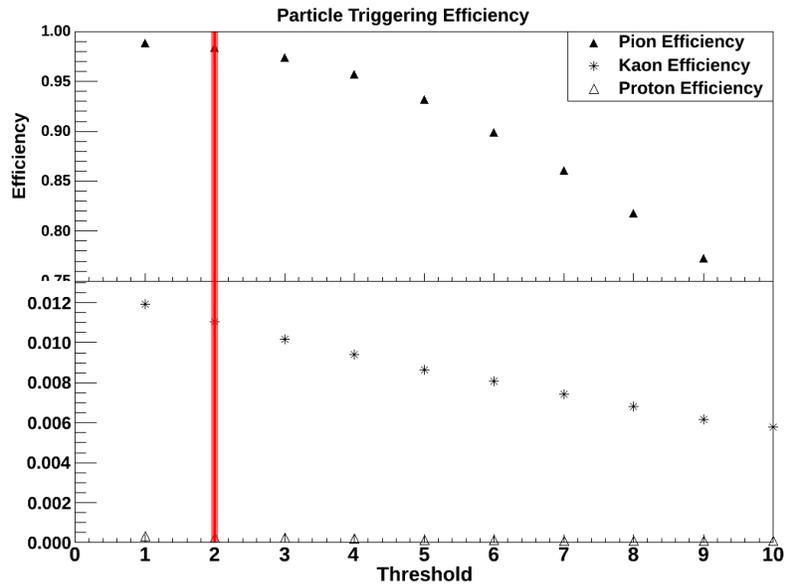}} \\
	\subfloat[7~GeV/c]{\includegraphics[width=0.75\textwidth]{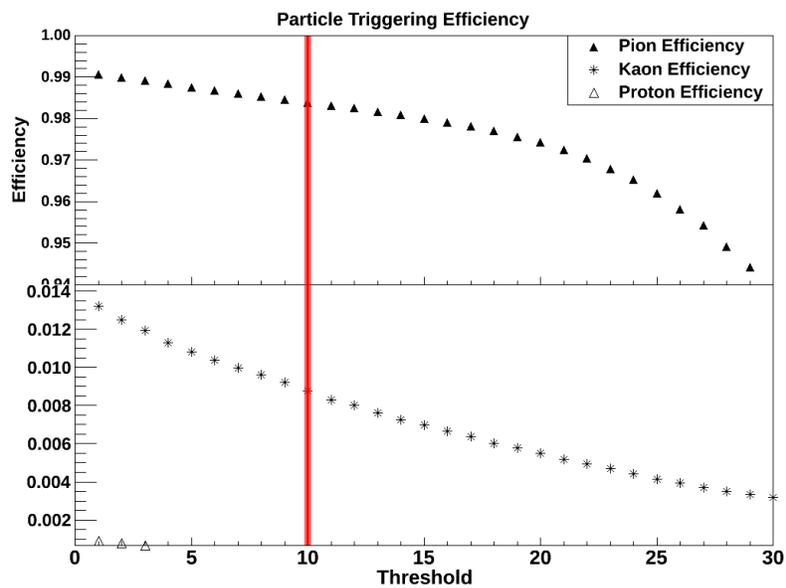}}
	\caption[Detection Efficiency Plots]{Particle detection efficiency vs. the threshold photo-electron plot at 3 and 7~GeV/c. The red solid lines indicate the signal threshold. Note the different x axis scales for plot (a) and (b). \oic}
	\label{pic_trigger_rate}
\end{figure}

\section{Oblate Ellipsoid Versus Spherical Mirrors}

All simulation results presented in this chapter are with the oblate mirrors. The performances for both oblate and spherical mirrors were studied and compared. Fortunately, both mirrors have the same photon detection efficiency at PMT-on mode $\eta_{pe}$=15.78\%, however, the pion identification efficiency for PMT-on mode for spherical mirrors is $\eta_{particle}=98.90$\%, which is 0.5\% higher than the oblate mirror efficiencies at 7~GeV/c momentum with the threshold set for 10 photo-electrons. This is due to the better focusing ability of spherical mirror, it is able to include larger area of the purple tail into the PMT active area, see Fig.~\ref{pic_xy_projection}.


\section{Remarks and Outlook}

The Hall~C collaboration has recently decided to shift the super-conducting dipole towards the $-x$ direction by 5~cm after some further optimization of the SHMS optics. Based on our preliminary simulation results using the new particle distribution corresponding to the changed SHMS optics, the detector and mirror positions need to be re-optimized to accommodate the new \v{C}erenkov radiation envelope. Since the PMT positions and angles have been fixed by the finalized detector vessel construction drawings, all adjustments have to be made on the mirrors. Further effort is needed to optimize the detector performance for the new beam and \v{C}erenkov envelope. We expect only small adjustments will be needed.

In principle, the current estimated photon detection efficiency could be lowered by as much as 30\%, due to reflection losses at each of the optical interfaces between the quartz window and the PMT. The current simulation model counts the survival photons after the quartz window, which does not include the photon loss from the quartz window to the quartz adaptor and from the quartz adaptor to the PMT cathode. As it was explained in the Section \ref{sec_pmt_host_assem}, silica optical gel will be used between the quartz window and the quartz adaptor; RTV compound will be used to make a UV transparent cookie between quartz adaptor and the PMT head. These precautions will hopefully reduce the photon loss to a minimum. The \v{C}erenkov photon loss rate will be determined experimentally with our cosmic ray measurement setup and then implemented in the simulation model.

}

%% file: truck/summary.tex
{
We are constructing a Heavy Gas \v{C}erenkov Detector for the Super High Momentum Spectrometer as part of the Hall C 12~GeV upgrade project at Jefferson Lab. The purpose of the HGC detector is to provide reliable $\pi/K$ separation between 3-11~GeV/c central momenta. The detector pressure is 0.95 atm between 3-7~GeV/c, and gradually reducing at higher momenta. The HGC consists of four sets of curved mirrors and PMTs, where each mirror focuses the \v{C}erenkov photons onto the active area of the corresponding PMT. The detector is designed to be a vacuum vessel which can with stand $10^{-6}$ atm pressure.

The HGC mirrors have dimension of 60 cm $\times$ 55 cm with radius of curvature of 110 cm. Based on our detailed mirror quality study, we conclude that all manufactured HGC mirrors have oblate elliptical curvature: $\kappa > $ 0, and their fitted radius of curvature are slightly larger than desired: R$>$ 110 cm. We were able to select four best mirrors with 0 $<\kappa<$ 1 and radius of curvature around 112 cm, where $\delta\kappa$=$\pm$0.15 and $\delta R$=$\pm$2 cm. The Geant4 simulation results suggest the difference in terms of optical performance between oblate and spherical mirrors is acceptably small.  

The \v{C}erenkov radiation generated when a high momentum charged particle passes through the HGC detector has wavelength in the upper visible to ultraviolet region. Thus, the mirror reflecting surfaces need to be coated with aluminum grain to reflect the UV photons. HGC Mirrors \#2 and \#8 were sent to Evaporated Coating Inc. (ECI) for a test aluminization. Based upon our reflectivity results obtained from the permanent reflectivity measurement setup at Jefferson Lab, we confirm that our reflectivity results are consistent within 2\% with the ECI witness sample reflectivity results between 200-400 nm. We believe the ECI aluminization quality meets our performance specifications and is sufficient for the aluminization of mirrors to be actually used in the HGC. Six more HGC mirrors were aluminized by ECI and have been recently sent to Jefferson Lab, further reflectivity tests will be carried out in Dec, 2012.

A Geant4 Monte-Carlo simulation was constructed according to the HGC design, to study the detector performances such as photon detection and particle identification efficiency. At 7~GeV/c, the HGC optics focuses 96.9\% of the generated \v{C}erenkov photons onto the PMTs active area. At 3 and 7~GeV/c central momenta, the signal threshold is set for two and ten photon electrons, and a pion identification efficiency $\eta>$ 98\% is predicted. The Hall C collabration has recently made further optimization on the SHMS vertical bending dipole elevation, so some modifications are required to our Geant4 model. The optical alignment will be re-optimized according to the new changes, and detector performances will be fine tuned.

The main vessel construction contract has been awarded to HAI Precision Waterjets Inc. \cite{HAI}, their estimated delivery time is November 2012. The HGC construction group will have a six month to assemble the detector before it is shipped to Jefferson Lab in the summer of 2013. We are confident that the HGC detector will function successfully along other particle identification detectors of Hall C for the next 10-20 years.

}

%% file: truck/bibliography.tex
{

}

%% file: truck/appendix.tex
{

\chapter{Reflectivity Data}
\thispagestyle{headings}

\section{AXUV-100G Detector Response Curve}
\begin{figure}[h!]
	\centering
	\includegraphics[width=0.6\textwidth]{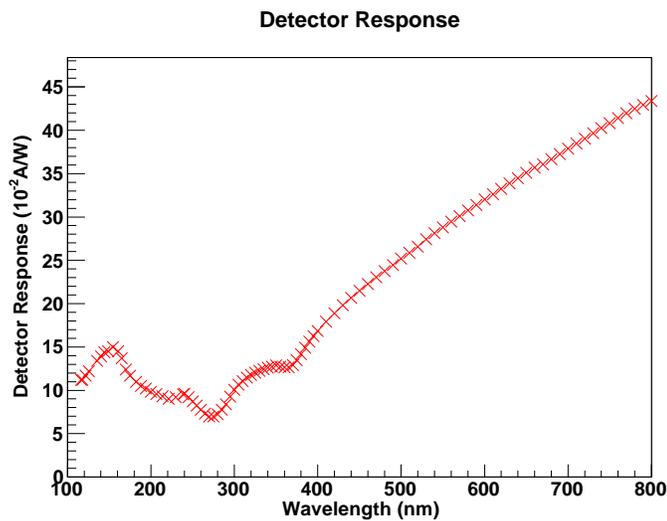}
	\caption[AXUV-100G Photo-diode Detector Response Curve]{IRD AXUV-100G photo-diode detector response curve \cite{ird}.}
	\label{pic_axuv-100G_response_curve}
\end{figure}
The graph above shows the signal response curve for the IRD AXUV-100G detector between 100-800nm. The numerical data are provided by the IRD company. 

\newpage
\section{Typical Melles Griots DUV Mirror Reflectivity Curve}

\begin{figure}[h!]
	\centering
	\includegraphics[width=0.9\textwidth]{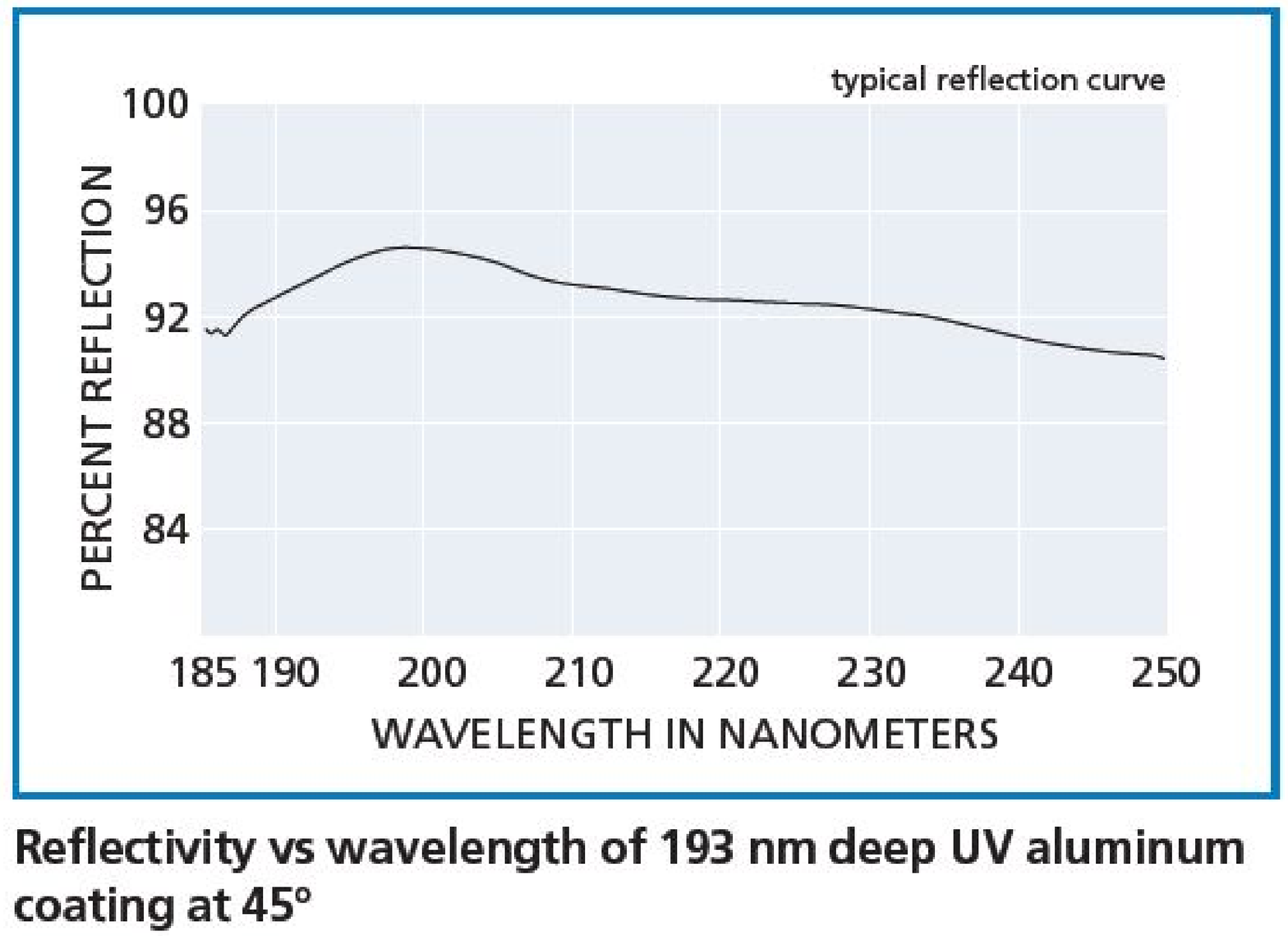}
	\caption[Melles Griots DUV Flipper Mirror Theoretical Reflectivity Curve]{Melles Griots DUV flipper mirror Theo critical reflectivity curve provided by the manufacturer \cite{melles}.}
	\label{duv_spec}
\end{figure}
The graph above shows the typical Melles Griots DUV mirror reflectivity between 185-250nm wavelength.


\newpage

\section{ECI Testing Sample Reflectivity Measurements}

\label{sample_results}


The ECI data sheets are attached in the following three pages, the actual placement of the witness samples with respect to the real mirror are shown in Fig. \ref{sample_pos}. The first page of the data sheet contains the theoretical reflectivity performance; 2nd and 3rd page record the reflectivity measurement of the witness samples.
 
\vspace{3cm}

\begin{figure}[h]
\centering
\includegraphics[trim = 10mm 30mm 20mm 30mm, clip, scale=0.75, angle=270]{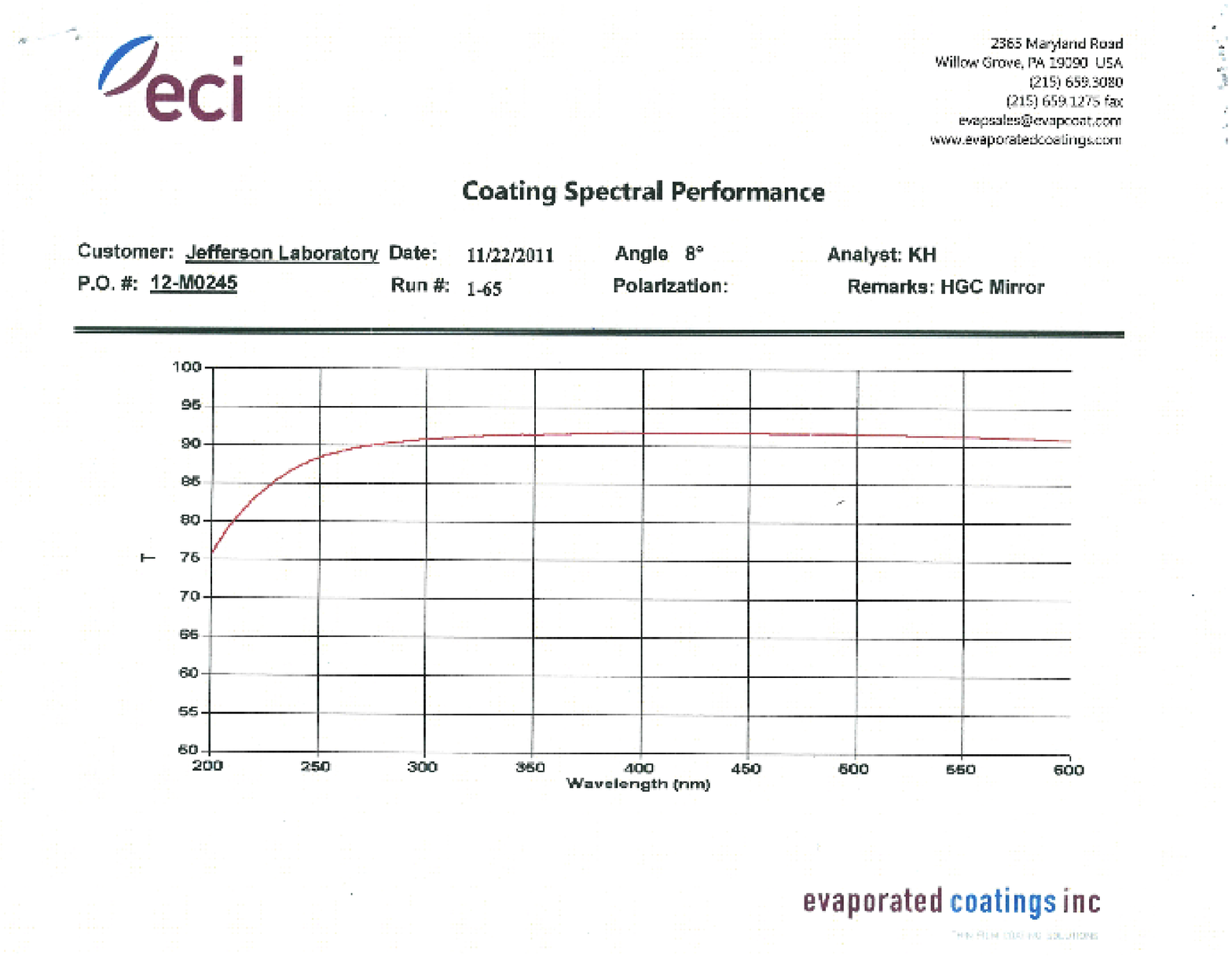}
\caption[ECI Theoretical Reflectivity Curve]{ECI theoretical reflectivity curve \cite{eci}.}
\end{figure}

\begin{figure}
\centering
\includegraphics[trim = 18mm 20mm 25mm 20mm, clip, scale=0.55, angle=270]{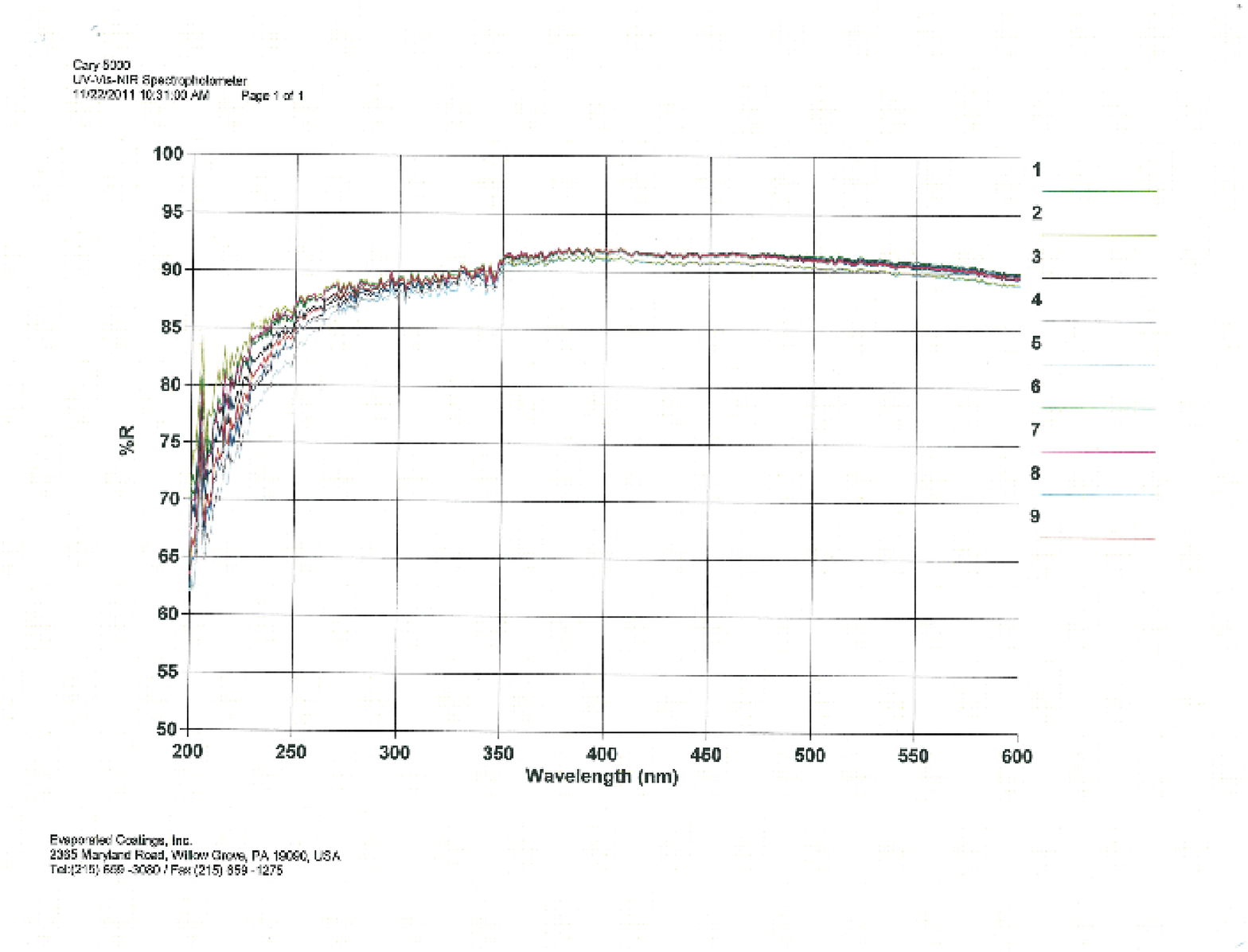}
\caption[ECI Witness Sample Reflectivity Measurements: 200-600 nm]{ECI witness sample reflectivity measurements between 200-600 nm \cite{eci}. \oic}
\end{figure}

\begin{figure}
\centering
\includegraphics[trim = 18mm 20mm 25mm 20mm, clip, scale=0.55, angle=270]{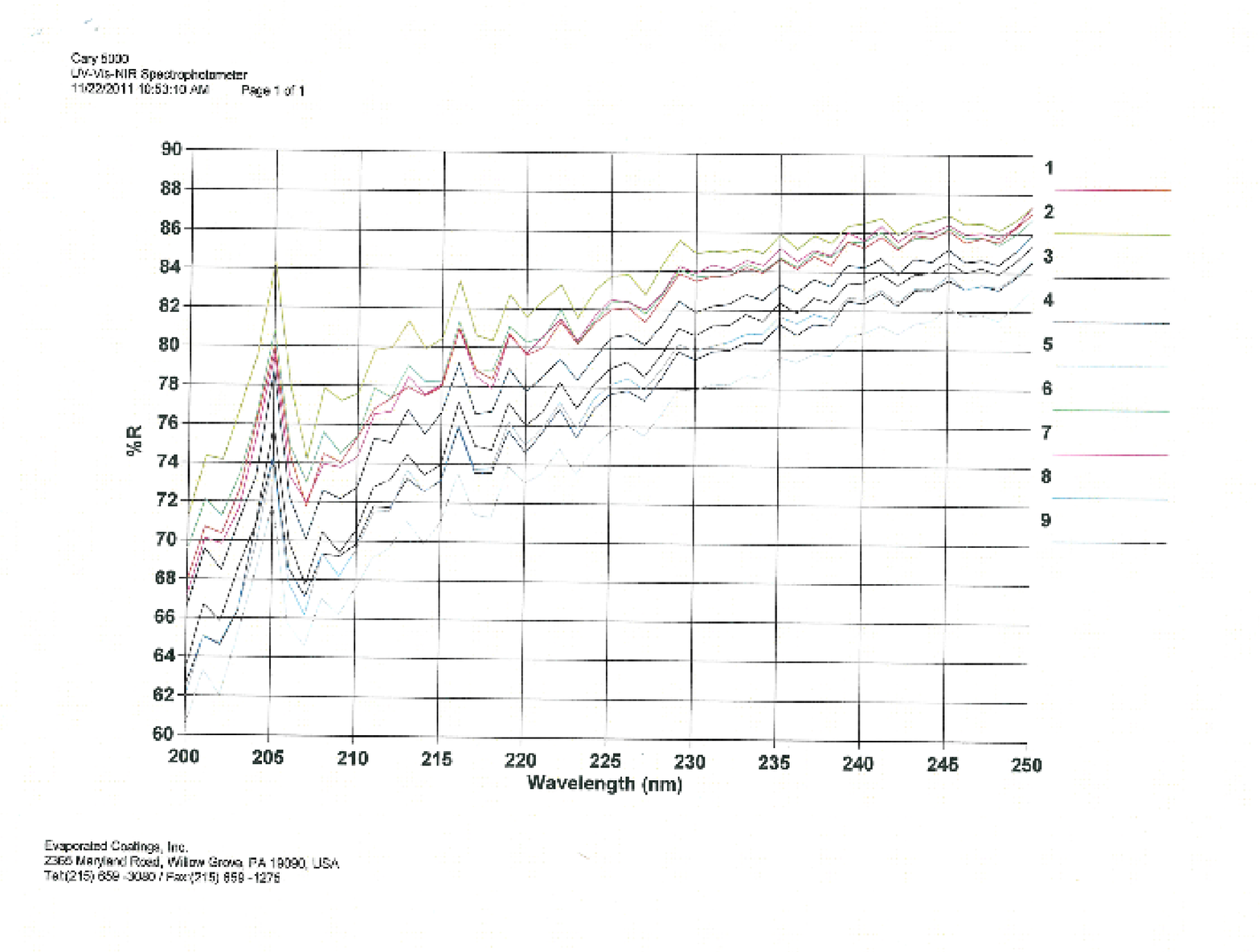}
\caption[ECI Witness Sample Reflectivity Measurements: 200-250 nm]{ECI witness sample reflectivity measurements between 200-250 nm \cite{eci}. \oic}
\label{ECI_sample_result_200-250}
\end{figure}

\chapter{Efficiency Documentation}
\thispagestyle{headings}

\label{app_quartz}

\section{Corning 7980 Quartz Transmissivity}
\begin{figure}[h!]	
\centering
\includegraphics[width=0.7\textwidth]{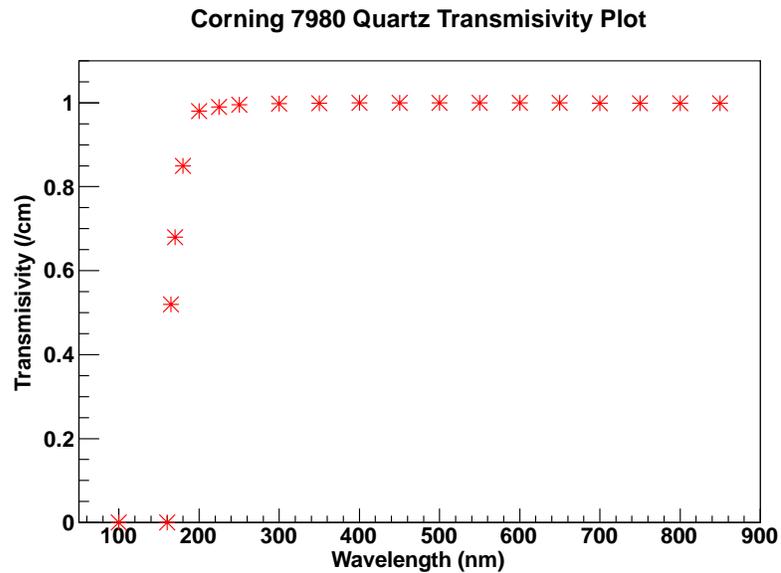}
\caption[Quartz Window Transmissivity]{Corning 7980 quartz wavelength dependent transmissivity \cite{corning}.}
\label{pic_corning_7980}
\end{figure}

Fig. \ref{pic_corning_7980} shows the Corning 7980 quartz wavelength dependent transmissivity. The plot data were converted into the absorption length using absorption equation
\begin{equation}
 \frac{I}{I_{0}} = P = e^{-x/\alpha}\\  
\end{equation}  
where $I_{0}$ is the initial beam intensity, $I$ is the beam intensity at depth $x$ and $P$ is the transmissivity. The absorption length $\lambda$ can be written as
\begin{equation}
\alpha = - x/\ln{P}\,.
\end{equation}

The absorption lengths were embedded into the Geant4 simulation and quartz material depth $x$ is set for 1 cm. Table \ref{tab_corning_quartz} shows the absorption length and transmissivity at different wavelength.

\begin{table}[t]
\centering
\caption[Corning 7980 Quartz Transmissivity and Absorption Length]{Corning 7980 quartz absorption length and transmissivity at different wavelength.}
\begin{tabular}{ccc}
\toprule%
$\lambda$   & Transmissivity &  Absorption Length $\alpha$    \\
(nm)	    &  (\%)          &       (m)                      \\
\toprule                   
100         &      00.00          &   0.00005                     \\     
160         &      00.00          &   0.00005                    \\
165         &      52.00          &   0.01530                     \\
170         &      68.00          &   0.02600                     \\
180         &      85.00          &   0.06200                      \\
200         &      98.00          &   0.49500                      \\
225         &      99.00          &   0.99500                     \\
250         &      99.57          &   2.32000                      \\
300         &      99.80          &   4.99500                      \\
350         &      99.93          &   14.2800                       \\
400         &      100.0          &   100.000                         \\
450         &      99.97          &   33.3280                      \\
500         &      99.96          &   24.9950                      \\
600         &      99.95          &   19.9950                      \\
700         &      99.92          &   12.4950                      \\
800         &      99.90          &   9.99500                       \\
\bottomrule
\end{tabular}
\label{tab_corning_quartz}
\end{table}

\newpage

\section{PMT Radial Dependent Efficiency}

\begin{figure}[h]
	\centering
	\includegraphics[width=0.7\textwidth]{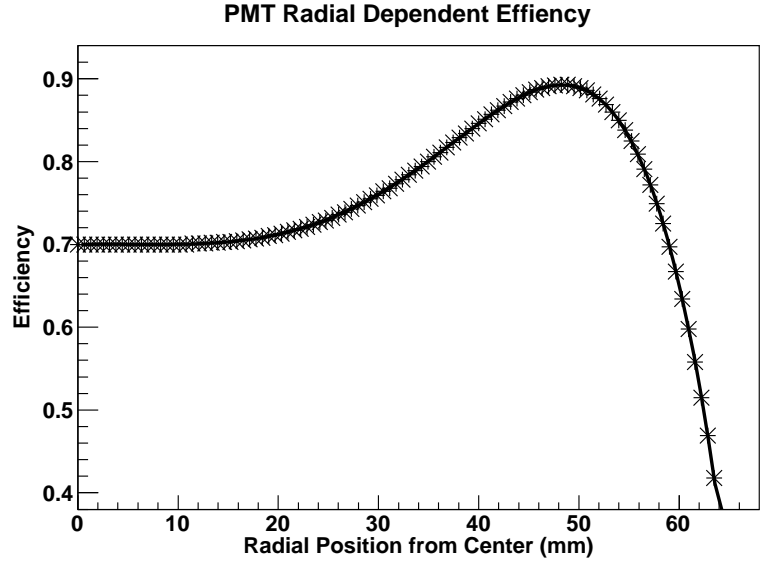}
	\caption[PMT Radial Dependent Efficiency]{PMT Radial Dependent Efficiency \cite{hamamastu}.}
	\label{pic_pmt_radial}
\end{figure}

Fig \ref{pic_pmt_radial} shows the radial dependent efficiency of Hamamatsu R1584 PMT. The data points are the average over 4 set of radial efficiency measurements provided by Hamamastu. A 6th degree polynomial function is used to fit the data and the efficiency can be written as
\begin{equation}
\eta =  P_0 + P_1 x + P_2 x^2 + P_3 x^3 + P_4 x^4 + P_5 x^5 + P_6 x^6
\label{eqn_radial_eff}
\end{equation}
where $x$ is the radial distance from the center of the PMT. The fitted parameter values are as given: 
\begin{center}
\begin{tabular}{ccc}
 $P_0$ &  = &  0.699858             \\
 $P_1$ &  = &  0.000199             \\
 $P_2$ &  = &  -4.8$\times10^{-5}$  \\
 $P_3$ &  = &  2.5$\times10^{-6}$   \\
 $P_4$ &  = &  3.4$\times10^{-8}$   \\
 $P_5$ &  = &  1.1$\times10^{-9}$   \\
 $P_6$ &  = &  -3.8$\times10^{-11}$ \\
\end{tabular}
\end{center}

(\ref{eqn_radial_eff}) is embedded into the HGC Geant4 simulation to provide the PMT radial efficiency.

\newpage

\section{PMT Quantum Efficiency}

\begin{figure}[h!]
	\centering
	\includegraphics[width=0.7\textwidth]{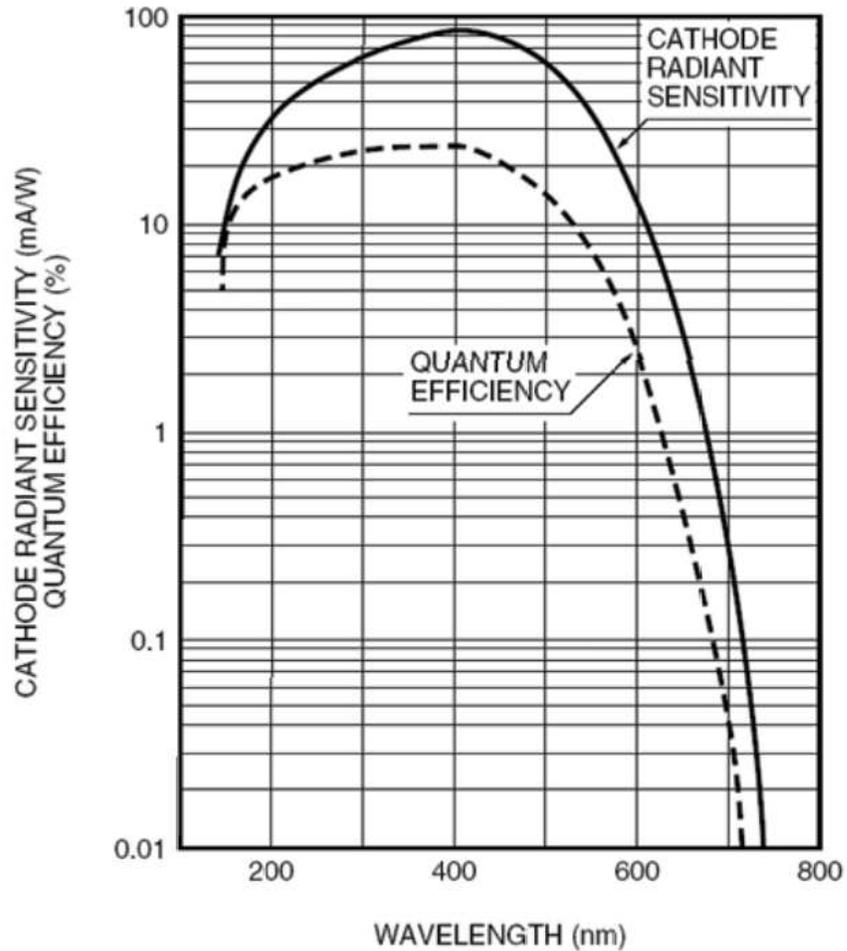}
	\caption[PMT Quantum Efficiency]{Hamamatsu R1584 PMT quantum efficiency \cite{R1250}.}
\end{figure}

The Hamamatsu R1584 PMT wavelength quantum efficiency from the catalog is shown in the figure above. We interpret this curve as the result of both quantum and position dependence in the PMT central region. Since we separately count for the position dependence and the PMT central region has position efficiency of 70\% as shown in Fig. \ref{eqn_radial_eff}, the quantum efficiency curve is corrected up by 43\%. The corrected efficiency curve was included in our Geant4 simulation.

}